\documentclass{pinchcr}   
\usepackage{amsmath}
\usepackage{amsfonts}
\usepackage{mathtools}   
\usepackage{leftidx}     
\usepackage{subfigure}  
\usepackage{float}    

\title{Slow Relaxations and Non-Equilibrium Dynamics in Classical and Quantum
Systems}
\author{Giulio Biroli}
\affiliation{IPhT, CEA/DSM-CNRS/URA 2306, CEA Saclay, F-91191 Gif-sur-Yvette Cedex, France}
\authors{2}

\begin{document}

\maketitle

\preface

These are notes of a series of lectures I gave in 2012 at the Les Houches Summer School of Physics {\it Strongly Interacting Quantum Systems Out of Equilibrium}. The aim of these lectures was to provide an introduction to several important 
and interesting facets of out of equilibrium dynamics. In recent years, there has been a boost in the research on quantum systems 
out of equilibrium. If fifteen years ago hard condensed matter and classical statistical physics remained rather separate research fields, now the focus on several kinds of out of equilibrium dynamics is making them closer and closer. The aim of my lectures 
was to present to the students the richness of this topic, insisting on the common concepts and showing that there is much to gain in considering and learning out of equilibrium dynamics as a whole research field. These notes are by no means self-contained. 
Each chapter is a door open toward a vast research area. I endeavoured to present the most striking or useful or important 
concepts and tools, often just in an informal and introductory way. I hope this will stimulate the interest of the readers that will then have to get specialised reviews and books (cited in the notes) to fully satisfy their curiosity and get a complete knowledge.

\acknowledgements

I wish to thank the organisers of the Les Houches Summer School of Physics {\it Strongly Interacting Quantum Systems Out of Equilibrium} for their invitation to lecture. It was a great opportunity to thoroughly think on several topics on out of equilibrium dynamics.\\ 
Much of what I learned on out of equilibrium dynamics, and that I retransmitted to the students, is due to collaborations and interactions with colleagues and friends. I will not list all their names here (a too long list \dots I had a lot to learn) but I thank them all. \\
Last but not least, special thanks are due to T. Giamarchi and L.F. Cugliandolo for their encouragements (and patience!). 

\tableofcontents

\maintext

\chapter{Coupling to the Environment: What is a Reservoir?}
Depending on the problem one is facing and the fundamental questions one is addressing
a physical system can be considered {\it closed} or {\it open}, i.e. decoupled or not decoupled from the environment. 
Actually, no system is truly isolated; however, some can be considered so on a limited range of timescales over which the coupling to the environment 
does not take place. In the uprising field of non-equilibrium quantum dynamics, which is the focus of this summer school, examples are provided by cold atoms and electrons in solids. In both cases there are situations in which the timescale over which dissipative effects take place is larger 
than the one corresponding to relaxation so that they can be considered isolated to some extent, see Altman's and Perfetti's lecture notes in this book. \\
In statistical physics or condensed matter, one generically deals with open systems that exchange energy and particles with the environment, that henceforth we shall call {\it reservoir} (or bath). Just to cite a few examples, think to solvent molecules for colloidal particles or phonons for electron systems. Of course any open system can always be considered as a subpart of a closed one, 
so actually the real choice whether considering open or closed systems depends on the kind of questions one wants to address.\\
In these lectures we will mainly focus on open systems except for discussing some fundamental issues related to thermalization. 
When studying open systems one declares all the rest of the  
world but the system "the environment" and changes the fundamental dynamical (Newton or Schr\"odinger) laws to take 
into account the role of "the environment" on the system. This {\it reductionist} step 
from closed to open systems is a very subtle one. The aim of this chapter is to unravel the main physical ideas and technical steps that lie below it.
\section{What does a reservoir do to a system?}
Let us start our analysis with some informal considerations. We shall focus on classical physics just to keep the discussion as easy as possible. The dynamics of an isolated 
system is governed by Netwon's law: 
\begin{equation}\label{newton}
m \frac{d^2{ x}}{dt^2}={ F}
\end{equation}
where for simplicity we have written Netwon's equation for a one dimensional particle of mass $m$, position ${x}$ and subject to the external force ${ F}$. There are two main effects 
due to environment:
\begin{itemize}
\item {\bf Dissipation}. Throw a ball in the air and model your experiment with eq. 
(\ref{newton}). What do you get? The ball accelerates (you exert a force on it) and then, once you have thrown it, it should go straight with a constant velocity 
.... forever. What really happens is quite different even if you are a good baseball thrower! The ball inevitably falls down and stops\footnote{As a matter of fact, the longest ball throw was apparently done Gorbous, a Canadian minor leaguer, who in 1957 throw a baseball over 135.6m.}. The initial kinetic energy given to the ball is dissipated in the environment. This effect is general and often is taken into account by adding a friction term to the equation of motion, which in the simplest case reads:
\begin{equation}\label{newton-friction}
m \frac{d^2{ x}}{dt^2}+\eta \frac{d{x}}{dt}={ F}
\end{equation}
where $\eta$ is the strength of the dissipation.
\item{\bf Thermal noise}. There is something missing in the previous description. Statistical mechanics teaches you that the equilibrium distribution of a system is given by the Boltzmann probability law. In particular, for a particle in an external potential $V$, 
the Boltzmann law is $P(x)=\frac 1 Z \exp(-V(x)/T)$ (henceforth the Boltzmann constant will be put equal to one). Consider for example the simplest
case of a quadratic potential $V(x)=\frac 1 2 x^2$ for which $F=-V'(x)=-x$. Equation (\ref{newton-friction}) implies that at long time the particle will sit in the bottom of the potential not moving at all, no matter what the initial condition is. The Boltzmann law instead predicts that the particle position fluctuates around the bottom of the well over a distance of the order $\sqrt{T}$.  
What is missing in eq. (\ref{newton-friction}) is a term corresponding to thermal noise: an environment at temperature $T$ acts on the system with the net effect that energy is exchanged: sometimes the reservoir sucks energy out from the system and sometimes releases to it. Langevin proposed to take this effect into account by adding 
a {\it random force} to the previous equation,  which in the simplest case reads:
\begin{equation}\label{langevin}
m \frac{d^2{x}}{dt^2}+\eta \frac{d{ x}}{dt}={ F}+\xi(t)
\end{equation}
where $\xi(t)$ is a Gaussian field with zero mean and variance $\langle \xi(t) \xi(t') \rangle=2 T \eta \delta(t-t')$. This stochastic equation guarantees that the particle 
thermalizes at long times and that its probability distribution is given by the Boltzmann law for any initial condition. The proof is simple and will be shown in Chapter 3; as an exercise you can verify this result by explicitly solving eq. (\ref{langevin})  
in the quadratic case, $V(x)=\frac 1 2 x^2$, and averaging over the thermal noise. 

\end{itemize}
We have discussed the main effects of the environment in the case of a classical particle in one dimension, an arguably very simple case. It turns out that more general cases
may be more difficult technically---the dissipative term can be represented by retarded friction, the noise can be coloured and even multiplicative and non-Gaussian, and in the quantum case a formalism more involved than stochastic equations must be used---still, the main physical effects, dissipation and thermal activation, are the same and play the same role.\\
An important outcome of the previous analysis is that by taking into account these two effects and by changing the dynamical laws, the particle naturally thermalizes at long times at the temperature $T$ of the environment. 
The fact that the environment is 
at equilibrium at temperature $T$ is actually encoded in the relationship between fluctuation and dissipation. A generic non-equilibrated environment also leads to fluctuation and dissipation. 
However, only for an equilibrated one these are tightly related: the fact that the variance of the noise is 
$2 T \eta$ is an expression of this fundamental law that we shall discuss in more detail later.
\section{The simplest reservoir: linearly coupled Harmonic oscillators}

We now go over the "reductionist step" discussed before, showing how one can obtain 
an open system from a close one by integrating out the degrees of freedom that correspond
to the reservoir. We do it for a simple, instructive and also quite general model: a system coupled to a very large set of Harmonic oscillators. In order to simplify a bit the notation we take the system one-dimensional. \\
The total Hamiltonian reads:
\begin{equation}\label{totalH}
H=H_{syst}+H_I+H_{R}
\end{equation}
where 
\[
H_{syst}=\frac{p^2}{2m}+V(x)
\]
is the Hamiltonian of the system we focus on,
\[
H_R=\sum_k^N \frac{p_k^2}{2}+\frac{\omega_k^2}{2}Q_k^2
\]
is the Hamiltonian of the Harmonic oscillators and
\[
H_I=-\sum_k^N \gamma_k Q_k
\]
is the interaction between the system and the oscillators. The number of oscillators is very large: $N\gg 1$. Moreover, the coupling $\gamma_k$
is such that $\gamma_k=\tilde{\gamma_k}/\sqrt N$, where $\tilde{\gamma_k}$ is of the order of one, i.e. all oscillators interact with the system but very weakly. 
We will discuss the physics behind these two important assumptions later. For the moment, 
we just notice that these choices are the ones that come up naturally if one wants to study 
a system coupled to classical phonons.

\fbox{\parbox{12cm} {%

{\bf Classical Phonons}

\vspace{.1cm}

The phonons Hamiltonian is the one of Harmonic vibrations and reads: 
\[
H_R=\sum_{\ell=-N/2+1}^{N/2} \frac{p_\ell^2}{2}+\sum_{\ell=-N/2+1}^{N/2}\frac{(q_\ell-q_{\ell+1})^2}{2}
\]
where $\ell$ is the lattice site index that runs over $N$ different values (we take periodic boundary conditions). The phonon-system interaction is local and without loss of generality we consider the case in which the system is located in the origin. 
\[
H_I=-q_0x
\]
We shall take a linear coupling, which is correct at small temperature (small fluctuations both of $q_\ell$ and $x$), but more complicated one can be considered too.
It is easy to check that  this model can be rewritten as the previous one: indeed, by going to Fourier space, i.e. defining $Q_k=\frac{1}{\sqrt N}\sum_{\ell=-N/2+1}^{N/2} e^{-i k\ell}q_\ell$, 
all phonons are decoupled and are characterized by the frequency  
$\omega_k^2=2 (1-\cos k)$; each one of them interacts with the system via a coupling constant $\gamma_k=1/\sqrt N$.
}}\\\\\\
The equations of motion for the system read:
\begin{equation}\label{eq-sys}
m\frac{d^2x}{dt^2}=-V'(x)+\sum_k \gamma_k Q_k
\end{equation}
\begin{equation}\label{eq-phon}
\frac{d^2 Q_k}{dt^2}=-\omega_k^2 Q_k+\gamma_k x
\end{equation}

A reservoir that is formed by Harmonic oscillators is simple to treat because it can be integrated out exactly; the generic solution of equation (\ref{eq-phon}) can be readily found:

\begin{eqnarray}
Q_k(t)=&&\frac{\gamma_k}{\omega_k^2}x(t)+\left(Q_k(0)- \frac{\gamma_k}{\omega_k^2} x(0)\right) \cos (\omega_kt)+\nonumber\\
&&\dot{Q}_k(0)\frac{\sin(\omega_k t)}{\omega_k}
-\frac{\gamma_k}{\omega_k^2}\int_0^t \cos(\omega_k(t-s))\dot{x} (s) ds
\end{eqnarray}

By plugging this expression into eq. (\ref{eq-sys}) one obtains:
\begin{eqnarray}
m\frac{d^2x}{dt^2}+&&\int_0^t \underbrace{\sum_k \frac{\gamma_k^2}{\omega_k^2}\cos(\omega_k(t-s))\dot{x} (s) ds}_{Dissipation}=-V'(x)+\underbrace{\sum_k\frac{\gamma_k^2}{\omega_k^2} x(t)}_{Effective\,\,potential}\nonumber\\
&&+\underbrace{\sum_k \left[\dot{Q}_k(0)\gamma_k\frac{\sin(\omega_k t)}{\omega_k}
+\gamma_k\left(Q_k(0)- \frac{\gamma_k}{\omega_k^2} x(0)\right) \cos (\omega_kt)\right]}_{Noise}
\end{eqnarray}

We have found that once the reservoir is integrated out new terms appear in the equation 
of motion for $x$: we have called them dissipation, effective potential and noise for reasons that will be clear in the following. Before discussing them, let us rewrite the previous equation in a more appealing form:
\begin{equation}\label{langevin-full}
m\frac{d^2x}{dt^2}+\int_0^t  K(t-s)\dot{x} (s) ds=-V'(x)-V_{eff}'(x)+\xi(t)
\end{equation}
where $V_{eff}=\frac 1 2 \sum_k\frac{\gamma_k^2}{\omega_k^2} x^2$ and 
$$K(\tau)=\sum_k \frac{\gamma_k^2}{\omega_k^2}\cos(\omega_k \tau)$$
$$\xi(t)=\sum_k \left[\dot{Q}_k(0)\gamma_k\frac{\sin(\omega_k t)}{\omega_k}
+\gamma_k\left(Q_k(0)- \frac{\gamma_k}{\omega_k^2} x(0)\right) \cos (\omega_kt)\right]$$
Now the meaning of each new term should become clear. The one corresponding to dissipation has indeed the form of a retarded friction. It is expected on general grounds since the system releases energy by interacting with the phonons. In our previous discussion of the Langevin equation we didn't add an effective force, $-V'_{eff}(x)$, due to the environment but this of course has to be expected too and is indeed what we find.  
The final term---the noise---is more problematic than the others. As its name makes clear, 
it should be characterized by some kind of randomness. However, as its definition makes also clear is perfectly deterministic. Where is the subtlety? We postpone a complete discussion to the following section; for the moment we just notice that for typical values of $Q_K(0)$ and $\dot{Q}_k(0)$, i.e. values extracted from the equilibrium Boltzmann measure, the function $\xi(t)$ behaves so erratically that in practice is undistinguishable from a random function if $N$ is moderately large ($N>20$);  moreover $\xi(t)$ contains indeed some randomness since   
 $Q_K(0)$'s and $\dot{Q}_k(0)$'s are random.\\
In the following we assume that the reservoir, i.e. the phonons, are at equilibrium at time $t=0$ given the position $x(0)$ of the system. In consequence their probability measure reads:
\[
P\left[\{Q_k(0),\dot{Q}_k(0)\}|x(0) \right]={\mathcal N}\exp\left(-\sum_k\left[\frac{1}{2T} \dot{Q}_k(0)^2+\frac{\omega_k^2}{2T}{Q}_k(0)^2-\frac 1 T \gamma_k {Q}_k(0) x(0)\right]\right) 
\]
where ${\mathcal N}$ is the normalization constant. By redefining the normalization constant
the previous expression can be rewritten as:
\[
P\left[\{Q_k(0),\dot{Q}_k(0)\}|x(0) \right]={\mathcal N}'\exp\left(-\sum_k\frac{1}{2T} \dot{Q}_k(0)^2-\sum_k\frac{\omega_k^2}{2T}\left[{Q}_k(0)-\frac{\gamma_k}{\omega_k^2} x(0)\right]^2\right) 
\]
This simple rewriting reveals that $\dot{Q}_k(0)$ and ${Q}_k(0)-\frac{\gamma_k}{\omega_k^2} x(0)$ are independent Gaussian variables of zero mean and variance $T$ and $T/\omega_k^2$ respectively.
It is now straightforward to show that $\xi(t)$ is a Gaussian random function, since it is a sum of Gaussian variables; its mean and variance are given by:
\[
\langle \xi(t) \rangle =0 \qquad \langle \xi(t) \xi(t') \rangle=TK(t-t')
\] 
This completes the "reductionist" step from the close to the open system.  
The main results we found are a first principle derivation of the Langevin equation
and that dissipation and fluctuations (or noise) are related: the memory function appearing in the retarded friction term is exactly equal to the variance of the noise times the temperature\footnote{The reader could be surprised since we got a factor of two less with respect to the Langevin equation introduced in the first section. There is no contradiction actually. The subtlety is that in the limit where $K$ becomes a delta function, only half of the delta function contributes to the integrated friction. Thus, one has to take $K(\tau)\rightarrow 2 \eta \delta(\tau)$ to get a friction term $\eta \dot{x}$. In this case one recovers that the variance of the noise is indeed $2T\delta(\tau)$ as expected. }. 

\section{Noise and irreversibility}
Let us now discuss in more detail the noise term and to what extent it is random. 
In a real experiment the initial condition of the system an the environment is fixed, in particular $\dot{Q}_k(0)$ and ${Q}_k(0)$ are fixed; 
in consequence, $\xi(t)$ is a completely deterministic function. Hence, one could wonder 
to what extent $\xi(t)$ does represent random noise. Well, it is known that deterministic function can be identical for any practical purposes to a random one. 
Think for example to the random number generators one uses in a Monte Carlo simulation: 
these are deterministic functions producing numbers that one considers random for all practical purposes. In reality these are not truly random because the sequence one obtains repeats after a certain number of trials. The better is the random number generator the longest is the period. The "noise" $\xi(t)$ behaves in a similar way; the larger is $N$ the more it resembles to a random function. In the limit $N\rightarrow \infty$ it is indeed indistinguishable from a random function\footnote{The frequencies $\omega_k$ should be incommensurate in order for the mapping to a random function to hold.}. In order to better clarify this last statement, let us recall that there are two ways of doing averages:
\begin{itemize}
\item {\bf Time Average.} In this case, given an observable $O(x(t))$, one computes its average as
\[
\langle O \rangle_T = \lim_{\tau\rightarrow \infty} \frac 1 \tau \int_0^\tau O(x(t'))dt'
\]
This is what one does in an experiment assuming that the system has reached a steady state. 
\item {\bf Ensemble Average.} In this case, given an observable $O(x(t))$, one computes its average by averaging over all possible realizations belonging to the ensemble. In the case we are considering this means averaging over all possible initial conditions for the oscillators, 
i.e. over the values of $\dot{Q}_k(0)$ and ${Q}_k(0)$. This leads to an average over $\xi(t)$
which then indeed becomes a random function:
\[
\langle O \rangle_E = \lim_{\tau\rightarrow \infty} \int \prod dQ_k(0)
d\dot{Q}_k(0)O(x(\tau))P\left[\{Q_k(0),\dot{Q}_k(0)\}|x(0) \right]
\]
\end{itemize}
The previous statement about the indistinguishability of $\xi(t)$ from a random function 
means that time averages and ensemble averages coincide. In particular 
\[
\langle \xi(t) \xi(t') \rangle_T=\langle \xi(t) \xi(t') \rangle_E=TK(t-t')
\]
As stated previously, the parameter controlling how much the pseudo-noise $\xi(t)$ behaves like a true noise is $N$, the number of degrees of freedom in the environment, see \cite{zwanzig,mazur} for more details.\\
We arrived to the conclusion that for $N$ large enough our deterministic open system is indistinguishable from a stochastic one based on the Langevin equation (\ref{langevin-full}). 
It is easy to prove, as we shall show later on, that a system obeying such an equation thermalizes at long time, {\it i.e.} it goes to a steady state where the distribution is Maxwell-Boltzmann and all equilibrium relationships, like fluctuation-dissipation, are satisfied.\\
Just to highlight how subtle is this reduction from a closed to an open system, let us discuss an apparent paradox. 
Consider as a system a Harmonic oscillator, $V(x)=\frac 1 2 k x^2$, and think to what the previous statement means: the oscillator thermalizes at long time for any initial condition. Isn't it weird? The global closed system is just a collection of coupled Harmonic oscillators---the easiest integrable system one can think of! And, as you know, integrable systems do not thermalize. Where is the catch?
The subtle point is that one does not have to resort to any chaotic dynamics to get irreversibility and equilibration. There is nothing surprising once one realizes that the fact of having an environment with a very large number of degrees of freedom is the key ingredient: when $N=4$ one gets only quasi-periodic motion and no equilibration at all, but for $N\gg1$ 
the quasi-periodicity is pushed up to times that are exponentially large in $N$, see \cite{zwanzig}. \\
It is instructive to disentangle two processes that take place when a system reaches equilibrium: irreversible behavior and steady state properties given by Maxwell-Boltzmann distribution. The former is in general related to reaching a steady state and to damping. 
An environment formed by a very large number of degrees of freedom leads to irreversible behavior and generically to a steady state but only one that is at equilibrium leads to true thermalization (and this independently of its chaoticity, $N\gg1$ is enough). 
See \cite{zwanzig,vulpiani} for a general discussion on the relationship between irreversibility, thermodynamic limit and red herrings related to chaos.
\section{The environment}
By looking at the Langevin equation (\ref{langevin-full}) one realizes that the only characteristic of the environment one needs to know is $K(t)$. In order to describe $K(t)$, 
it is useful to introduce the spectral function 
$$J(\omega)=\frac 1 N \sum_k \delta(\omega-\omega_k) \frac{\tilde{\gamma}_k^2}{\omega_k}$$
and rewrite $K(t)=\int_0^\infty d\omega \frac{J(\omega)}{\omega}\cos(\omega t)$. Environments are classified on the basis of the low frequency behavior of $J(\omega)\simeq \eta \omega^s$: an Ohmic bath corresponds to $s=1$ whereas a sub(super)-Ohmic one
corresponds to $s<1$($s>1$) \cite{weiss}. Note that a purely Ohmic bath leads to $K(t)=\eta \delta(t)$. 
Of course, in reality, there is always a change of behavior at high frequency and, hence, the delta function acquires a finite width.\\
Before concluding this chapter on the role of the environment let us stress that the previous 
procedure of integrating out the environment can be repeated in the quantum case as well. Indeed it is at the root of first principle descriptions of quantum open systems: examples are small quantum systems coupled to phonons, quantum dots coupled to non-interacting leads, etc...\\
We postpone a complete discussion of quantum reservoirs to the following chapters. However, we anticipate that it is not possible in general to obtain a stochastic equation as one can do for classical systems. In the literature there are quantum versions of the Langevin equation (\ref{langevin-full}) which are identical to the classical one except for the fact that the dissipation and fluctuations are related by the {\it quantum} fluctuation-dissipation relation. These are only approximations valid for small quantum fluctuations as we shall explain later on.
For more discussion and details on open systems and the role of reservoirs, see the books \cite{zwanzig,vulpiani,weiss,gardiner}. 

\chapter{Field Theory, Time-Reversal Symmetry and Fluctuation Theorems}
In the following we present a brief introduction and a derivation of the field theories
used to study off-equilibrium dynamics in classical and quantum systems. We then go on studying a particularly important symmetry for equilibrium systems: time-reversal. 
Out of equilibrium systems violate this symmetry but, quite interestingly, this violation
leads to a variety of interesting consequences that go under the name of fluctuation theorems
and that received a lot of attention recently. We shall discuss them and conclude the section by addressing generalizations to the quantum case. 
\section{Martin-Siggia-Rose-DeDominicis-Janssen field theory}
We consider an open classical system whose dynamics is governed by a stochastic equation. 
For the sake of concreteness and for simplicity we focus on a simplified version of the Langevin equation (\ref{langevin-full}):
\begin{equation}\label{langevin-msr}
\eta \dot{x}(t)=-V'(x(t))+\xi(t) \qquad \langle \xi(t) \xi(t') \rangle=2T\eta \delta(t-t')
\end{equation}
We neglected inertia and memory effects. This is reasonable in certain physical systems
such as e.g. colloids in which the system (colloidal particles) are much bigger and heavier (and hence much slower) than the particles forming the environment (solvent molecules). We present the derivation in this case, but also discuss the results we would have obtained in more general cases at the end. Note that  we still focus on one dimensional systems just to keep the notation as less involved as possible (to this end we also set $\eta=1$, i.e. we reabsorb $\eta$ by redefining the unit of time). It is straightforward to generalize the results to higher dimensions. \\
Studying the dynamics of a system means computing (or measuring) averages, correlation and response functions. Imagine we want to compute the average value of the observable
$O(x(t))$, where $O(x)$ is a generic function of $x$:
\begin{equation}\label{scaling}
\langle O(x(t)) \rangle=\int D[\xi]dx_0O(x_\xi(t))\exp\left(-\frac{1}{4T}\int_0^t\xi^2(t')dt'\right)P(x_0)
\end{equation}
What the previous functional integration means in words is that one has to solve eq. (\ref{langevin-msr}) for a given $\xi(t)$ and initial condition $x_0$, compute $O(x(t))$ and then average the result over all possible Gaussian thermal noises and initial conditions\footnote{Note that the average in the previous equation is the ensemble average, cf. the previous chapter. Henceforth we shall neglect the subscript $E$ to lighten the notation.}. 
Formally, we can insert in the previous integral a useful representation of the unity by integrating over all paths $x[t]$ the functional Dirac $\delta$ which imposes eq. (\ref{langevin-msr}):
\[
\int_{x(0)=x_0} D[x] \delta[\dot{x}+V'(x)-\xi]=\int_{x(0)=x_0} D[x] D[\hat{x}]e^{-\int_0^t dt' \hat{x}(t')\left(\dot{x}(t')+V'(x(t'))-\xi(t')\right)}=1
\] 
Actually, a Jacobian should be also present but since it can be proved to be a constant we just absorb in the normalization of the functional integral \footnote{This is actually a tricky business, see e.g. \cite{cardybook,Zinn,kamenev}. In order to compute the Jacobian one has to specify the discretization of the Langevin equation. We choose the Ito discretization $x_{t+\Delta t}-x_t=(-V'(x_t)+\xi_t)\Delta t$. In this case it is easy to show that the Jacobian is indeed a constant. Other discretizations lead to different results. As long as the noise is not multiplicative, different discretizations lead to different field theories but physical averages are the same. In case of multiplicative noise instead different discretizations correspond to different physical processes. }. This insertion is useful because it allows one to integrate out the noise---just a functional Gaussian integral---and obtain the famous Martin-Siggia-Rose-DeDominicis-Janssen (MSRDJ) field theory:
\begin{equation}\label{MSR}
\langle O(x(t)) \rangle=\int D[x]D[\hat{x}]dx_0O(x(t))\exp(S)P(x_0)
\end{equation}
where the action $S$ reads:
\begin{equation}\label{MSR}
S=-\int_0^t dt' \hat{x}(t')\left[\dot{x}(t')+V'(x(t'))-T\hat{x}(t')\right]
\end{equation}
Using this field theory one can also compute correlation functions $\langle O(x(t_1))  O(x(t_2)) \rangle$ and response functions as discussed below. 
\section{Meaning of the fields}
The MSRDJ field theory contains two fields: $x[t]$ and $\hat{x}[t]$. The meaning of the former is clear: it represents the fluctuating degrees of freedom of the system.  The latter is called response field, see \cite{cardybook}, because it intervenes in response functions. To see how this works let's add an external field coupled to $x(t)$: $V(x)\rightarrow V(x)-hx$. By repeating the previous derivation one finds that the action gets an extra term: 
 \begin{equation}\label{MSR-R}
S=-\int_0^t dt' \hat{x}(t')\left[\dot{x}(t')+V'(x(t'))-T\hat{x}(t')\right]+\int_0^t dt' \hat x(t') h(t')
\end{equation}
The response function measuring the variation of the average of $x(t)$ due to an infinitesimal
and istantaneous external field acting at time $t'$ can be expressed in term of a correlation function of the fields $x$ and $\hat x$:
\[
R(t,t')=\left.\frac{\delta\langle x(t)\rangle}{\delta h(t')}\right|_{h=0}=\langle x(t) \hat{x}(t')\rangle
\]
Note that from this identity one can immediately obtain that the average of $\hat x$ is zero since the average of one does not change if one applies a field! \footnote{$\langle \hat x \rangle=\left.\frac{\delta \langle 1\rangle}{\delta h(t')}\right|_{h=0}=0$} Similarly, one can find that correlation functions with only $\hat x$ or in which the field with largest time is a response field vanish by causality. 
\section{Generalizations} 
We now report the resulting MSRDJ field theory one would have obtained starting from the 
full Langevin equation (\ref{langevin-full}). The derivation, which is a straightforward generalization of the previous one, is left as an exercise for the reader. One still ends up with a field theory on $x,\hat x$; the difference is in the form of the actions which reads:
\begin{eqnarray}\label{MSRDJ-full}
 S&=&-\int_0^\infty dt' \hat{x}(t')\left[m\ddot{x}(t')+V'(x(t'))\right]-\\
 &&\int_0^\infty dt  \int_0^t dt' 
 \hat x(t) \eta (t-t') \dot{x}(t')+\frac 1 2 \int_0^\infty dt  \int_0^t dt' 
 \hat x(t) \nu (t-t') \hat{x}(t')\nonumber
\end{eqnarray} 
where $\eta (t-t')=K(t-t')$ is the retarded friction and $\nu (t-t')=TK(t-t')$ the variance of the noise. Because of the inertia one has now to specify both $x(0)$ and $\dot{x}(0)$ as initial 
conditions. For more details see \cite{kamenev,aronS,kurchan}.
\section{Relationship with Schwinger-Keldysh field theory}
We now want to make a relationship between the MSRDJ field theory used to study the dynamics of classical open systems and the Schwinger-Keldysh one used for open quantum systems. As we have shown in the previous chapter the Langevin equation that we used as 
a starting point to derive MSRDJ is obtained starting from a sub-system coupled to an environment, the combination of the two being a closed system characterised by the Hamiltonian (\ref{totalH}). The starting point to derive the  Schwinger-Keldysh field theory is to write down a path integral representation for the dynamics of the sub-system coupled to the reservoir. Then, at the level of the functional integral, one can integrate out the reservoir degrees of freedom and obtain an action for the sub-system only. This is shown in the lectures by Aleiner and Berges so we shall not repeat it, see also \cite{kamenev}. We will only quote the main results. \\
In the closed contour representation of the functional integral one has two fields $x^+(t)$
and $x^-(t)$ corresponding to the two forward and backward time paths. In order to make the relationship with MSRDJ explicit, it is useful to make a change of variable and define:
\[
x(t)=\frac{x^+(t)+x^-(t)}{2}\qquad \hat x(t)=i\frac{x^+(t)-x^-(t)}{\hbar}
\]
In the literature these are also called classical and quantum field respectively. In terms of these fields the action of the Schwinger-Keldysh field theory can be written as:
\begin{eqnarray}\label{SKFT}
S&&=\frac{i}{\hbar}\left({\mathcal L}[x-i\hbar \hat x /2]- {\mathcal L}[x+i\hbar \hat x /2]\right)\\
&&\int_0^\infty dt  \int_0^t dt' 
 \hat x(t) \eta (t-t') \dot{x}(t')+\frac 1 2 \int_0^\infty dt  \int_0^t dt' 
 \hat x(t) \nu (t-t') \hat{x}(t')\nonumber
\end{eqnarray}
where $\mathcal L$ is the Lagrangian of the sub-system:
\[
{\mathcal L}[x]=\int_0^t m\frac{\dot{x}^2}{2}-V(x)
\]
and $\eta (t-t')$, which is equal to $K(t-t')$ as in MSRDJ, has the interpretation of a retarded friction; the other term, $\nu (t-t')$, plays the role of the variance of the thermal noise and 
is related to $\eta$ by the quantum fluctuation dissipation theorem \cite{weiss}. In Fourier space their
relationship reads:
\begin{equation}\label{QFDT}
\nu(\omega)=\hbar \coth \left( \frac{\hbar \omega}{2T}\right) \Im \eta(\omega)
\end{equation}
The properties of the environment are all encoded in the specific form of $\eta(\omega)$.
It is easy to check that MSRDJ field theory defined by the action (\ref{langevin-full}) is the classical limit of the Schwinger-Keldysh field theory. Indeed by taking the limit $\hbar \rightarrow 0$ the quantum fluctuation-dissipation relation
becomes the classical one, {\it i.e.} $\nu(\tau)\rightarrow TK(\tau)$ so that the second part of the action (\ref{SKFT}) coincides with the corresponding one of MSRDJ. Concerning the first part, by taking the limit $\hbar \rightarrow 0$ only the term linear in $\hat x$ survive and one 
indeed finds:
\[
\lim_{\hbar \rightarrow 0}\frac{i}{\hbar}\left({\mathcal L}[x-i\hbar \hat x /2]- {\mathcal L}[x+i\hbar \hat x /2]\right)=-\int_0^\infty dt' \hat{x}(t')\left[m\ddot{x}(t')+V'(x(t'))\right]
\]
Thus, the quantum nature of the Schwinger-Keldysh field theory is encoded in: (1) the relationship between $\eta$ and $\nu$, (2) the existence of terms with possibly all odd powers 
of $\hat x$. In general one cannot interpret the action (\ref{SKFT}) as one originating from a physical stochastic equation. However, if one disregards all odd terms beyond the linear one then the corresponding field theory is MSRDJ-like for a Langevin equation (\ref{langevin-full}) with a noise related to the friction via the quantum dissipation relation. This stochastic equation is the so-called quantum Langevin equation. It corresponds to the first term in the expansion in the quantum field $\hbar\hat x$ and, hence, is valid only in the regime of small quantum fluctuations. See \cite{weiss} for more material on quantum Langevin equations. 
\section{Time-reversal symmetry}
Time-reversal is {\it the symmetry} that characterizes an equilibrium system: being at thermodynamic equilibrium means being unable to determine the arrow of time.\\
 In the following we shall show how this symmetry arise within the MSRDJ formalism. 
 Consider a trajectory $x[t]$ starting from $x_0$ at time $t=0$ and reaching $x_f$ at time 
 $\tau$. The time-reversed trajectory of $x[t]$ is defined as 
 $x_R(t)=x(\tau-t)$. Although the response field is not a physically measurable field, it is useful to also define its time-reversed counterpart: $\hat x _R(t)=\hat x(\tau-t)+\frac 1 T \frac{dx(\tau-t)}{d(t)}$. Let us study now how the action transforms under time-reversal. First, we shall rewrite it as:
  \begin{eqnarray}\label{MSR2}
S(x[t],\hat x[t])&=&-\int_0^t dt' \hat{x}(t')\left[\dot{x}(t')+V'(x(t'))-T\hat{x}(t')\right]=\\
&&-\int_0^t dt' \hat{x}(t')V'(x(t'))+T\left(\hat{x}(t')-\frac{\dot x^2}{2T}\right)^2+\frac{\dot x^2}{4T}\nonumber
\end{eqnarray}
The second and third term are clearly invariant under time reversal whereas the first time 
leads to an extra constribution:
\[
S[x_R[t],\hat x_R[t]]=S[x[t],\hat x[t]]+\frac{1}{T}\int_0^\tau \dot x(t')V'(x(t'))dt'
\] 
The probability to measure the time-reversed trajectory for the system, given the initial condition $x_f$ at time 0 reads:
\[
P(x_R[t]|x_R(0)=x_f)=\int D[\hat{x}_R]\exp(S[x_R[t],\hat x_R[t]])
\]
By applying the time-reversal transformation which amounts just to a functional change of variables (whose Jacobian is one) one finds:
\[
P(x_R[t]|x_R(0)=x_f)=\int D[\hat{x}]\exp\left(S[x[t],\hat x[t]]+\frac{1}{T}\int_0^\tau \dot x(t')V'(x(t'))dt'\right)
\]
By separating the two terms in the exponential, performing the integration over time for the second and formally integrating over $\hat x$, we find
\begin{equation}\label{trseq}
P(x_R[t]|x_R(0)=x_f)=P(x[t]|x(0)=x_0)e^{\frac{V(x_f)-V(x_0)}{T}}
\end{equation}
By dividing both terms by the partition function $Z=\int dx e^{\frac{V(x)}{T}}$, 
bringing $e^{\frac{V(x_f)}{T}}$ on the RHS and recalling that the equilibrium probability measure is $P_{eq}(x)=e^{\frac{-V(x)}{T}}/Z$ we finally find the expression of time-reversal symmetry
\[
P(x_R[t]|x_R(0)=x_f)P_{eq}(x_f)=P(x[t]|x(0)=x_0)P_{eq}(x_0) 
\]
which simply means that the probability of starting from $x_0$ in equilibrium and following a given path to $x_f$
coincides with the probability of starting from $x_f$ in equilibrium and following the time-reversed path. 
This is a general property of equilibrium dynamics. We showed it for Langevin dynamics, i.e. 
a classical system coupled to a bath, but it is generically valid for all physical equilibrium dynamics, 
in particular Netwon and Schr\"odinger evolutions. \\
A general principle in physics is that the existence of symmetries 
implies non-trivial identities for correlation functions and observables: the most famous example being momentum and angular momentum conservations laws and invariance with respect to translations and rotations. 
We thus expect that time-reversal symmetry leads to important consequences. Indeed three fundamental relationships that are generically taken as fundamental signature of equilibrium dynamics can be derived from it: time-reversal symmetry of correlation functions, fluctuation-dissipation relations and Onsager reciprocity relations. 
\\\\\\
\fbox{\parbox{12cm} {%

{\bf Consequences of time reversal symmetry}

\vspace{.1cm}
\begin{itemize}
\item {\it Time-reversal symmetry of correlation functions}: $\langle O_1(t_1)\cdots O_n(t_n)\rangle=\langle O_1(-t_1)\cdots O_n(-t_n)\rangle$ where $O_i(t_i)$s are observables at time $t_i$.  
This is a direct implication of the fact that it is not possible to identify microscopically an arrow of time: the direct and time-reversed path have the same probability. 
\item {\it Fluctuation dissipation relations}: $R(t-t')=\frac 1 T \partial_{t'}C(t-t')$ where 
$C(t-t')=\langle x(t) x(t')\rangle$ and $R$ is the response of $\langle x(t)\rangle$ to a field applied at time $t'$.
\item {\it Onsager reciprocity relations}: $R_{AB}(t-t')=R_{BA}(t-t')$ where $R_{AB}$ denotes the response of $\langle A(t)\rangle$ when a field conjugated to the observable $B$ is applied at time $t'$.
\end{itemize}
}}\\\\\\
These three identities can be easily proven following the procedure outlined above and performing the change of path 
in the functional integral. 
They are identities between correlation functions implied by the time-reversal symmetry (formally the response function is a correlation function between the physical and the response fields). Their proof is left as an exercise to the reader. 
Note that one also needs to use that the system is in a steady state, i.e. it is time-translation invariant, which means that indeed it is at equilibrium (the only steady state possible for a finite system coupled to an equilibrium reservoir is equilibrium dynamics, as we shall discuss in the next chapters).  
\section{Time-reversal symmetry breaking and fluctuation relations out of equilibrium}
In the last fifteen years it has been realised that studying the way in which time-reversal symmetry is broken out of equilibrium leads to new kind of fluctuation relations. These are valid out of equilibrium and belong to what is now called
the field of stochastic thermodynamics, see e.g. \cite{seifert}. This topic deserves a series of lectures on its own, which is certainly out of the scope 
of these notes. On the other hand, one cannot certainly avoid to mention it. With this in mind in the following I shall just discuss one of the most known relations, discuss its physical origin in a nutshell and then point out
complete reviews for the interested readers. As the identities discussed before at equilibrium, these out of equilibrium relations hold for all types of physical dynamics. Our analysis will be performed using Langevin dynamics.  \\
As setting for out of equilibrium dynamics we choose: 
\[
\dot x(t)=F(t)+\xi(t) \qquad \langle \xi(t) \xi(t')\rangle=2T\delta(t-t') 
\]
where $F(t)$ is a generic time dependent force that we write as $F(t)=-V'(x(t),\lambda(t))+f(t)$. 
This representation means that the system is out of equilibrium for two reasons: a control parameter of the 
external potential is changed during the dynamics and moreover there is a non-conservative force applied 
to the system\footnote{The force is non-conservative because it depends explicitly on time 
and, moreover, it is generically assumed to not derive from a potential (for this latter statement one needs to consider spatial dimensions higher than one since in one dimension a generic function of $x$ can always be written as the derivative of a potential).}. This is a very simple model but it contains the essence of the problem 
(treatments for more complicated models are straightforward generalizations). Repeating the derivation that lead to eq. (\ref{trseq}) one now finds:
\begin{equation}\label{trsoeq}
P(x_R[t]|x_R(0)=x_f)=P(x[t]|x(0)=x_0)e^{-\int_0^\tau dt \frac{\dot x F}{T}}
\end{equation}
The term $\int_0^\tau dt {\dot x F}$ has the interpretation of the heat exchanged between the system and the 
bath \cite{seifert}. To see where this interpretation comes from, remember the first law of thermodynamics (applied in a small interval of time $dt$): $dw=dV+dq$, the work done on the system is equal to the change in energy plus the heat exchanged. The work done on the system reads from the mechanic point of view: $dw=\frac{\partial V}{\partial \lambda} \dot \lambda dt+f dx$. Using the chain rule   
 $dV=\frac{\partial V}{\partial \lambda} \dot \lambda dt+\frac{\partial V}{\partial x} dx$ we can eliminate $dw$ from the two previous equations and obtain $dq=\left( -\frac{\partial V}{\partial x}+f\right) dx=Fdx$. \\
 From the relationship above all fluctuation relations valid out of equilibrium can be derived straighforwardly. 
In the following we show how this works in a specific case and derive the Jarzinsky-Crooks identities.  \\
Consider the specific out of equilibrium protocol where the system is at equilibrium at $t=0$ and is brought out of equilibrium by the changing of the external parameter $\lambda(t)$. We take $f=0$, i.e. $F=-V'(x(t),\lambda(t))$. The term in the exponential of eq. (\ref{trsoeq}) can be rewritten in this specific case as:
\[
-\frac 1 T \int_0^\tau dt \dot x F= \int_0^\tau dt\left(\frac{dV}{dt}- \frac{\partial V}{\partial \lambda} \dot \lambda\right)=
\frac{V(x_f,\lambda(\tau))-V(x_0,\lambda(0))}{T}-\frac{W}{T}
\]
where $W=\int_0^\tau \frac{\partial V}{\partial \lambda} \dot \lambda dt$ is the work performed on the system between time $0$ and $\tau$. Note that $W$ is a stochastic quantity 
since it is a functional of the dynamical trajectory. Multiplying eq. (\ref{trsoeq}) by the Boltzmann measure that the system would have reached if it was equilibrated at $\tau$, i.e. $e^{-V(x_f,\lambda(\tau))/T}/Z_f$ and multiplying the RHS by $Z_i/Z_i$ we finally get the equation:
\begin{equation} \label{trs2oeq}
P(x_R[t]|x_R(0)=x_f)P_{eq}(x_f)=P(x[t]|x(0)=x_0)P_{eq}(x_0) e^{-\frac{W}{T}+\frac{\Delta F}{T}}
\end{equation}
where $\Delta F=-\log (Z_f/Z_i)/\beta$ is the difference in the free energies corresponding to the two equilibrium states characterized by value of $\lambda$ equal to  $\lambda(\tau)$ and $\lambda(0)$. \\
Let's now integrate eq. (\ref{trs2oeq}) over all paths. This leads to the celebrated Jarzynski identity, see \cite{Jarzynski}:
\begin{equation}\label{jarz}
\langle e^{-\frac{W}{T} }\rangle =e^{-\frac{\Delta F}{T}}
\end{equation}
which is remarkable since it express the average over an out of equilibrium process (LHS) 
in terms of a pure equilibrium quantity (RHS). This is at first sight a surprising result and even lead to debates at the beginning, since on the LHS there is an average over a completely out of equilibrium process whereas on the RHS we only find the equilibrium free energies for $\lambda=\lambda(\tau)$ and $\lambda=\lambda(0)$.
Using Jensen's inequality one derives that $\langle W \rangle \ge \Delta F$
which is an expression of the well known Clausius inequality of classical thermodynamics: the dissipated part of the work,
$\langle W \rangle - \Delta F$, has to be positive or equal to zero. 
This is one example of a main result of stochastic thermodynamics: a microscopic statistical mechanics identity, eq. (\ref{jarz}),  
from which macroscopic ones characteristic of classical thermodynamic can be derived. \\
Another interesting relation can be obtained by integrating  (\ref{trs2oeq}) over all paths that correspond to a given 
value of the work $W$. This leads to the Crooks identity \cite{Crooks}:
\[
P_R(-W)=P(W)e^{-\frac{W}{T}+\frac{\Delta F}{T}}
\]
where $P(W)$ is the probability to observe the work $W$ during the non-equilibrium process and $P_R$ its time-reversed counterpart. It is interesting to consider a cyclic process, which means that $\lambda(0)=\lambda(\tau)$. 
In this case, the previous identity is a constraint on the form of $P(W)$:
\[P(-W)=P(W)e^{-\frac{W}{T}}\] 
If the process is adiabatic $\langle W \rangle =\Delta F= 0$, i.e. no work is done on the system globally. On the other hand a typical out of equilibrium process leads to dissipation and to an average value $\langle W \rangle =W_{diss}> 0$. The previous
equation implies that even though $P(W)$ is centred around $W_{diss}>0$ it necessarily has tails extending to negative values: 
the probability of measuring a {\it negative} work in a cycle--an apparent violation of thermodynamics--is finite and given by $\int_0^\infty dW P(W)e^{-\frac{W}{T}}$.\\
This example contains the essence of stochastic thermodynamics and its main consequences. First, it allows to understand, actually derive, the laws of thermodynamics as identities on averages. 
Second, it also shows that since for large systems  the quantities like $W, \Delta F, É$ are extensive, i.e. they scale as the number of degrees of freedom, the exponential weights in these 
out of equilibrium identities are extremely small. Thus for large systems the probability of observing violations 
of the laws of thermodynamics is extremely small. It vanishes in the thermodynamic limit as expected, which is a very nice way to show the emergence of thermodynamics as a theory of macroscopic systems. The previous observation also implies that these fluctuation identities are in practice  meaningful only for small systems, i.e. in situations where the terms in the exponential 
are not too negative and hence probability of "violations" of thermodynamic laws are not too rare events. 
There are several experimental systems where this is the case and these laws have been tested and/or usefully applied, e.g. colloidal particles, proteins, etc. We refer to the review by Seifert \cite{seifert} for a comprehensive introductions to this field.  
\section{Quantum systems}
Time-reversal symmetry and its implications discussed in the two previous sections also hold for quantum systems, even though derivations are sometimes more involved.\\
In the case of equilibrium quantum dynamics time-reversal symmetry implies, as in the classical case time-reversal symmetry of correlation functions, fluctuation-dissipation relations and Onsager reciprocity relations. The only difference being the form of the fluctuation-dissipation relation.\\
 In the case of quantum out of equilibrium dynamics, identities 
 analogous to the ones obtained for classical system can be derived. Sometimes issues related to process measurements in quantum mechanics arise.  This makes the quantum version a bit more subtle. \\
We discuss, as an example, the Jarzynski-Crooks equalities for which the issue of providing a correct definition of work arise. The easiest derivation is obtained for an isolated system, so we focus on this case to keep things as simple as possible. In the classical case, since the system is closed and does not exchange heat, the work is simply the difference between the final and the initial  energy. In the quantum case, one has to take into account the
 measurement process in order to recover Jarzynski-Crooks equalities, see \cite{kurchan,esposito}. 
 The correct measurement protocol was defined in this way: one starts at $t=0$ from the Boltmann density matrix, measures the energy $E_i$, does the non-equilibrium process and measures the energy again, $E_f$, at the end. The measurements lead to a collapse of the wave function, hence at the beginning of the process the system is in an eigenstate of the initial Hamiltonian 
 $H(\lambda(0))$ and at the end in an an eigenstate of the final Hamiltonian 
 $H(\lambda(\tau))$. With this in mind it is easy to write down the probability distribution of the work:
 \[
 P(W)=\sum_{n,m}\frac{e^{-E_n/T}}{Z_i}\left|\langle m|Te^{-\frac i \hbar \int_0^\tau H(s) ds} |n\rangle  \right|^2 \delta(W-(E_m-E_n))
 \]
where $T$ denotes time-ordering.  
The probability for the reversed process reads:
 \[
 P_R(W)=\sum_{n,m}\frac{e^{-E_m/T}}{Z_f}\left|\langle n|\tilde Te^{-\frac i \hbar \int_{\tau}^0 H(s) ds} |m\rangle  \right|^2 \delta(W-(E_n-E_m))
 \]
 where $\tilde T$ denotes reverse time-ordering. By renaming the indices $n,m$ in $m,n$ in the last equation 
and by noticing that the amplitude squared term is the same between the two expression one recovers the Crooks identity that we derived previously in the classical case.\\
As discussed for classical systems, fluctuation relations out of equilibrium are useful for small systems. In the context of quantum systems the recent interest in nano-devices provides a natural ground to study and develop these new relations, see \cite{esposito} for a comprehensive review on the subject.

\chapter{Thermalization}
In this chapter we shall discuss in some detail the issue of thermalization, which is the process by which a system 
initially out of equilibrium relaxes to the thermal Maxwell-Boltzmann distribution, the cornerstone of equilibrium statistical mechanics. We shall discuss first systems coupled to a bath and then focus on the more difficult problem 
of isolated systems. Understanding how and why physical systems thermalize is a fundamental problem which is at the root of statistical physics. Note that not all systems do it, actually several physical systems are found to be out of equilibrium. Some of them are driven out of equilibrium by shear, currents, dissipation, etc. hence they cannot equilibrate; others instead just fail to equilibrate at least on experimental time-scales. Before discussing these latter cases in the following chapter, we now explain why equilibrating is more the rule than the exception,  
which is also the reason why avoiding equilibration is so interesting and puzzling. 
\section{The meaning of thermalization}
A macroscopic system, classical or quantum, is said to have reached thermalization if all the time-averaged 
local observables and correlation functions coincide with the ones obtained by the ensemble average over the Maxwell-Boltzmann distribution:
\[
\lim_{\tau\rightarrow \infty} \frac 1 \tau\int_0^\tau  O(t) dt=\langle O \rangle_T
\]  
where $O$ denotes a local observable, which means an observable whose value depends on degrees of freedom 
that are locally close in real space: for example a products of spin operators contained in a finite subsystem 
or products of local densities in a liquid\footnote{In the quantum case $O(t)$ is the expectation value under Schr\"odinger evolution.}.
The average on the RHS is the usual ensemble average of statistical mechanics: it is performed using the Maxwell-Boltzmann distribution at temperature $T$. The value of $T$ is fixed by the dynamical evolution: for an open system it is given by the temperature of the bath, for an isolated system it is the value of the micro-canonical temperature leading to an average intensive energy equal to the one of the system (which is conserved during the dynamics since the system is isolated)\footnote{We assume that there are no conserved quantities other than the energy. If this is not the case one has to
add more control parameter conjugated to the extra-conserved variables, e.g. the pressure for the volume, etc.}. 
The reason for  insisting that observables have to be local will be discussed at the end of this chapter. 
It is important for macroscopic systems only. \\
What makes thermalization mind-boggling at first sight is that the LHS of the previous equality depends on a lot of things, in particular on the initial condition and on the bath evolution if the system is open. On the RHS instead the only parameter that matters is the temperature of the bath or the initial energy if the systems is isolated:  
all the extra knowledge (or information) contained on the LHS has to be lost irreversibly during the dynamical evolution. 
This is the reason that makes equilibrium statistical mechanics so powerful. We don't need to know 
almost anything on the past history of the system to predict its equilibrium properties.\\
As anticipated before, thermalization is expected to be the rule for macroscopic systems. However, proving that a system does thermalize can be a tough problem especially for isolated systems. In the following we shall first discuss the case of open classical systems. 
\section{Classical systems coupled to a bath}     
Showing thermalization for classical systems coupled to a bath is the most easy case, and for this reason we shall discuss it in detail in the following. As we have shown in the first chapter, the {\it deterministic} dynamics of a classical system coupled to a bath reduces to a {\it stochastic} damped dynamics for the system alone. Within this framework  time-averages coincide with averages over the stochastic noise. Thus proving thermalization means showing that at long times
averages over the stochastic noise converge to Maxwell-Boltzmann averages:
\[
\lim_{t \rightarrow \infty} \langle O(t) \rangle=\langle O \rangle_T
\]
We shall consider finite systems, so there is no difference between local and non-local observables. 
For simplicity, we focus again on a one-dimensional system undergoing over-damped Langevin dynamics:
  \begin{equation}\label{langevinoverdamped}
 \frac{d{ x}}{dt}={-V'(x)}+\xi(t)\,,\qquad \langle \xi(t) \xi(t') \rangle=2T\delta(t-t')
\end{equation}
where the friction $\eta$ has been reabsorbed in a re-definition of the unit of time and $V(x)$ is an external potential.  
Although this is the simplest setting one can choose, it is very instructive and not so reductive after all, since 
all other cases can be treated technically as simple generalisations. Indeed, introducing inertia, 
considering non-white noise and systems with more than one degrees of freedom lead to more complicated 
stochastic equations which can always be re-written as sets of coupled Langevin equations introducing extra-variables. For example in the case of inertia one introduces the momentum $p=\dot x$ and write the second-order stochastic equation as a two first-order coupled Langevin equations of the type described above. \\
In the classical case, an observable is a function of the configuration of the system; in the simple case 
we are considering just a function of $x$. Introducing $P(x,t)$, the probability distribution of $x$ at time $t$, thermalization can be expressed as:
\[
\lim_{t \rightarrow \infty} \int dx P(x,t) O(x) =\int dx \frac{e^{-V(x)/T}}{Z} O(x)\qquad \forall O(x)
\]
since this has to hold for any function $O(x)$, it implies that:
$$\lim_{t \rightarrow \infty}  P(x,t)=e^{-V(x)/T}/Z$$ where $Z$ is the partition function. \\
The way to proceed to show that this is indeed what happens consists in deriving an equation for the evolution of $P(x,t)$. This is quite standard 
and can be found in several textbooks, see e.g. \cite{kamenev,kurchan}. The starting point is the identity:
\[
\frac{d}{dt} \langle O(x(t)) \rangle=\int dx \partial_t P(x,t) O(x) \qquad \forall O(x)
\] 
The LHS can be re-written in a discretised way as 
\[
\frac{1}{dt} \left(\langle \frac{dO}{dx} \dot x dt+\frac 1 2 \frac{d^2O}{dx^2} (\dot x dt)^2 \rangle+\cdots\right)
\]
In order to go further and understand why the second term has been retained one has to specify the discrete form of 
the Langevin equation. We choose the so-called Ito-prescription\footnote{Another one would lead to the same results but 
to a different derivation. Strictly speaking the 
stochastic equation we wrote is ill-defined in the continuum limit and the only correct mathematical way is to define it through its integrated version.} which corresponds to the discretization 
\begin{equation}\label{langevindiscretized}
 \frac{x(t+dt)-x(t)}{dt}={-V'(x)}+\xi(t)\,,\qquad \langle \xi(t) \xi(t') \rangle=2T\frac{\delta_{t,t'}}{dt}
\end{equation}
where $\delta_{t,t'}$ is the Kronecker delta. Now that everything is well defined we can evaluate the terms obtained above. \\
 \[
 \langle \frac{dO}{dx} \dot x \rangle= \langle \frac{dO}{dx} ({-V'(x)}+\xi(t)) \rangle =
  \langle \frac{dO}{dx} ({-V'(x)}) \rangle = -\int P(x,t) \frac{dO}{dx} {V'(x)} dx 
 \] 
 In the first equality we have used the expression of the Langevin equation and in the second one 
 that $x(t)$ is uncorrelated from the noise at the same time $t$, see eq. (\ref{langevindiscretized}).
 As for the second term:
\[ \frac 1 2 \langle \frac{d^2O}{dx^2} (\dot x dt)^2 \rangle=
\frac 1 2 \langle \frac{d^2O}{dx^2} ({-V'(x)}+\xi(t))^2 \rangle\]
only the term containing the square of the thermal noise is of the order $dt$. It is equal to 
\[\frac 1 2 \langle \frac{d^2O}{dx^2} \xi(t)^2 \rangle=T dt\langle \frac{d^2O}{dx^2} \rangle 
=Tdt\int P(x,t) \frac{d^2O}{dx^2} dx\]
where we have used again the same tricks than before. Collecting all the pieces together and dividing by $dt$ we finally reach the equation:
\[
\int  \partial_t P(x,t) O(x)dx=\int O(x)\frac{d}{dx} \left(V(x) P(x,t)\right) dx+\int O(x) T\frac{d^2P(x,t)}{dx^2} dx
\]
Since this has to be true for any function $O(x)$ the distribution $P(x,t)$ satisfies the so-called Fokker-Planck equation:
\begin{equation}\label{FP}
\partial_t P(x,t)=\frac{d}{dx}\left( V'(x)+T\frac{d}{dx}\right)P(x,t)
\end{equation}

By staring at this equation two observations immediately comes to mind. First, it is clear that the Maxwell-Boltzmann distribution is stationary since the RHS vanishes for $P(x,t)=e^{-V(x)/T}/Z$. Second, it looks much like an imaginary time Schr\"odinger equation:
\[
\partial_t P(x,t)=-H_{FP} P(x,t) \,,\qquad H_{FP}= -\frac{d}{dx}\left( V'(x)+T\frac{d}{dx}\right)
\]
The analogy is however not as straightforward as it seems at first sight since thermal averages cannot be written 
as quantum averages. In fact, introducing the bra-ket notation,
formally solving the Fokker-Planck equation and denoting by $\langle 1|$ the bra corresponding 
to the function equal to one for all $x$, i.e. $\langle 1| x\rangle=1$, we get:
\[
\langle O(x(t)) \rangle=\langle 1|Oe^{-H_{FP}t}|P_0\rangle=\langle 1|e^{H_{FP}t}Oe^{-H_{FP}t}|P_0\rangle
\]
where we have explicitly used that $\langle 1| H_{FP}=0$ and $|P_0\rangle$ represents the initial probability distribution. For a fixed initial condition $x_0$, one has $\langle x|P_0\rangle=\delta(x-x_0)$. The expression
above differs from its quantum mechanic counterpart for three main reasons: (1) the time is imaginary, (2) $H_{FP}$
is not a Hermitian operator, (3) the bra  $\langle 1|$ is not the transposed of the ket $|P_0\rangle$. Nevertheless the analogy is useful as we discuss below. Before let us list some general properties of $H_{FP}$. Their derivation is easy and left as an exercise to the reader. \\\\\\
\fbox{\parbox{12cm} {%

{\bf Properties of $H_{FP}$}

\vspace{.1cm}
\begin{itemize}
\item $H_{FP}$ is not Hermitian but the operator $H_S=e^{V/2T}H_{FP}e^{-V/2T}$ is Hermitian as it can be checked 
by simple algebra. 
\item $H_S$ is a positive operator, i.e. all its eigenvalues $\lambda_i\ge0$.
\item The eigenvalues of $H_{FP}$ coincide with the ones of $H_S$ and, hence, are all positive. 
\item Under very general hypothesis, i.e. if the potential $V(x)$ grows fast enough when $|x|\rightarrow \infty$,
all  $\lambda_i$ are positive but one. The single eigenvalue $\lambda_1=0$ has the Boltzmann distribution as right eigenvector and $\langle 1|$ as left eigenvector.  
\end{itemize}
}}\\\\\\

These properties imply that the Maxwell-Boltzmann distribution is the only stationary distribution.
Moreover, by decomposing the initial distribution $|P_0\rangle$ and using right and left eigenvectors of $H_{FP}$ (which form
a basis of the Hilbert space): $|P_0\rangle=\sum_i c_i |i\rangle_R$ where $c_i=
\leftidx{_L}{\langle i|}P_0\rangle$, we find:
\[
P(x,t)=\sum_i e^{-\lambda_i t} c_i \langle x|i\rangle_R\]
In consequence, since all exponentials but one vanish at long times and $c_1=1$ by normalisation, we have proven the convergence to the Maxwell Boltzmann distribution for any initial condition:
 \[P(x,t)\xrightarrow {t\rightarrow \infty} \frac{e^{-V/T}}{Z}
\]
Along the way, we have also found that the the eigenvalues of $H_{FP}$ have the interpretation of inverse relaxation times. The time to reach equilibrium 
is given by $1/\lambda_2$.\\
Several other interesting properties come out from this analogy with quantum mechanics both for classical and 
quantum systems. From the classical point of view, using spectral analysis of  $H_{FP}$ and WKB methods 
it is possible to show that the right and left eigenvectors allows one to define very precisely and quantitatively metastable states and their basins of attraction, see \cite{gaveauschulman,biku,bovier}. From the quantum point of view, the analogy was fruitful to construct variational wave-functions \`a la Jastrow \cite{mahan} and also in the analysis of quantum spin liquids. In this context, quantum 
dimer models were shown to display special points in their phase diagram, the so-called Rokshar-Kivelson point, at which 
the quantum Hamiltonian maps exactly on a Fokker-Planck operator $H_{FP}$ \cite{RK}. Thus allowing one to obtain several results on the ground state 
properties from the knowledge of the corresponding classic stochastic dynamics \cite{RK2}. 
\section{Quantum systems coupled to a bath}
\begin{figure}[t]
\begin{center}
 \includegraphics[width=0.8\textwidth]{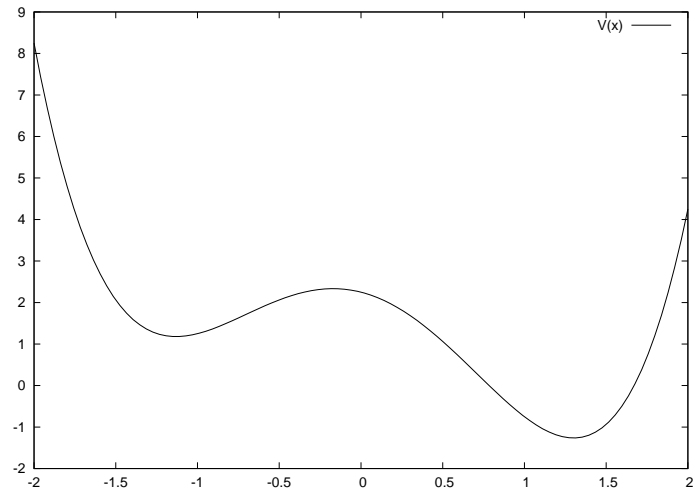}
\caption{Double well potential}
\label{doublewell}
\end{center}
\end{figure}

Showing thermalization is more difficult for open quantum systems, however general results have been obtained recently. 
One has to show in full generality
that the Schwinger-Keldysh field theory for a quantum system coupled to a bath 
leads to thermalization at long times. Although this has been done  in the classical limit for the MSRDJ field theory
\cite{gozzi}, up to now it has been only performed on specific, albeit quite general, settings in the quantum case\footnote{If one approximates the dynamics of a quantum open system using Lindblad operators then proving thermalization is 
rather straightforward \cite{gardiner}.} \cite{doyon}. The arguments that have been used are too involved to be presented here and we refer to the original publications \cite{gozzi,doyon}.\\
Despite these technical difficulties, from the physical point of view, there is no reason to doubt that a finite quantum system coupled to a bath do not thermalize. Actually, equilibration is even more likely for a quantum system than for a classical one 
because of quantum fluctuations. An instructive case is provided by the double well potential in Fig. \ref{doublewell}
which models a situation in which there is a metastable state that can trap the dynamics for some time. A classical system starting at $t=0$ in the left well with small kinetic energy needs to receive energy from the bath in order to 
 jump across the barrier and reach equilibrium. This is a less and less likely process the smaller 
is the temperature and leads to a life-time of the metastable state that follows the Arrhenius law $e^{\Delta E/T}$, 
where $\Delta E$ is the height of the energy barrier that the system has to overcome, see e.g. \cite{kurchan} for a derivation of this result.   
Strictly at zero temperature a classical system starting in the left well looses kinetic energy, absorbed by the bath through friction, and remains trapped forever in that well, never reaching the right well which corresponds to the true ground state of the system at $T=0$. Instead in the quantum case, because of quantum tunnelling the system would never be trapped and eventually relax to equilibrium at $T=0$.  \\
Only if the two wells have exactly the same height then the quantum system could avoid thermalization. 
Depending on the strength of the coupling to the bath and the spectral properties of the bath, localisation in one well can occur at $T=0$. This is the famous Caldeira-Leggett problem \cite{caldeiraleggett}.   
\section{Isolated systems}
The theory of thermalization of isolated system is one of the most venerable part of statistical mechanics. 
For classical systems is by now a textbook subject (good textbooks only though), instead for quantum systems
it is a topic on which current research is still focusing on. My aim in the following is certainly not giving a presentation
of it, which is well beyond the scope of these notes. It would deserve a series of lecture on its own. I would like instead to discuss some important points and to describe briefly the results on quantum systems. 
\subsection{Classical systems}
\begin{figure}[t]
\begin{center}
 \includegraphics[width=0.8\textwidth]{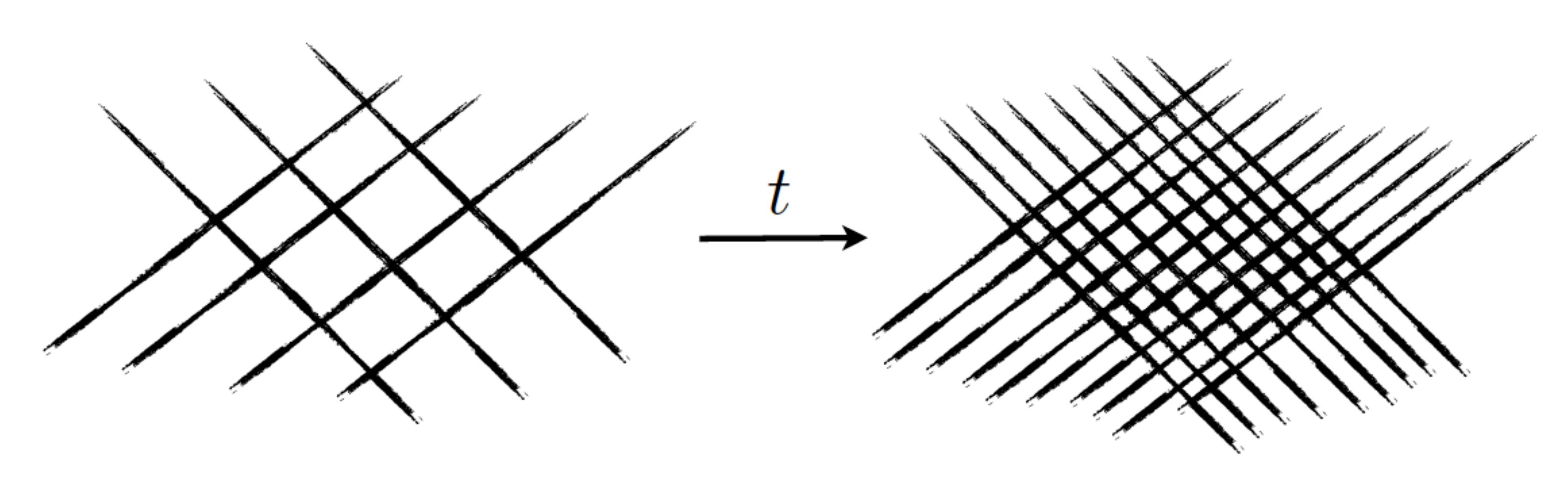}
\caption{Sketch of the progressively blurring of the probability measure over time. Usual physical observables are unable to probe the granularity of the measure, whereas unphysical ones such as the characteristic function corresponding to 
the being in one given many-body configuration do.}
\label{fig:CG}
\end{center}
\end{figure}

In the case of classical systems, the theory of thermalization is intertwined with the one of ergodic systems. 
The bottom line is that a system whose dynamics is chaotic enough (mixing) thermalises. Roughly (very roughly), these two properties mean that if one starts with a distribution of initial conditions strongly peaked around a certain point ${\mathbf x}_0$ of configuration space, after a certain (equilibration) time the distribution obtained by following the dynamical trajectories will be almost uniformly spread on the hyper-surface corresponding to all points with an energy 
equal to $E({\mathbf x}_0)$. \\
We shall not spend more time on ergodic theory and refer for the interested reader to modern textbooks such as \cite{vulpiani}. We would like just to mention two important points: first proving that a system is ergodic is extremely difficult and it has been achieved in academic cases only. On the other hand it is thought that generically a macroscopic interacting system  
is ergodic. This actually is routinely observed in molecular dynamics simulations of interacting particle systems that thermalize quickly. Integrable systems are an exception to this whole scenario. They are not ergodic and, by KAM theory \cite{vulpiani}, perturbations of such systems are not ergodic if the perturbation is small enough. On the other hand it is currently believed and sometimes shown that the KAM region corresponding to the values of the perturbation where the system remains non-ergodic shrinks to zero in the thermodynamic limit,  see for example the studies of the Fermi-Pasta-Ulam model \cite{casetti}.  
 In summary all macroscopic interacting systems should thermalize except an ensemble of very fine-tuned ones, say of measure zero in the space of models,   
which correspond to integrable systems. If a macroscopic system is found numerically or experimentally to avoid thermalization this has to do with an underlying phase transition and the emergence of very long time-scales but certainly not with a loss of chaoticity. \\
One point that is often not stressed enough is that there is more to thermalization than just ergodic theory.  The other very important aspect is considering macroscopic systems and focusing on special observables, in particular local observables because these are naturally what we look for in simulations and experiments. Why is this important? Let us do an analogy with Monte-Carlo dynamics. Imagine to have $N$ Ising spins that flip with a rate $\lambda=1/\tau_0$ and to start the dynamics from the configuration with all spins up. 
This Monte-Carlo dynamics is the one characteristic of $T=\infty$ since spins flip independently of the initial and final configuration. Hence, the Monte-Carlo dynamics is just a random walk in the configuration space (which is an hyper-cube in $N$ dimensions). It is easy to show (exercise!) that it thermalises toward the flat distribution which, as expected, corresponds to the Maxwell-Boltzmann distribution for $T=\infty$ ($e^{-\beta H}=1$).  Spins flip independently with a rate $1/\tau_0$, hence the time-average of local correlation functions or observable converge to their thermalised values within a time-scale $\tau_0$, which is therefore the relaxation time. On the other hand on the time-scale $\tau_0$, all spins have flipped a finite number of time in a given run, so the system has visited a number of configuration which is of the order of $N$ only, much much smaller than the $2^N$ configurations available. 
How is possible then that the system seems thermalised as if the measure was flat in configuration space and at the same time the number of configurations visited is extremely small?  The way out, sketched pictorially in Fig. \ref{fig:CG}, is 
that typical dynamical trajectories visit the configuration space in such a way that for the observables we usually focus on time-averages are essentially identical to the ones obtained by the Maxwell-Boltzmann measure: the random walk very quickly spreads over the configuration space. 
Observables local in real space cannot probe the granularity of this spreading and, as far as they are concerned, the distribution obtained after a time-scale $\tau_0$ is identical to the flat one even though this is clearly not the case. For instance, imagine that one focus on the observable equal to one if and only if the configuration is the one with all spins down. This is clearly not a local observable, actually it corresponds to a $N-$point correlation function. Its thermalised value is $1/2^N$. In order to reach thermalization for this observable one has to wait a time long enough to be sure that this configuration is encountered at least once during the dynamics. An estimate of this time, more a lower bound actually, is given by $\tau_0$ times the number of configurations available divided by the number of configuration visited in a time $\tau_0$. The resulting time scale is $\tau_0 2^N/N\gg \tau_0$ which is much larger than the relaxation time $\tau_0$ and actually enormous for realistic values of $N$. This example is very instructive since it contains the essence of the issue we wanted to clarify. Although we pointed it out for 
Monte-Carlo dynamics, the same ideas hold for for isolated systems. Also in this case it is crucial to take into account that we always focus on (1) macroscopic systems and (2) observables which are local. Non-local observables can avoid thermalization or display it on a time-scale which is gigantic and anyhow completely irrelevant for all practical purposes.   
\subsection{Quantum systems}
\begin{figure}[t]
\begin{center}
 \includegraphics[width=0.6\textwidth]{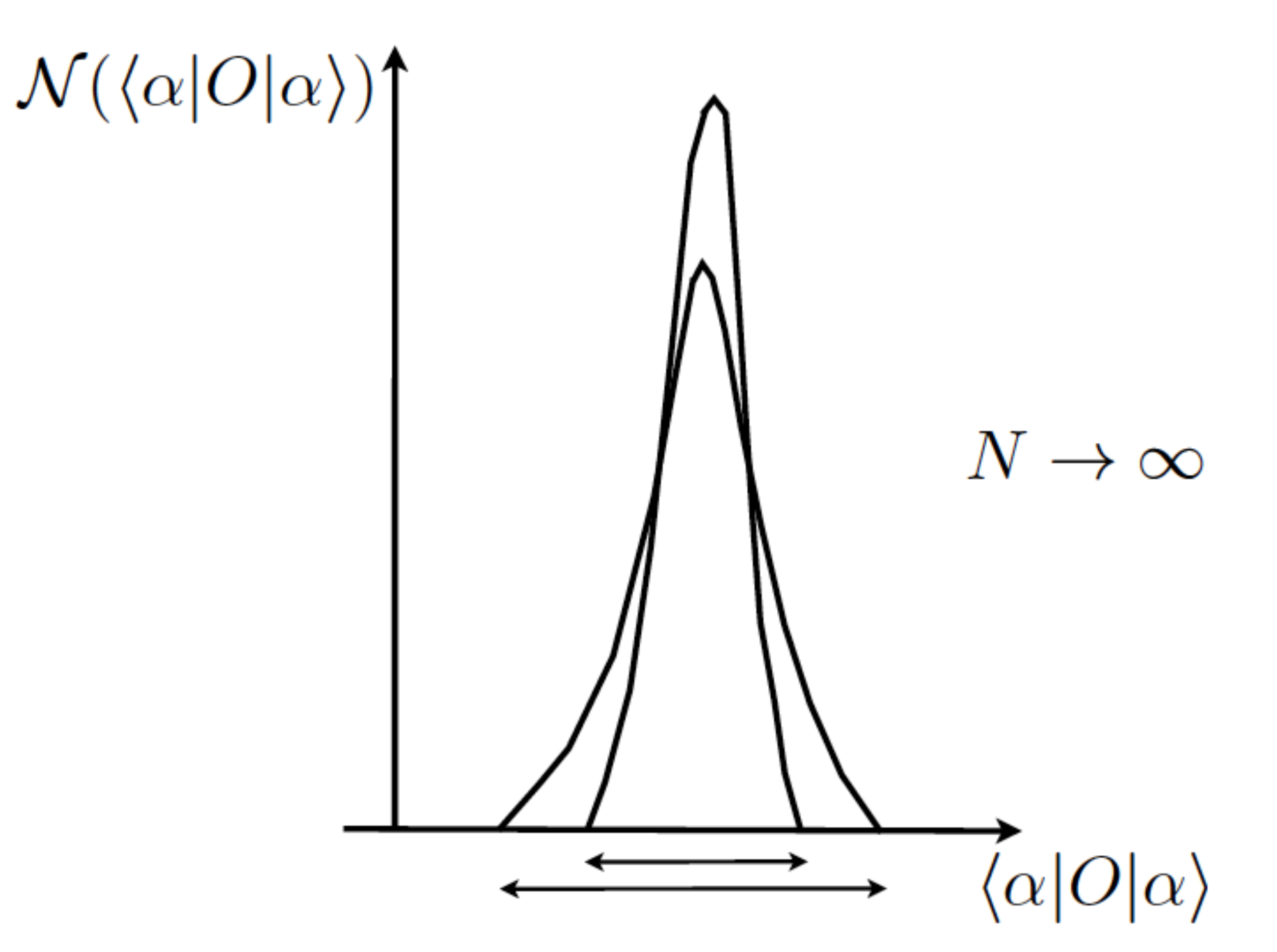}
\caption{Cartoon of how the distribution of $\langle \alpha|O|\alpha\rangle$ evolves by increasing $N$. In the strong version of ETH the support shrinks to zero so that indeed in the thermodynamic limit
all states verifies ETH.}
\label{fig:ETHS}
\end{center}
\end{figure}

The problems mentioned above for classical systems are even more striking for quantum ones. In particular, it is clear that it does not make much sense to talk generically about thermalization of finite isolated quantum systems. In fact, many such systems display a discrete spectrum and, starting for a given initial condition, they show Rabi-like oscillation and do not even reach a steady state. \\
Let us then focus from the beginning on macroscopic systems.  
The initial condition for the dynamics is a given wave-function $|\psi\rangle$. Thermalization means that 
\[
\lim_{\tau\rightarrow \infty} \frac 1 \tau\int_0^\tau  \langle \psi| e^{iHt/\hbar}O e^{-iHt/\hbar} |\psi \rangle dt=\langle O \rangle_e
\]
where we have used on the RHS the micro-canonical average at intensive energy $e$, which equals the average energy of the initial condition $\langle\psi|H|\psi\rangle/N$. Decomposing $|\psi\rangle$ on the eigenvectors $|\alpha\rangle$ of $H$ we find:
\[
\lim_{\tau\rightarrow \infty} 
\sum_{\alpha \ne \beta} c^*_\beta c_\alpha \langle \beta|O|\alpha\rangle \frac{1-e^{i(E_\beta-E_\alpha)\tau/\hbar}}{i(E_\alpha-E_\beta)\tau/\hbar}+\sum_\alpha |c_\alpha|^2 \langle \alpha|O|\alpha\rangle=\langle O \rangle_E
\]
Under very general hypothesis, see \cite{BS}, the first term on the LHS, dies off in the large $\tau$ limit. Hence the 
formal expression of thermalization reads:
\[
\sum_\alpha |c_\alpha|^2 \langle \alpha|O|\alpha\rangle=\frac{1}{\mathcal N}\sum_{\alpha_e} {\langle }\alpha_e|O|{\alpha_e}\rangle
\] 
where $\alpha_e$ are the eigenstates with intensive energy $e$ and $\mathcal N=\sum_{\alpha_e}$ is the micro-canonical normalisation factor. Again, as at the beginning of this chapter, we face the riddle of thermalization: the LHS 
contains a lot of information on the initial condition since the $c_\alpha=\langle \alpha|\psi \rangle$ depend on the initial wave-function $|\psi \rangle$, the RHS instead 
depends on $e$ only. How is possible that by changing the initial condition but keeping the average intensive energy fixed all the $c_\alpha$s change but in such a way that the sum above does not? Clearly this is generically impossible for a finite system. 
As stressed before, considering very large systems is mandatory. Only in the thermodynamic limit, the distribution 
of the  $c_\alpha$s can reach a limit in which this becomes true. \\
The solution of this riddle was started to be elucidated by Deutsch and Srednicky \cite{DR1,DR2} with the introduction of the so-called Eigenstate Thermalization Hypothesis (ETH). The main idea is that a {\it given eigenstate is thermal}, i.e. the average of any local observable in an eigenstate must coincide
with the one obtained by equilibrium thermal average: if $\langle \alpha|O|\alpha\rangle$ depends on 
$\alpha$ only through its intensive energy then it is clearly equal to the micro-canonical average which is given by the flat sum 
over all eigenstates characterized by the same intensive energy.  
Again, this cannot be true in a finite system and can become true only as a limiting process for $N\rightarrow \infty$. 
Indeed, it is easy to show, see \cite{birolikollath}, that the distribution of the values of  $\langle \alpha|O|\alpha\rangle$ for a given local observable becomes peaked around the micro-canonical value in the thermodynamic limit. 
These however is not enough to have an argument for thermalization since even though the majority of eigenstates 
lead to the micro-canonical average if there exist rare ones for which this does not happen then thermalization could be avoided in principle if the initial conditions have a projection large enough on them. \\
Two solutions to this 
impasse were discussed in the literature. The first, that we called strong version of ETH \cite{rigol} is sketched in Fig. \ref{fig:ETHS}: not only the distribution of  $\langle \alpha|O|\alpha\rangle$ becomes peaked but also its support shrinks to zero so that effectively in the thermodynamic limit
all states verifies ETH. The second is that the support does not shrink to zero but physical initial conditions are spread in configuration space in such a way that rare states never give a contribution, i.e. the $|c_\alpha|^2$ are rather uniform (see the discussion in \cite{birolikollath} for more details). 
What is the correct 
solution is not clear yet, since the values of $N$ available in numerical diagonalization are not large enough to solve this issue clearly, neither rigorous proofs are available. What is becoming clear, however, is that for integrable systems, that indeed do not thermalize, the distribution of $\langle \alpha|O|\alpha\rangle$ has a finite support in the thermodynamic limit and the failure of thermalization is due to sampling of rare states.  \\
In summary, macroscopic isolated quantum systems generically are expected to thermalize similarly to classical systems,. There is however an important exception which is due to the presence of quenched disorder: quantum disordered systems can show Many Body Localization, i.e. localisation in configuration space and avoid thermalization \cite{reviewMBL1,reviewMBL2}. This is a pure quantum effect with no classical counterpart. Contrary to classical systems, in the quantum world there could be a new kind of 
breaking of ergodicity which is not due to any underlying thermodynamic phase transition.  


\chapter{Quenches and Coarsening}
There are two ways of putting a system off equilibrium: by an external 
drive (e.g. a shear, a
voltage difference, etc\dots ) or by changing 
a control parameter $C$ (e.g. the magnetic field, the temperature, the pressure, etc\dots ).
In the former case the system is maintained out of equilibrium, whereas in the latter 
it starts to evolve towards the new equilibrium state corresponding
to the new value of $C$. This may take a long time. 
Sometimes, so long, that the system remains out of equilibrium forever. 
In this chapter we focus on this latter kind of slow 
relaxation and out of equilibrium dynamics. We shall consider systems coupled to a bath 
since this is the common situation and discuss at the end the case of isolated systems.\\
As we discussed in the previous chapter, the relaxation time of a 
an ensemble of interacting classical or quantum degrees of freedom coupled to a bath is generically finite\footnote{A possible exception is provided by zero temperature baths, see Chapter 3.}. 
Thus, one expects that after the change of the control parameter $C$ a {\it finite} system reaches 
equilibrium in a {\it finite} time. Only in the thermodynamic limit one can observe never-ending 
off-equilibrium relaxation, which is therefore, as phase transitions, a collective phenomenon
with at least one growing correlation length. A simple argument to see why this must be the case is 
that if all correlation
lengths were bounded during the relaxation, then one could roughly divide the system
in independent {\it finite} sub-systems each one having a {\it finite} relaxation time  $t_{B}$. 
In consequence, the system would have to relax on timescales not larger 
than $t_{B}$ which contradicts the starting hypothesis of never reaching equilibrium 
on any finite timescales.

\section{Thermal Quenches}
\begin{figure}[h]
\centering
\includegraphics[width=0.6\textwidth]{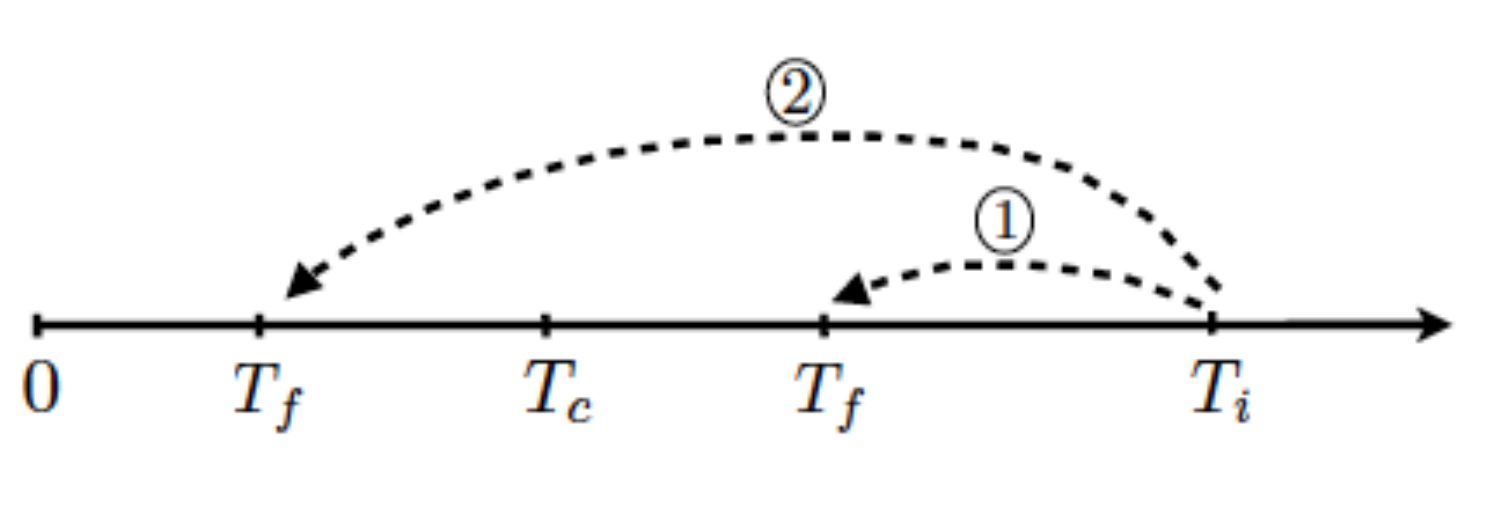}
\caption{The two different protocols discussed in the main text. In the first the quench is toward a temperature 
$T_f>T_c$ whereas in the second is across the phase transition.}
\label{fig:quench1}
\end{figure}

\begin{figure}[H]
\centering
\includegraphics[width=0.6\textwidth]{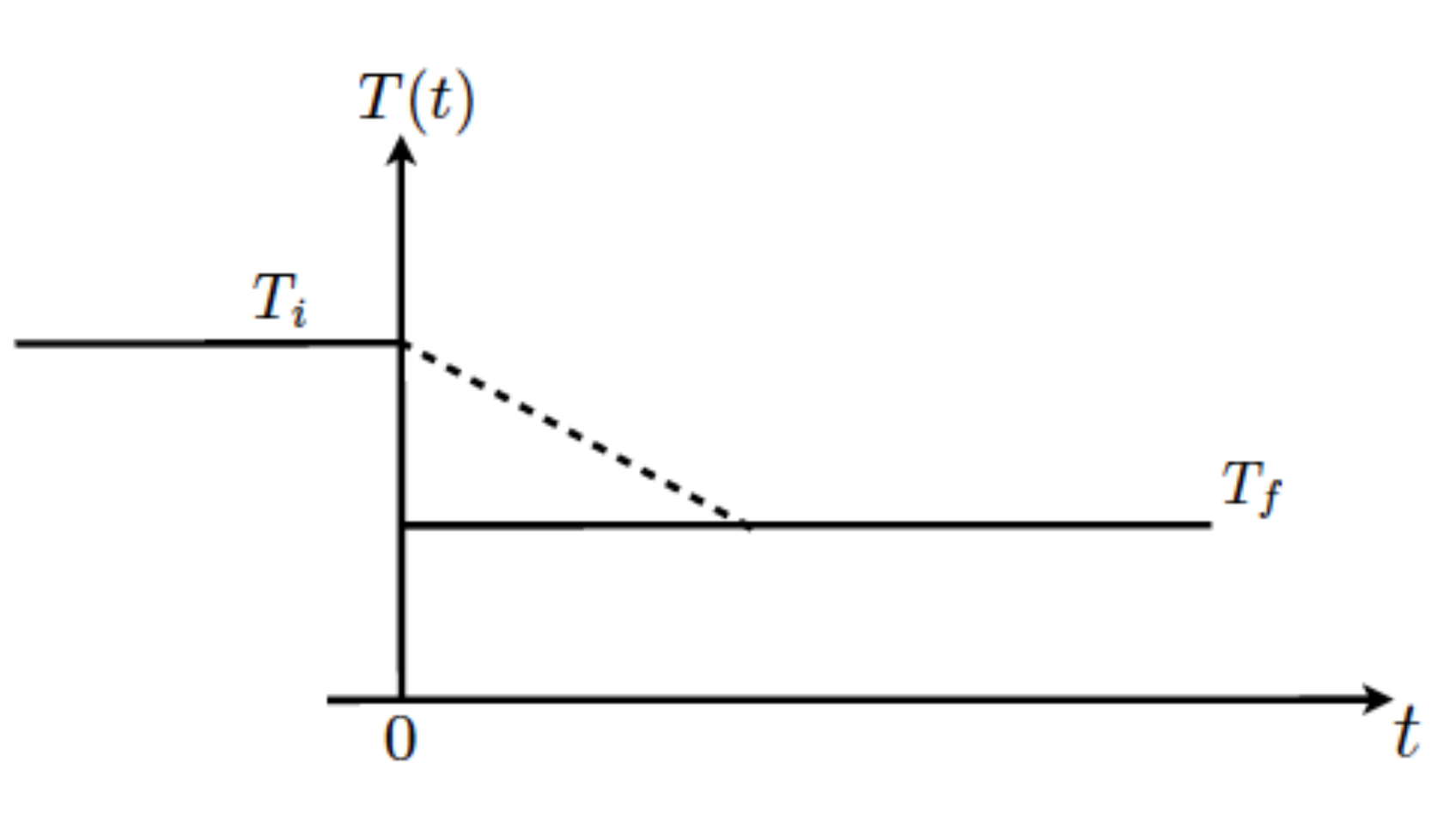}
\caption{Temperature time-dependence during a quench. The dashed line corresponds to realistic situations whereas
the continuous line to the sudden quench studied theoretically.}
\label{fig:quench2}
\end{figure}

\begin{figure}[H]
\centering
\includegraphics[width=0.7\textwidth]{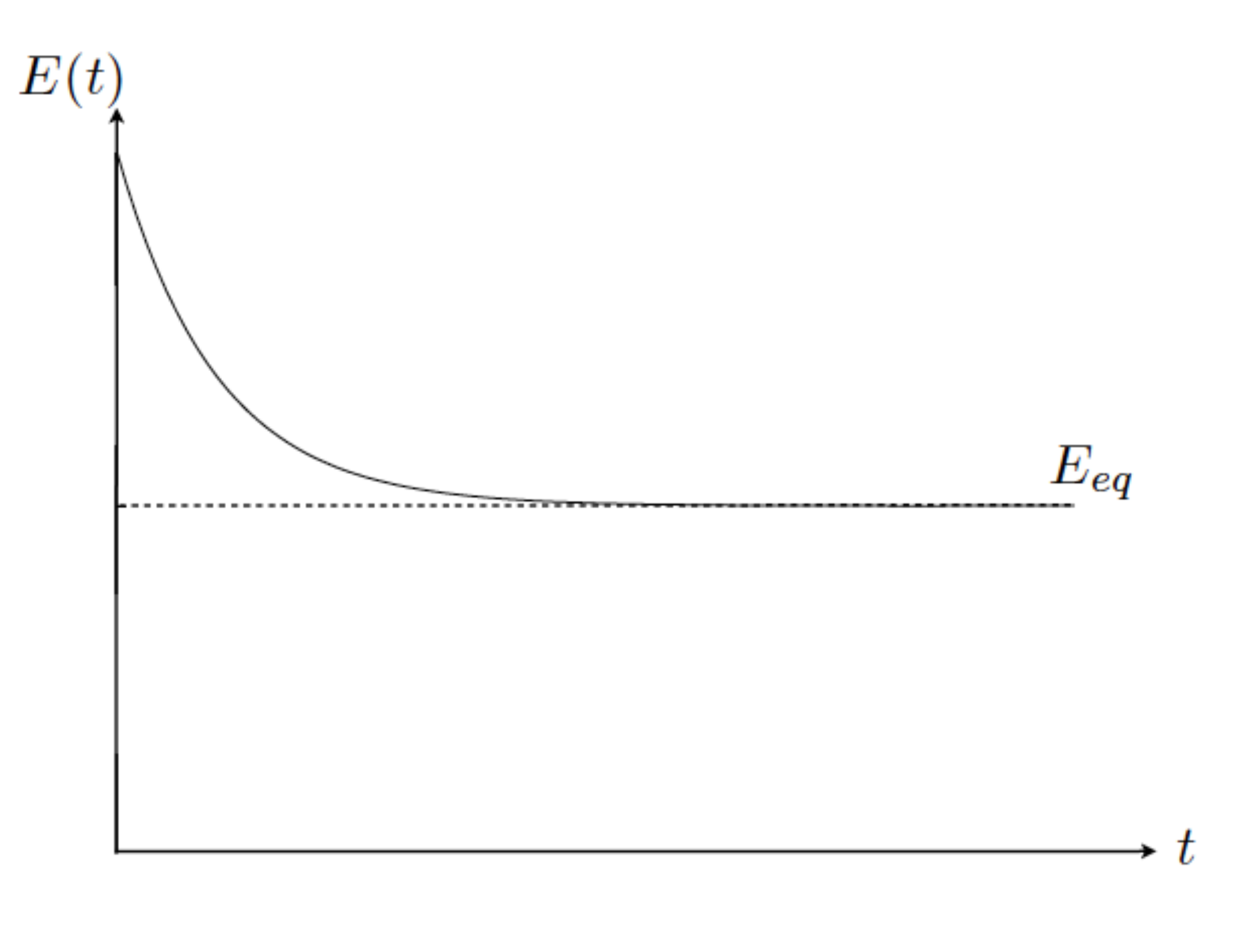}
\caption{Sketch of the time-evolution of the energy after a quench with protocol (1). The relaxation to the equilibrium value is fast and often exponential in time.}
\label{fig:eeq}
\end{figure}

\begin{figure}[H]
\centering
\includegraphics[width=0.7\textwidth]{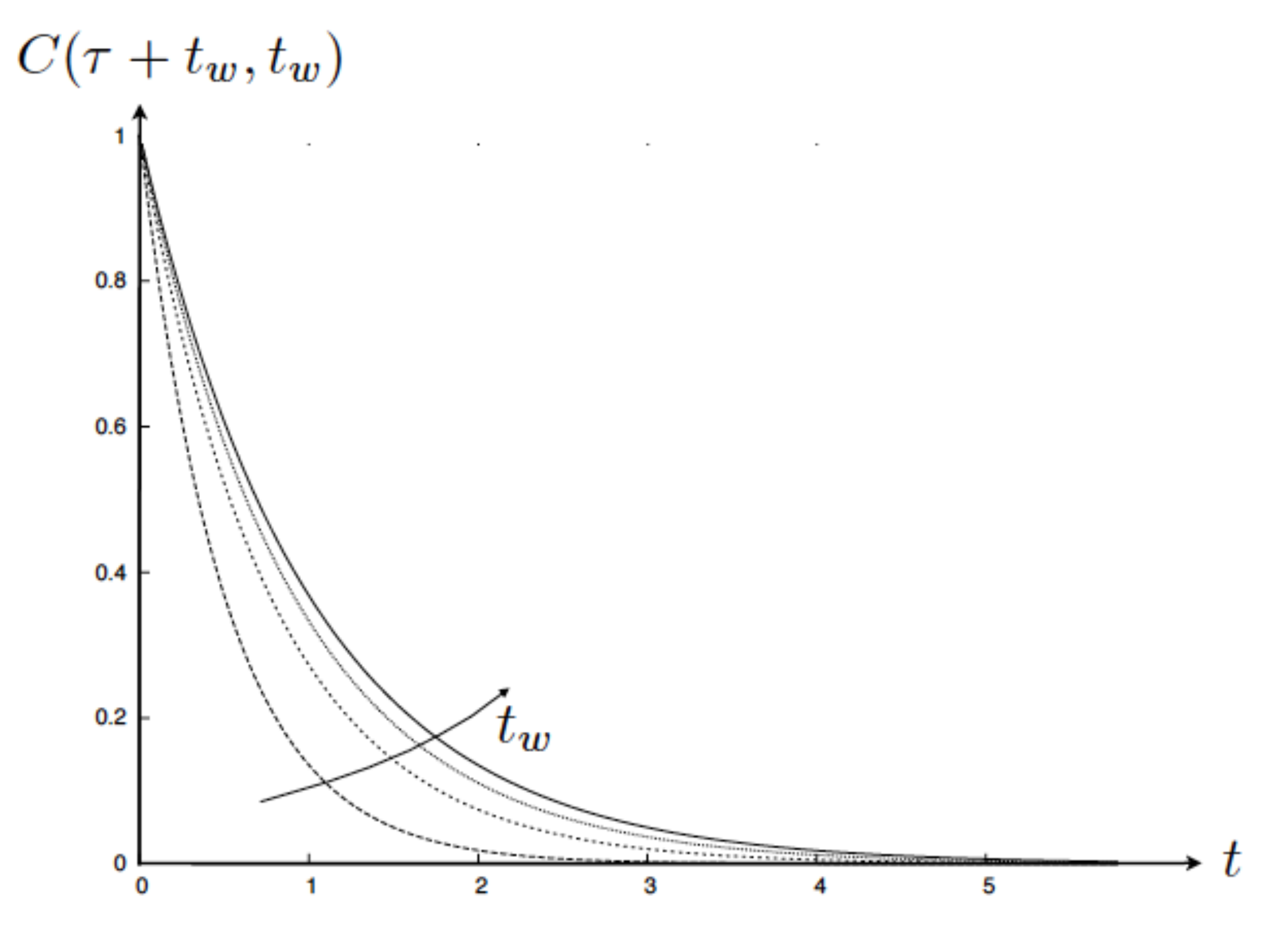}
\caption{Sketch of the time-evolution of the two-time correlation after a quench with protocol (1). Consider for instance 
the case of the spin-spin correlation function in the Ising model: $C(\tau+t_w,t_w)=\frac 1 N \sum_i \langle \sigma_i(\tau+t_w)\sigma_i(t_w)\rangle$. Once $t_w$ becomes of the order of the relaxation time the different curves converge and the two-time correlation function becomes time-translation invariant.}
\label{fig:correq}
\end{figure}

\begin{figure}[H]
\centering
\includegraphics[width=0.7\textwidth]{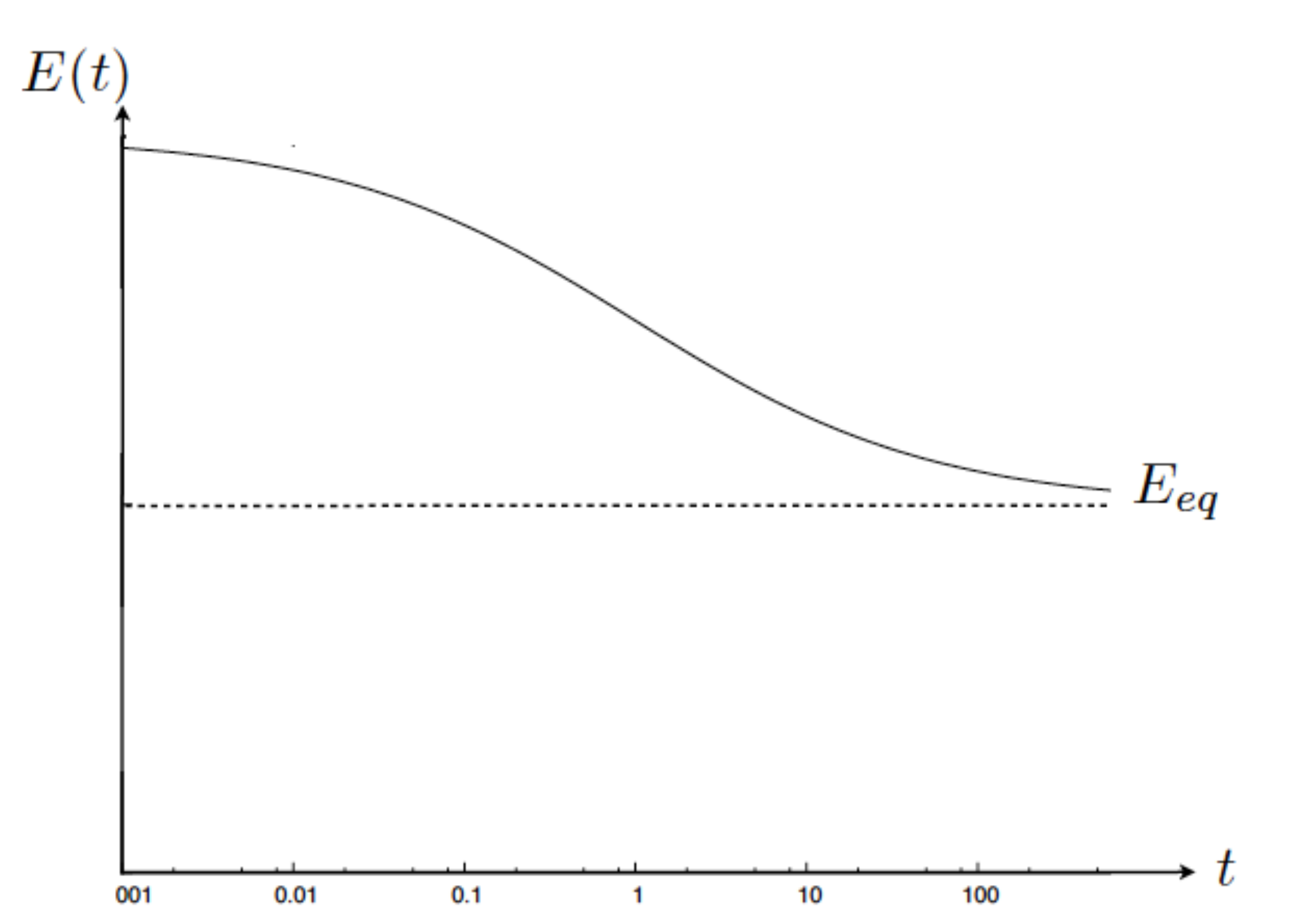}
\caption{Sketch of the time-evolution of the energy after a quench with protocol (2). The relaxation to the equilibrium value is slow and often power law in time.}
\label{fig:eaging}
\end{figure}

\begin{figure}[H]
\centering
\includegraphics[width=0.7\textwidth]{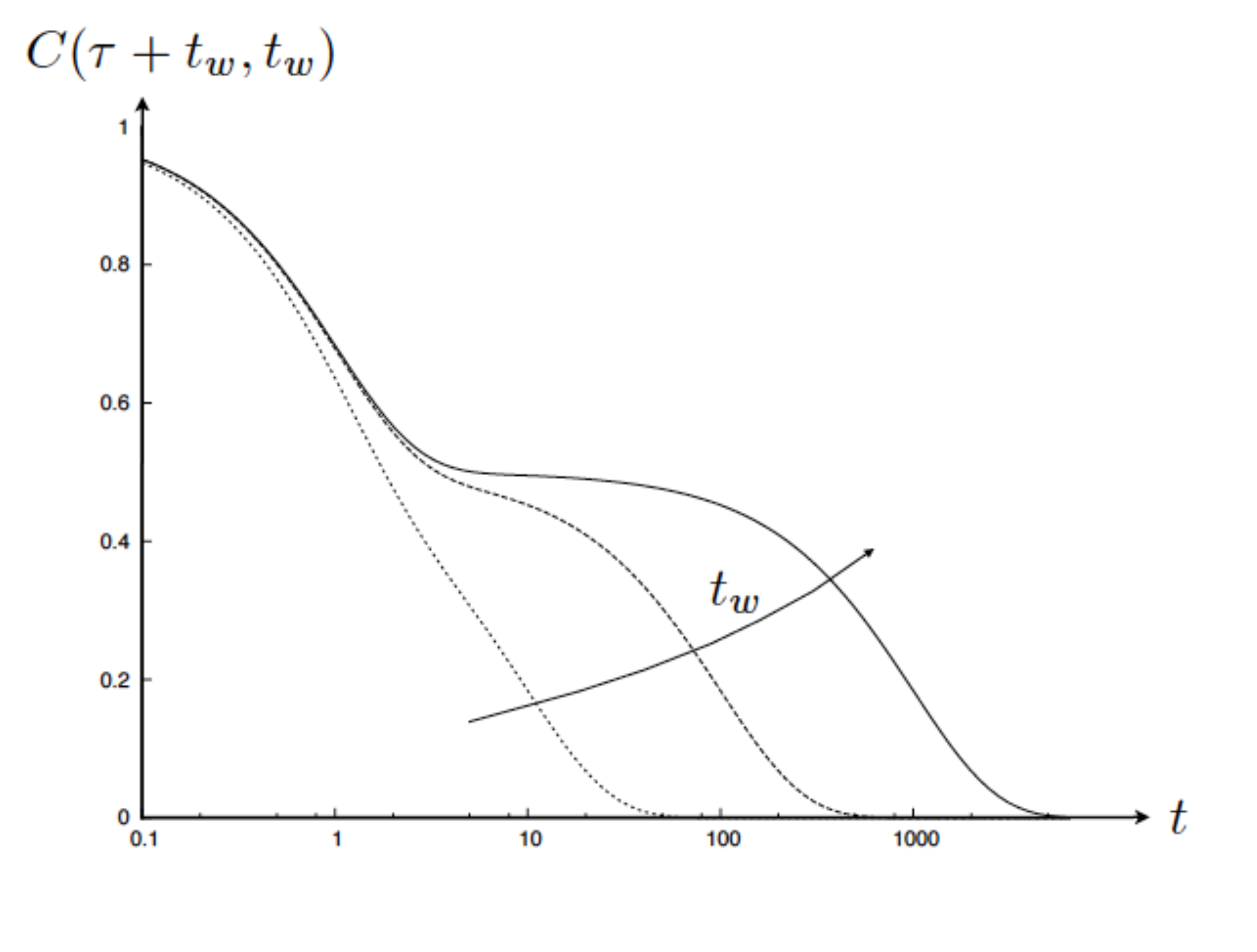}
\caption{Sketch of the time-evolution of the two-time correlation after a quench with protocol (2). Consider for instance 
the case of the spin-spin correlation function in the Ising model: $C(\tau+t_w,t_w)=\frac 1 N \sum_i \langle \sigma_i(\tau+t_w)\sigma_i(t_w)\rangle$. The different curves never converge. The time to escape from the plateau becomes larger the larger is the time spent after the quench.}
\label{fig:corraging}
\end{figure}

Let's consider a system that displays a second order phase transition at a temperature $T_c$. 
We consider the following protocol which is called thermal quench: the system starts at equilibrium 
at a temperature $T_i$ at time $t=0$, see. Fig. \ref{fig:quench1}. The temperature of the heat bath is then suddenly changed at 
a value $T_f$. Theoretically, we shall study the limiting case where this happens instantaneously. In 
reality the rate is finite, and actually often quite large, e.g. a fraction of Kelvin per minute, see Fig. \ref{fig:quench2}. 
In the literature the time spent after the quench is called waiting time and denoted $t_w$. In our case this coincides with $t$ since the quench occurs at time $t=0$.
Our argumentation is general
but for concreteness it is useful to have in mind an example, e.g. the ferromagnetic Ising model undergoing some kind of Monte-Carlo dynamics. The Hamiltonian of the Ising model reads:
\[
H=-J\sum_{\langle i,j \rangle} \sigma_i \sigma_j \qquad \sigma_i=\pm1 \,\,\,\forall i=1,\dots,N 
\] 
where the sum is over nearest neighbours on the three-dimensional cubic lattice. This model has a phase transition between a high temperature paramagnetic phase and a low temperature ferromagnetic phase characterized by 
long-range ferromagnetic order and a non-zero intensive magnetisation $m$.\\
We consider first the protocol in which the final temperature, $T_f$, is larger than $T_c$. In this case the system 
equilibrates after a characteristic time-scale, $\tau$, that depends on $T_f$ (and possibly other control parameters and microscopic quantities).
This means that one-time quantities like the
magnetization, the energy, etc. approach for $t\gg \tau$ their equilibrium values
and, hence, they become time independent, whereas for example 
the two-time correlations and response functions become invariant under 
translation of time, i.e. they depend just on the time difference.
Very often this approach to equilibrium is 
exponential in time. See Figs. \ref{fig:eeq} and \ref{fig:correq}. 
Finally, typical equilibrium relations like fluctuation-dissipation
relations between correlation and response are verified. \\
In the second case, where $T_f<T_c$, one-time quantities still approach their equilibrium values 
at long times but slowly, with a power-law time dependence, see Fig. \ref{fig:eaging}. A first signal that the dynamics changes 
crossing the phase transition. A more striking behaviour is found in two-times correlation functions, 
which display a first rapid relaxation to a plateau value over a time-scale $\tau$, that depends on $T_f$, 
and then a second relaxation from the plateau to zero which takes place on an increasingly larger time-scale, see Fig. \ref{fig:corraging}. 
Two things are striking in this behaviour: (1) the first relaxation toward the plateau is the one characteristic of systems equilibrated at $T_f$, i.e. it coincided with what one would observe starting
the dynamics from equilibrium at $T=T_f$. For example, in the case of the Ising model, the spin-spin correlation function converges to a plateau value equal to $m^2$; (2) no matter how long is the time spent after the quench 
the system does not equilibrate. Correlation functions do eventually escape from their plateau value and never become time-translation invariant. The time-scale on which these non-equilibrium effects take place is not an intrinsic time-scale fixed by $T_f$ (or other parameters) but instead depends on the age of the system itself: it is larger the larger is the time spent after the quench. For this reason this behaviour is called {\it aging} and emerges generally in quenches across phase transitions.   
\section{Coarsening}
\begin{figure}[H]
\centerline{
\includegraphics[width=0.8\textwidth]{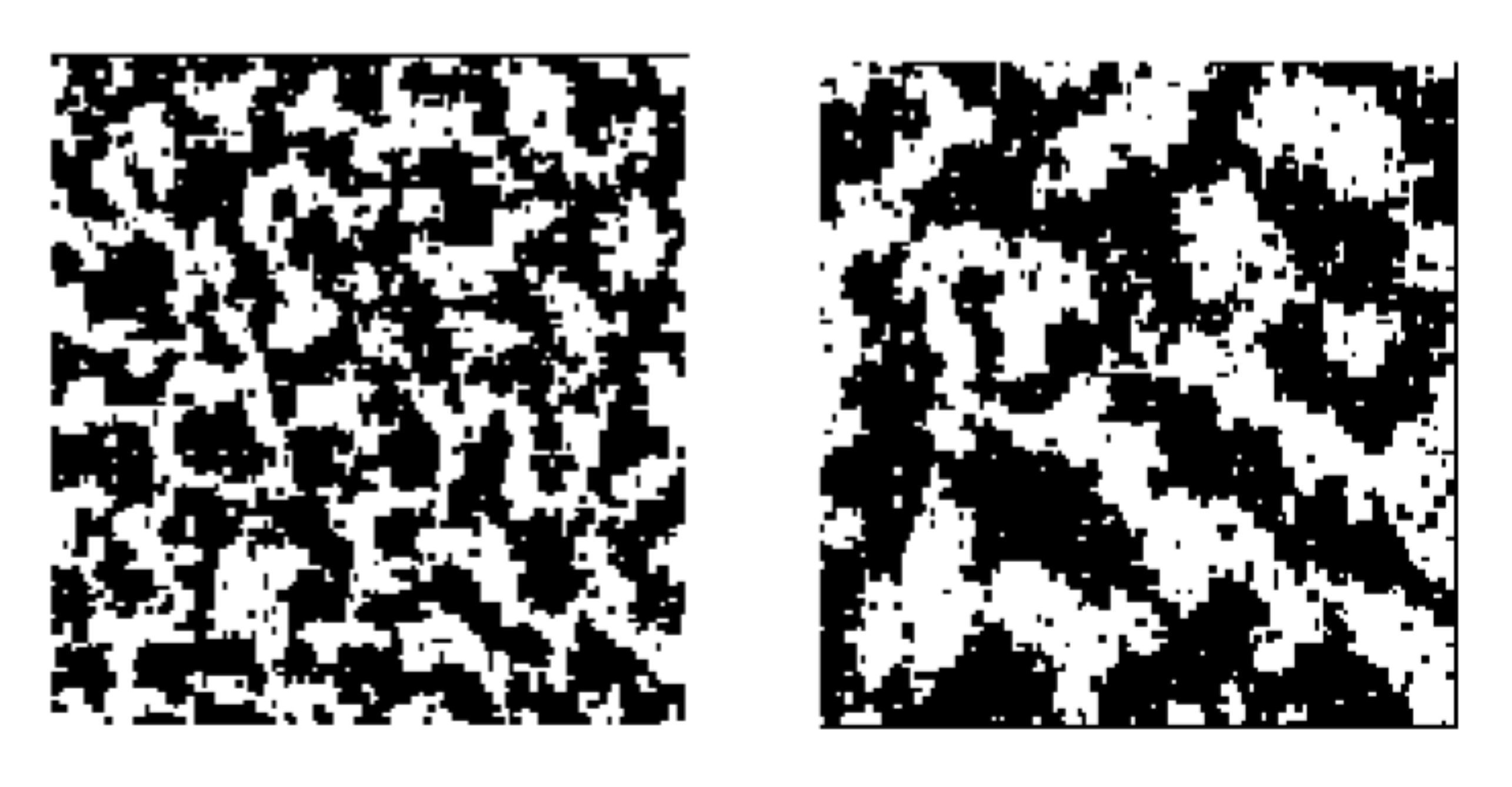}
}
\caption{Two snapshots of a 2d cut of a 3d ferromagnetic Ising model evolving 
with a Glauber dynamics after a quench at time $t=0$ from high temperature to
$T<T_{c}$. Black color is used for minus spins and white for plus
spins. On the left the configuration reached after $t_{w}=10^{3}$
MonteCarlo steps. On the right the configuration reached after $t_{w}=10^{5}$
MonteCarlo steps. Statistically the two configurations looks the same 
after a length rescaling. }
\label{aging-ising}
\end{figure}

To understand what's going on and what is the physical origin of the out of equilibrium dynamics, 
the most useful thing is to look to the results of a numerical simulation, a picture actually. 
We shall focus on our preferred example: the aging (Monte-Carlo) dynamics of the three
dimensional ferromagnetic Ising model after a quench from the high disordered temperature phase to a temperature at which, in equilibrium, the system is ordered. 
In Fig. \ref{aging-ising}  two snapshots of a 2D cut of the 3D Ising model
after a quench from the high to the low temperature phase are shown
(black color is used for minus spins and white for plus
spins). 
What do we see? Four main features. 
\begin{itemize}
\item The system wants to break the symmetry and does it locally over a length-scale $\xi(t)$ which grows 
with time. 
\item The figure looks self-similar, i.e. after rescaling the unit of length using $\xi(t)$ right and left panel would look
the same. 
\item The system reaches local equilibrium over domains of size $\xi(t)$.
\item The locally equilibrated domains do not persist forever: negative magnetised regions become positively magnetised on larger times (and so on and so forth).
\end{itemize}
These features are generically found in quenches across second-order phase transitions. The results of numerical simulations, experiments and theoretical analyses lead to the emergence of a description of the resulting 
out of equilibrium dynamics which is called {\it coarsening} and it is based on the following well verified assumptions:  
the system breaks the symmetry locally and forms domains inside which a temporary local equilibrium is reached. 
Inside each domain one of the possible low-temperature values of the order parameter is displayed, in the case of the
Ising model these correspond to $\pm m$. The characteristic size of the domains, $\xi(t)$, grow with time generically in a power law way. The out-of equilibrium dynamics is characterized by scaling with respect to $\xi(t)$. For instance, two-time correlation functions can be expressed in terms of scaling functions and power laws: 
\begin{equation}\label{scalingeq}
\langle O_x(\tau+t) O_y(t) \rangle=\frac{1}{|x-y|^\eta}f\left(\frac{|x-y|}{\xi(t)},\frac{\xi(\tau)}{\xi(t)} \right)\qquad \xi(t)\simeq \xi_0 t^z
\end{equation}
where $O_x$ is a generic observable evaluated at site $x$, e.g. $\sigma_x$ for the Ising model. This scaling description only holds on large time and large length-scales. On finite times and finite lengths, less than $\xi(t)$, 
correlation functions display their equilibrium form. This description is confirmed by all results obtained so far
and is considered to hold generically. What changes from one case to another, depending on the type of second-order phase transition that is crossed by quenching, is the growth law of $\xi(t)$ and the specific form of the scaling functions. These need a case by case study. In the following we shall explain in detail the analysis of the Ising model, or more generically quenches across phase transitions in which the order parameter acquires two different values.\\
Before doing that, we conclude this section pointing out that only in the thermodynamic limit the out of equilibrium dynamics goes on forever. For a finite system, instead, when $\xi(t)$ reaches the size of the system there remain few domains and, after a while, only one survives. 
Which values of the order parameter it corresponds to, e.g. plus or minus $m$ for the Ising model, it is a random event that depends on the previous (stochastic) dynamics. After the timescale $\tau_L$, defined by $\xi(\tau_L)\simeq L$ where $L$
is the linear system size, another behaviour sets in: the system remain for a very long time, much larger than $\tau_L$, in one of the ordered states until a very rare random fluctuation creates a system spanning domain corresponding to another ordered state
that then take over. In the case of the Ising model this process corresponds to the creation of an interface between the plus and minus states and it takes place on a time-scale scaling as $\tau\simeq \exp\left(K L^2/T_f \right)$ where 
$K$ is a constant.    
\section{Curvature-driven Domain Growth}
\begin{figure}[H]
\centerline{
\includegraphics[width=0.5\textwidth]{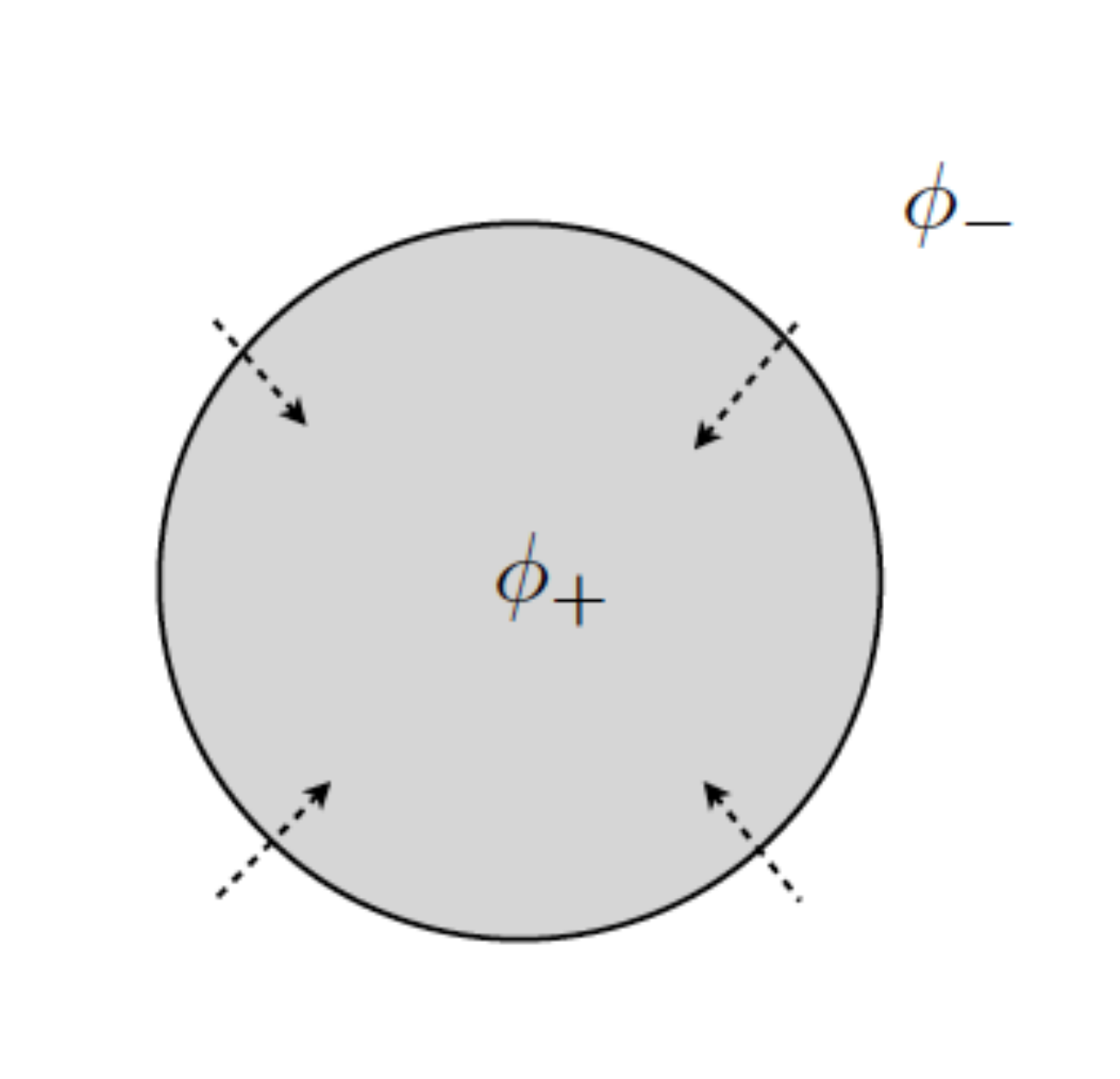}
}
\caption{Sketch of the spherical domain whose evolution is analysed in the main text.}
\label{fig:droplet}
\end{figure}

As explained above, the out of equilibrium dynamics emerging after a quench across a phase transition is due to the evolution of the domains structure. In order to understand the main physical reason for this evolution we shall focus on quenches across phase transitions in which the order parameter acquires two different values and on the simplest case of a single spherical domain.  \\
We model the dynamics using a Langevin Equation for the ordering field (the magnetisation for ferromagnets):
\[
\partial_t \phi ({\mathbf{x}},t)=-\frac{\delta F}{\delta \phi({\mathbf{x}},t)}+\xi({\mathbf{x}},t)\qquad F=\int d{\mathbf{x}} \left(
\frac 1 2 (\nabla \phi({\mathbf{x}},t))^2 +V(\phi({\mathbf{x}},t))\right)
\]  
The potential $V(\phi({\mathbf{x}}))$ has a double well structure with two minima at $\phi_+$ and $\phi_-$
of equal heights. The thermal noise corresponds to a heat bath at temperature $T$, that we take equal to zero 
at first. One can of course wonder how much the results we will find depend on the large amount of assumptions and approximations already done. The answer is that as for critical phenomena, the physics on large time and length-scales, is independent of these assumptions which just determine a few constants but not the scaling laws and the scaling functions.  \\
The initial condition for the dynamics correspond to a spherical domain of radius $R$ corresponding to $\phi({\mathbf{r}})=\phi_+$ for $r\ll R$,  $\phi({\mathbf{r}})=\phi_-$ for $r\gg R$ and a sharp drop from 
$\phi_+$ to $\phi_-$ at $r\simeq R$, see Fig. \ref{fig:droplet}.
 Because of spherical symmetry the zero temperature Langevin equation reads:
\[
\partial_t \phi({\mathbf{r}},t)=\nabla^2 \phi({\mathbf{r}},t)-V'(\phi({\mathbf{r}},t))=\frac{d^2\phi(r,t)}{dr^2}+\frac{d-1}{r}\phi(r,t)-
V'(\phi(r,t))
\]
As we have discussed, the dynamics of large domains takes place on long time scales. In consequence, for $R\gg1$
we search for a scaling solution of the previous equation. It is natural, and as we shall show correct, to focus 
on a travelling wave form: $\phi(r,t)=f(r-R(t))$. Since $\phi(r,t)$ is expected to maintain the shape corresponding to a domain, the function $f(x)$ equals $\phi_+$ for $x\ll 0$, $\phi_-$ for $x\gg 0$ and has a drop between these two values at $x=0$.
By plugging this Ansatz in the evolution equation we obtain:
\[
f''(r-R(t))+\left(\frac{2}{r}+\dot R(t)\right) f'(r-R(t))-V'(f(r-R(t)))=0
\]
Since $f'(r-R(t))$ vanishes for $|r-R(t)|\gg1$ one can replace $\frac{2}{r}+\dot R(t)$ 
by $\frac{2}{R(t)}+\dot R(t)$ up to sub-leading corrections of the order $1/R(t)$.   
The only way for the previous equation to be consistent is that the middle term, which violates the travelling wave form, cancels out. This leads to two equations
\begin{eqnarray}
&&\dot R(t)=-2/R(t)\\
&&f''(x)- V'(f(x))=0
\end{eqnarray}
The solution of the first reads $R(t)=\sqrt{R^2(0)-4t}$. The second is equivalent to the Netwon equation for a
particle moving in the inverted potential $-V(f)$ starting at time $-\infty$ at $\phi_+$ and arriving at infinite time 
at $\phi_-$ (in this analogy $f$ and $x$ respectively denote the particle position and the time).
Energy conservation imposes $f'(x)^2+V(\phi_+)=V(f)$ at all times. From this equation one obtain the solution 
in terms of the implicit equation
\[
\int_0^{f(x)} \frac{df'}{\sqrt{V(f')-V(\phi_+)}}=x
\]
The final result is that a large spherical domain shrinks and is characterized by an interface whose shape 
is given by $f(x)$. The equation $\dot R(t)=-2/R(t)$ is a special form of a more general equation valid for non-spherical
domains \cite{bray} which express the fact the local speed of the interface of a domain is proportional to its curvature. 
In consequence wavy domains tend to flatten out and small compact domain tend to shrink. In particular a spherical
domain of size $R$ shrinks and disappears in a time $t\propto R^2$. This relationship between time and length, that follows from the equation on $R(t)$, is very general and implies that small domains disappear faster than large ones. The net result is that after a time-scale $t$ domains of size less than $\sqrt t$ have disappeared typically. What remains
are mainly domains of size $\sqrt t$ with a typical radius of curvature of the order of $\sqrt t$ \cite{cugliandolobray}. \\
This is a way to justify the scaling assumption: indeed we just obtained that after a time the characteristic length-scale $\xi(t)$ is of the order $\sqrt t$. To fully derive the scaling eq. (\ref{scalingeq}) one needs to show also self-similarity, but this is beyond the scope of these notes. For a  
thorough discussion and an exact derivation of the scaling laws see \cite{cugliandolobray}. \\
Let's now lift the the simplifying assumption of considering quenches to zero temperature. Qualitatively the results
remain the same for any $T_f<T_c$. Inside the domains there is a local equilibration at temperature $T_f$ and although there are now stochastic thermal fluctuations the evolution of domains is in average  still driven by 
the curvature. The only main change is in the pre-factor of the growth law which is renormalised by thermal fluctuations. One finds that the typical length-scale $\xi(t)$ grows as $\simeq D(T_f) \sqrt t$ where $D(T_f)$ vanishes when $T_f\rightarrow T_c$ as a power law (see next section for a derivation of this result). The scaling functions are expected to remain unchanged. \\
Combining all previous results we can now explain the behaviours sketched in Figs \ref{fig:eaging} and \ref{fig:corraging}. The energy approaches as a power law its equilibrium behaviour because the only contribution comes from the boundary of the domains since the regions inside the domains 
are equilibrated. Hence they do not give any extra contribution compared to the equilibrium case, whereas instead the boundaries of the domains cost extra-energy. This extra surface energy is of the order of the surface of the domains $\xi(t)^{d-1}$. Therefore the extra-energy per unit of volume scale as $1/\xi(t)\propto1/\sqrt t$, which leads to a power law approach to the asymptotic value for the energy as a function of time. Concerning the behaviour of correlation functions, Fig \ref{fig:corraging}, the first rapid decrease to the plateau is related to the equilibration inside the domains (toward the value $m(T_f)^2$), whereas the secondary evolution is due to the slow motion of the domains and described by the scaling theory. \\
A natural question is how general is this description of the out of equilibrium dynamics after a thermal quench across
a phase transition. The answer is very general. The main underlying mechanism is always the same: the system breaks the symmetry locally over a length-scale $\xi(t)$. Below this length the system is equilibrated in one of the possible 
symmetry-breaking state, whereas the slow out of equilibrium dynamics is due to the motion of the topological defects that break the symmetry. These are domain walls for the $\mathbb{Z}^2$ symmetry we considered previously, but they can be more complicated objects, e.g. vortices for the $X-Y$ model in two dimensions. In order to obtain a description 
of the out of equilibrium dynamics one has to understand the evolution of the defects. This depends on the kind of symmetry breaking and the nature of the resulting topological defects: it is curvature-driven for 
domain walls as we explained whereas instead is (to a good approximation) diffusion for vortices. This actually leads again a growth law $\xi(t)\propto \sqrt t$ which has however a very different physical origin. In general the growth law is power law for quenches across second-order phase transitions, $\xi(t)\propto t^{1/z}$, and the value of the exponent depends on the dynamics and the kind of symmetry breaking. As in critical phenomena $z$ depends on the dynamical conservation law, for example quenches across the ferromagnetic transition or equivalently spinodal decomposition is characterized by $z=2$ if the dynamics does not conserve the order parameter, as we have 
derived before, and by $z=3$ in the conserved case \cite{bray}. Contrary to critical phenomena $z$ is in general dimension independent and, hence, there is no upper critical dimension. \\  
A final comment concerns quenches exactly at $T_c$, also called critical quenches. These are characterized by growth laws which are different from the previous ones and are related to the critical behaviour characterising the {\it equilibrium} phase transition \cite{calabresegambassi}. 
In particular, the typical lenth-scale over which the system is out of equilibrium, $\xi(t)$, grows as $t^{1/z_d}$ where 
$z_d$ is the exponent relating time and length-scale in equilibrium critical dynamics \cite{calabresegambassi} ($z_d$ and $z$ 
are generically different, e.g. for the 2D Ising model $z_d\simeq 2.17$ and $z=2$ for non-conserved dynamics).
There are no well defined domains in this case; instead the system should be considered a patch-work of equilibrated critically correlated regions of length $\xi(t)$. There are still scaling functions but different from the ones obtained for $T_f<T_c$. 
In summary critical quenches and thermal quenches below $T_c$ both lead to out of equilibrium dynamics, which are however quite different.  

\section{Kibble-Zurek Mechanism and Slow Thermal Quenches}

In the previous secton we studied very sudden quenches. Here we focus instead on very slow ones. This is interesting for several reasons. First, real quenches are generically slow: temperature can be decreased at rates of a fraction of Kelvin per minute, i.e.  on time-scales which are 
very large compared to the typical microscopic ones (picoseconds or less). Moreover, theoretically, studying slow quenches is useful to understand the connection between critical and standard coarsening. \\
Consider a situation in which the order parameter inducing the quench, for instance the temperature, is changed slowly over the time-scale $\tau_Q$:
\[
T(t)=T_i+(T_f-T_i)\frac{t+\tau_Q/2}{\tau_Q}\qquad t\in \left(-\frac{\tau_Q}{2},\frac{\tau_Q}{2} \right)
\] 
We consider a quench across a phase transition, i.e. $T_i>T_c$ and $T_f<T_c$. 
Clearly, if $\tau_Q$ is much larger than the characteristic microscopic time scales then, at least at the beginning of the evolution,
the system follows adiabatically the temperature change. Since by decreasing $T$ the phase transition 
is approached, the system develops long-range equilibrium correlations over a length-scale $\xi(T(t))\propto |T(t))-T_c|^{-\nu}$. 
\begin{figure}[H]
\centerline{
\includegraphics[width=0.8\textwidth]{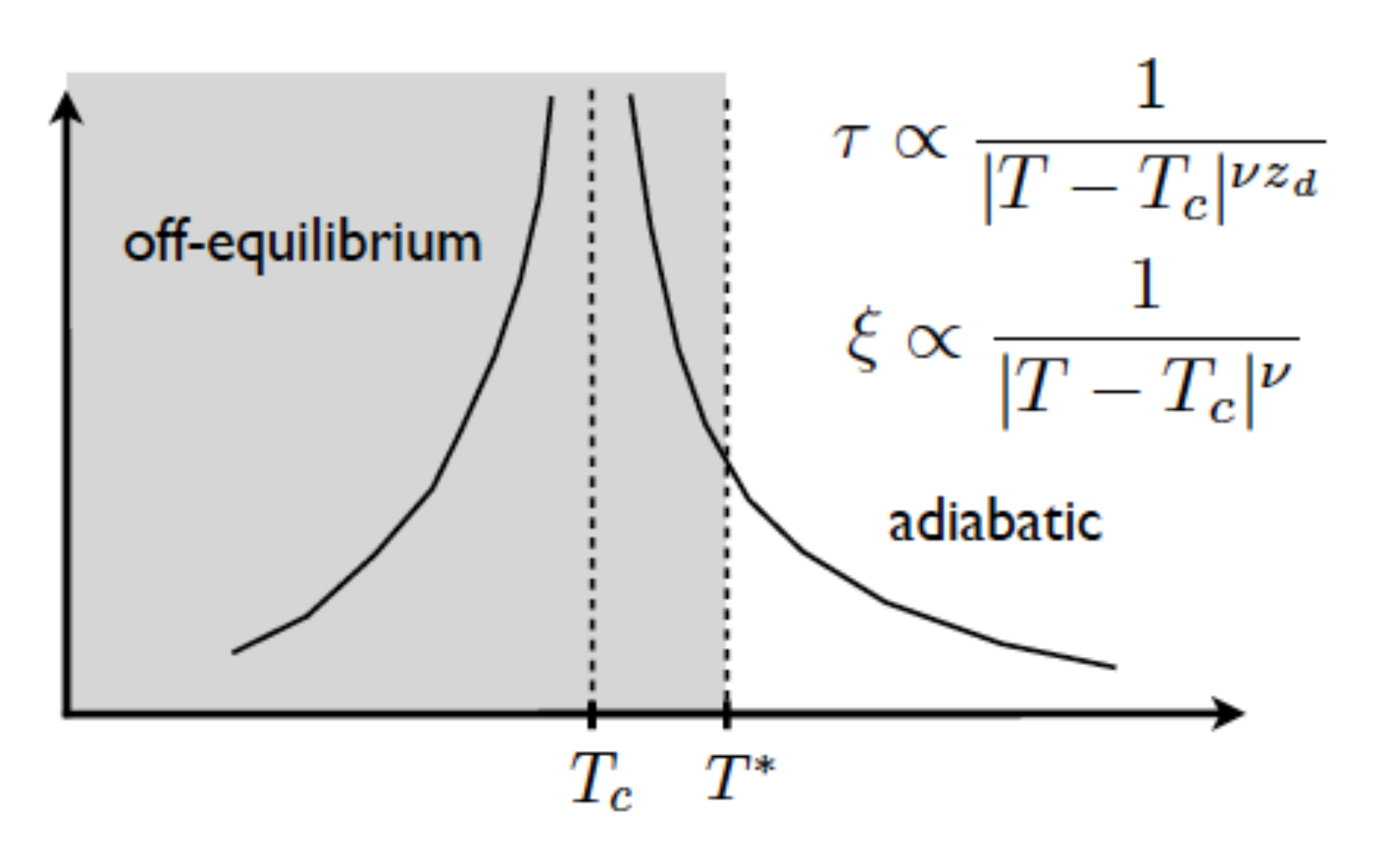}
}
\caption{Sketch of the phase diagram for slow quenches. Until $T^*$ the evolution is adiabatic and the system develops critical 
long-range correlations.}
\label{fig:KZ}
\end{figure}
Because of this "stiffness" due to the spatial correlation the characteristic 
time-scale over which the equilibrium dynamics takes place also increases as $\tau(T(t)) \propto \xi^z_d\propto  |T(t))-T_c|^{-\nu z_d} $, where $\nu$ and $z_d$ are the equilibrium critical exponents, see Fig. \ref{fig:KZ}. This divergence implies that the adiabatic evolution inevitably breaks down when $T(t)$ approaches $T_c$ too closely. In order to estimate when this happens we consider a system that reaches the temperature $T(t)$ in equilibrium, i.e. following adiabatically the temperature change, and compare the interval of time $\Delta t$ that remains before crossing $T_c$. If this is smaller than the equilibrium relaxation time $\tau(T(t))$ then the system has to fall out of equilibrium. In this way we obtain an equation for the temperature $T^*$ at which adiabatic evolution breaks down:
\[
\Delta t= \frac{T^*-T_c}{T_i-T_f}\tau_Q\simeq \tau(T^*) \qquad \Rightarrow \qquad (T^*-T_c)\propto \tau_Q^{-1/(\nu z_d+1)}
\]    
As expected, the smallest is the cooling rate, i.e. the largest is $\tau_Q$, the closest the system can approach $T_c$ remaining in equilibrium. Nevertheless, for any large but finite value of $\tau_Q$, the adiabatic evolution eventually breaks down. Our main aim is to find the growth of the characteristic length-scale $R(t)$ over which the system is ordered as a function of time.\\
Physically, there are three regimes: first the system follows adiabatically the temperature change and develops critical correlations. 
When it goes out of equilibrium at $T^*$, it is a collection of critically correlated regions that start to grow by critical coarsening. There exist domains in this regime, but they are fractal and characterized by a magnetisation which is not extensive. 
Eventually, a cross-over from critical to standard coarsening takes place: domains become compact, the magnetisation inside
becomes extensive and the evolution starts to be dominated by the curvature of the boundary of the domains. In the following we assume scaling and use that $R(t)$ tracks the equilibrium correlation length during the adiabatic regime and at very long times we have to recover the coarsening growth law $R(t)\propto t^{1/z}$. Since the 
characteristic time-scale governing the cross-over from adiabatic to off-equilibrium evolution is $\tau(T(t))$ the scaling law reads:
\[
R(t)=\xi(T(t))f\left(\frac{t}{\tau(T(t))}\right)
\]
where the two regimes, adiabatic and standard coarsening, respectively corresponds to the $-\infty$ and $+\infty$ limit of the scaling variable 
$x=\frac{t}{\tau(T(t))}$. In order to recover the expected growth law in these regimes $f(x)$ has to go to one for $x\rightarrow -\infty$
and to diverge as $x^{1/z}$ for $x\rightarrow +\infty$. The latter result allows us to obtain the pre-factor of the coarsening growth law: one finds $R(t)\propto (T(t)-T_c)^{\nu (-1+z_d/z)}t^{1/2z}$ and, hence, $D(T)\propto   (T_f-T_c)^{\nu (-1+z_d/z)}$ if the temperature is kept constant at $T_f$ at long times. \\
Kibble and Zurek originally studied this problem in cosmology, largely independently from coarsening in statistical physics. They were mainly interested in the density of topological defects created out of equilibrium 
when the temperature $T(t)$ is below $T_c$ and symmetric respect to $T^*$ (the end of the so-called impulse regime). 
In the time interval between the adiabatic regime and the time at which $T(t)$ becomes equal to $T_c-(T^*-T_c)$ the length $R(t)$ grows following the critical coarsening law. As it can be explicitly checked, this increase just leads to a value that is greater than the one
at the end of the adiabatic regime by a numerical factor greater than one. In consequence the density of topological defects $N$ is proportional to $R(t^*)^{-d}\propto \tau_Q^{-d/z}$. Because of the original works by Kibble and Zurek, this result and some of the arguments
explained above go under the name of Kibble-Zurek mechanism for the formation of topological defects, see the reviews\cite{KZ1}. The 
explanations presented above are based on \cite{sicilia}.  The Kibble-Zurek mechanism has been recently the focus on an intense research activity in the context of quantum quenches for isolated quantum systems. On this point, 
see E. Altman's lectures in this book. 
 \section{Quantum fluctuations}
\begin{figure}[H]
\centerline{
\includegraphics[width=0.7\textwidth]{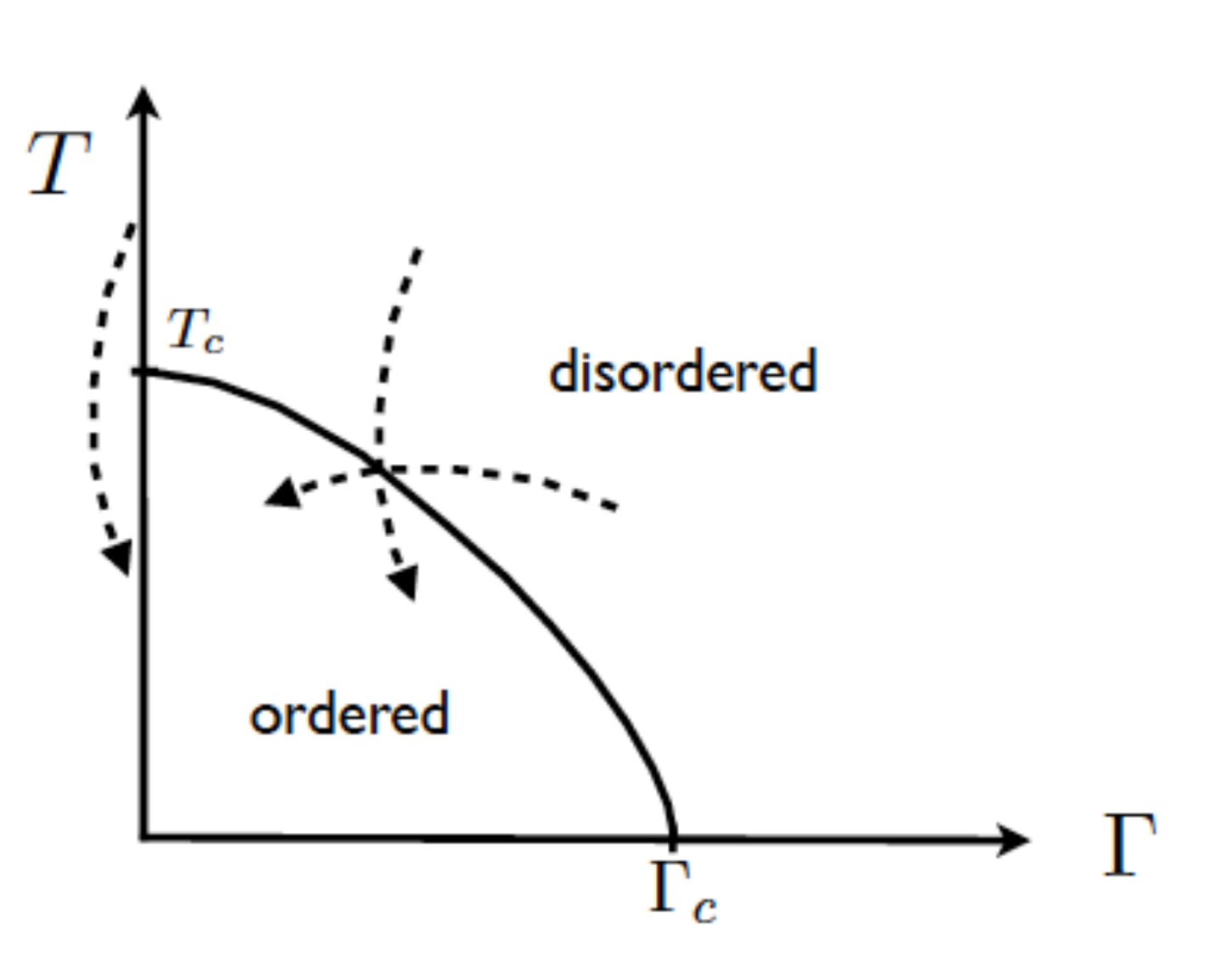}
}
\caption{Typical form of a phase diagram for a quantum system displaying a second-order phase transition. The dashed line indicates several different quenching protocols.}
\label{fig:QCP}
\end{figure}

\begin{figure}[H]
\centerline{
\includegraphics[width=0.8\textwidth]{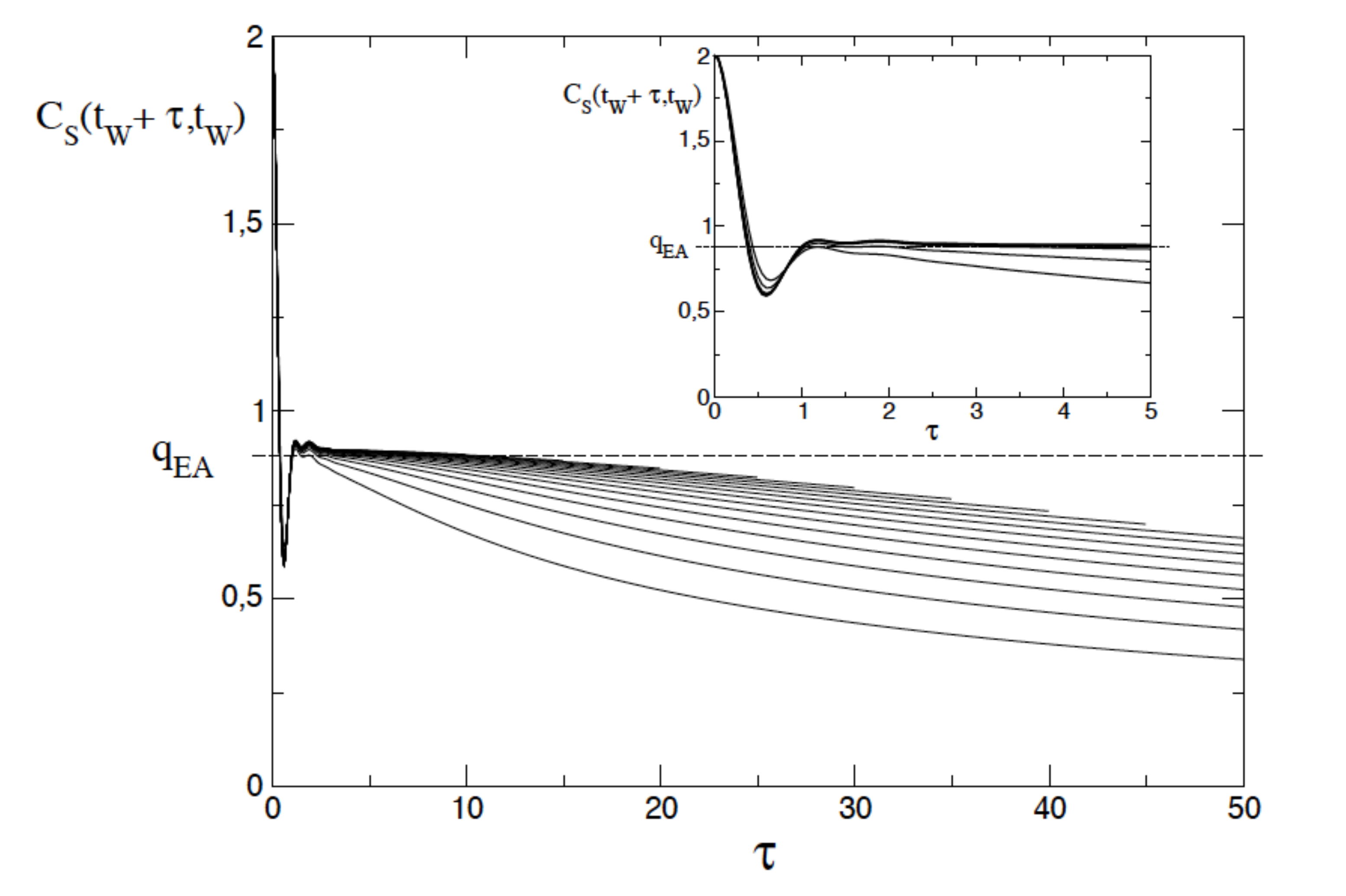}
}
\caption{Typical form of a correlation function after a quench in quantum systems (result from [Biroli and Parcollet 2002]
). The inset shows the evolution for times $\tau \ll t_w$: the system displays equilibrium dynamics. In the main panel, for $\tau\propto t_w $, ageing behaviour is manifest.}
\label{fig:corragingquantum}
\end{figure}

 Let us now take into account quantum fluctuations. A generic phase diagram for a quantum system undergoing a second order 
 phase transition is the one shown in Fig. \ref{fig:QCP} in which $\Gamma$ represents the strength of quantum fluctuations \cite{sachdev}. At zero temperature a quantum critical point (QCP) takes place at $\Gamma_c$ from which originates the critical line separating the disordered 
 and the ordered phases. This line ends for $\Gamma=0$ at the classical value of the critical temperature. Consider now quenches
 where the control parameters, $T$ or $\Gamma$, are changed in order to cross the transition line, see Fig.  \ref{fig:QCP}
 (quenching $T$ is experimentally feasible by changing the bath temperature, changing $\Gamma$ is a more theoretically oriented protocol). 
The resulting out of equilibrium dynamics is very similar to the one described in the classical case: the system equilibrates locally over a length-scale of size $\xi(t)$ that grows with time as a power law.  In particular, for a quantum ferromagnets or generically for systems with only two competing low temperature states, one finds again domain growth. 
Correlation functions display a behaviour similar to the one discussed before. Equilibration seems to takes place on short-time scales toward a plateau value which is related to the value of the order parameter at $T_f,\Gamma_f$. However, ageing behaviour, which is related to the motion of topological defects that break the long-range order on the scale $\xi(t)$, eventually sets in. As it is clear from Fig. \ref{fig:corragingquantum}, although quantum fluctuations show up in the first part of the relaxation (for instance the correlation function displays oscillating behaviour due to quantum coherence) they seem to be absent in the out of equilibrium regime. This is indeed the case since this regime takes place at very long time. Hence, dissipation and decoherence inevitably kick in and quantum coherence is lost. In consequence the motion of the topological defects is 
classical and, accordingly, the out of equilibrium is governed by {\it classical coarsening laws}. Whether exceptions to this 
conclusion can be found is not clear. In particular quenches across QCP, i.e. by changing $\Gamma$ and keeping the bath temperature at $T=0$, are still not well studied. Since in this case, depending on the bath density of states, the coherence effects can be much 
long-ranged in time there could be room for a more important influence of quantum fluctuations on out of equilibrium ageing dynamics.\\
Recently, there has been an intense research activities on quantum quenches, i.e. protocols in which an isolate quantum system  
initially in the ground state is put of out equilibrium by changing suddenly a coupling in the
Hamiltonian. The discussion of the resulting dynamics is out of the scope of these lectures and, moreover, is an issue still not completely settled. By quenching the coupling one effectively injects energy inside the system since for the new Hamiltonian the old ground state is a wave-function consisting in a mixture of excited states. Hence, we expect is that the rapid degrees of freedom 
equilibrate and provide a bath for the slow degrees of freedom that then undergo classical coarsening at long times. 
This is indeed what happens for classical systems in analogous situations \cite{chate}. 
 An unexpected behaviour could result from the non-thermal fixed point discussed in Berges' lectures. 
Indeed it was found that depending on the type of initial condition isolated quantum systems could approach a quasi stationary 
critical behaviour reminiscent of turbulence. The competition between this behaviour and coarsening is presently not understood. 
Another effect found in the literature are the so-called dynamical transitions, see e.g. \cite{DT} and refs therein. These could be actually related to the non-thermal fixed point just discussed above. As the reader has certainly understood from these few lines: the research on quantum quenches is still very much ongoing. Surprises and new explanations are to be expected.     
 
\chapter{Quenched Disorder and Slow Dynamics}
The presence of quenched disorder can induce very slow dynamics and lead to new kind of out of equilibrium phenomena.
In this chapter we present some of them. First we discuss the effect of disorder on equilibrium dynamics and then 
focus on out of equilibrium ones. This chapter is just an introduction, a comprehensive presentation
would deserve an entire series of lectures on its own, see for instance the Les Houches lecture notes \cite{leticialeshouches}. Some topics will be left 
out for lack of space, in particular glasses and spin-glasses that would need quite a long discussion.     
\section{Broad distribution of relaxation times}
One of the most relevant phenomenological consequences of quenched disorder is the presence of several relaxation times: the heterogeneity induced by the disorder leads to the simultaneous presence of regions that relax fast and other that relax slow. Instead of having a characteristic time-scale $\tau$, disordered systems are characterized by a distribution
of relaxation times $P(\tau)$. A broad $P(\tau)$ can have an important impact on the dynamics leading to rare regions that 
take a very large time to relax. This leads to a strong non-exponential relaxation in contrast to non-disordered systems that usually display an exponential relaxation\footnote{When they have no quantities conserved by the dynamics.}. \\
In the following, by focusing on the example of random ferromagnets, we shall present a general phenomenon taking place in disordered systems, which goes under the name of Griffiths phase. The Hamiltonian of a simple model of a random ferromagnet
reads:
\[
H=-\sum_{\langle i,j \rangle}J_{ij}\sigma_i^z \sigma_j^z
\]
where the couplings are independent and identically distributed random variables with the distribution
 $P(J_{ij})=(1-p)\delta(J-J_1)+p\delta(J-J_2)$ and $J_1$ and $J_2$ are both positive ($J_2>J_1$). This system displays a ferromagnetic 
 phase transition at a temperature $T_c$ which depends on the values of $p,J_1,J_2$. It is clear that $T_c$ has to be a increasing function of  $p$. In particular $T_c(p,J_1,J_2)$ is bounded between the temperatures, $T_c^{J_1}$ and $T_c^{J_2}$, at which a system with all couplings equal to $J_1$ or $J_2$ orders. Numerical simulations of the equilibrium dynamics show that at high temperature, roughly above $T_c^{J_2}$, correlation functions display exponential relaxation whereas below they become more and more stretched.\\
 This result can be understood \cite{braygriffiths} (and actually proven rigorously \cite{martinelli}) by noticing that there are regions inside the system that have all couplings equal to $J_2$. Below $T_c^{J_2}$ they are "ordered", or more correctly they flip very slowly between the two low temperature magnetised states and, hence, induce a very slow dynamics. The temperature regime below $T_c^{J_2}$ is called Griffiths phase\footnote{The name was chosen in relation with the singularities that Griffiths found studying 
 the thermodynamics of disordered systems \cite{griffiths}.}.\\
 Let's be precise and recall that for an ordered ferromagnet of size $L$ the time to flip between the  two low temperature magnetised states is of the order $\tau_L\simeq\exp(2J_{eff}(T)L^{d-1}/T)$, where $J_{eff}(T)$ plays the role of a surface tension. 
 It equals $J$ at zero temperature, it is a decreasing function of $T$ and vanishes at $T_c$. The  spatially averaged equilibrium correlation function
 can be easily shown to be self-averaging, hence:
 \[
 C(t)=\frac 1 N \sum_i \langle\sigma_i^z(t) \sigma_i^z(0) \rangle=\overline{\langle\sigma_0^z(t) \sigma_0^z(0)\rangle}
 \] 
 If the origin belongs to a region of linear size $L$ where all couplings are equal to $J_2$ its relaxation is given by $e^{-t/\tau_L}$. 
 Therefore considering only cubic regions centred around the origin we can obtain easily the bound:
 \[
 C(t)\ge \sum_L p^{L^d} \left(1-p^{2^dL^{d-1}} \right)e^{-t\exp(-2J_{eff}L^{d-1}/T)}
 \]
 where the first term is the probability that in the whole cubic region of size $L$ couplings are equal to $J_2$ and on the boundary of the 
 cube are equal to $J_1$. This sum can be performed by the saddle-point approximation when $t$ is large: one finds that the 
 value $L(t)$ that dominates the sum is proportional to $\left(\frac{T\log t}{J_{eff}}\right)^{1/(d-1)}$ and that:
 \[
 C(t)\ge K_1\exp \left(-K_2 (\log t)^{d/(d-1)} \right)
 \]
 where $K_1,K_2$ are two positive constants that depend on $p,T,J_1,J_2$. This result shows that even in the high-temperature regime, before the phase transition takes place, equilibrium correlation functions 
 relax much slower than exponentially. This is due to slow and rare Griffiths regions. We focused on equilibrium dynamics but of course 
 this effect is also present if one quenches the temperature in the high temperature phase ($T_f>T_c$). 
 
   \section{Broad distribution of low energy excitations}
   The Griffith phase studied in the previous section has a counterpart in quantum disordered systems. Analogously to the 
   existence of rare slow relaxing regions, quantum disordered systems contain rare regions that give rise to very low energy excitations.    
   In particular this leads to a density of states that vanishes as a power law at low frequency contrary to non-disordered quantum systems that instead display a gap whose value is related to the typical energy scale. 
    As before, let us 
   describe this general phenomenon, that goes under the name of quantum Griffiths phase \cite{sachdev}, in a simple example which is  the transverse 
    random Ising ferromagnet whose Hamiltonian reads:
    \[
    H=-\sum_{\langle i,j \rangle}J_{ij}\sigma_i^z \sigma_j^z-\Gamma \sum_i \sigma_i^x
    \]
    where $\sigma^x,\sigma^z$ are Pauli matrices and $J_{ij}$ positive and independent random variables with the same distribution defined in the previous section. At zero temperature 
    this model displays a phase transition between an ordered ferromagnetic phase for $\Gamma<\Gamma_c$ and a disordered 
    paramagnetic phase for $\Gamma>\Gamma_c$, see \cite{sachdev}.\\
    $\Gamma_c$ plays the same role of $T_c$ before: it depends on $p,J_1,J_2$ and is bounded between the values $\Gamma_c^{J_1} $ and $\Gamma_c^{J_2}$. The phase below $\Gamma_c^{J_2}$ is called quantum Griffiths phase and is characterized by the existence by very low energy excitations. The main arguments parallel the one presented in the previous sections. The system contains regions of linear size $L$ inside which all couplings are equal to $J_2$. Their density is 
of the order of $p^{L^d}$. For $\Gamma< \Gamma_c^{J_2}$ these regions are "ordered" hence the lowest excitation energy 
scales as $\omega_L\simeq\exp(-KL^d)$. This corresponds to a multitude of virtual processes that allow to go from one ordered state to the other and, hence, to lift the degeneracy between them. \\
The density of states associated to the spin-spin correlation function is given by:
\[
\rho(\omega)=\overline{\sum_\alpha |\langle \alpha|\sigma_0^z|GS\rangle|^2 \delta(\omega-[E_\alpha-E_{GS}])}
\] 
where the index $\alpha$ is associated to the many body eigenstates and GS denotes the ground state. The contribution to the above sum due to Griffiths regions is of the form: $\sum_L \exp(-[\log p]L^d)\delta(\omega-\exp(-K'L^d))$. Evaluating this sum for large $L$ one finds that the Griffiths regions contribution leads to a power law in the density of states which, when $\omega \rightarrow 0$, behaves as:
\[
\rho(\omega)\simeq \frac{\omega^{\frac{K}{K'}-1}}{(\log \omega)^{(d-1/d)}}
\]
Thus, below $\Gamma_c^{J_2}$, the density of states is not gapped and instead display non-universal power laws at small frequency. This in turn implies that the zero-temperature equilibrium dynamics is very slow below $\Gamma_c^{J_2}$.  

\section{Activated dynamics scaling and zero temperature fixed points}
\begin{figure}[H]
\centerline{
\includegraphics[width=0.7\textwidth]{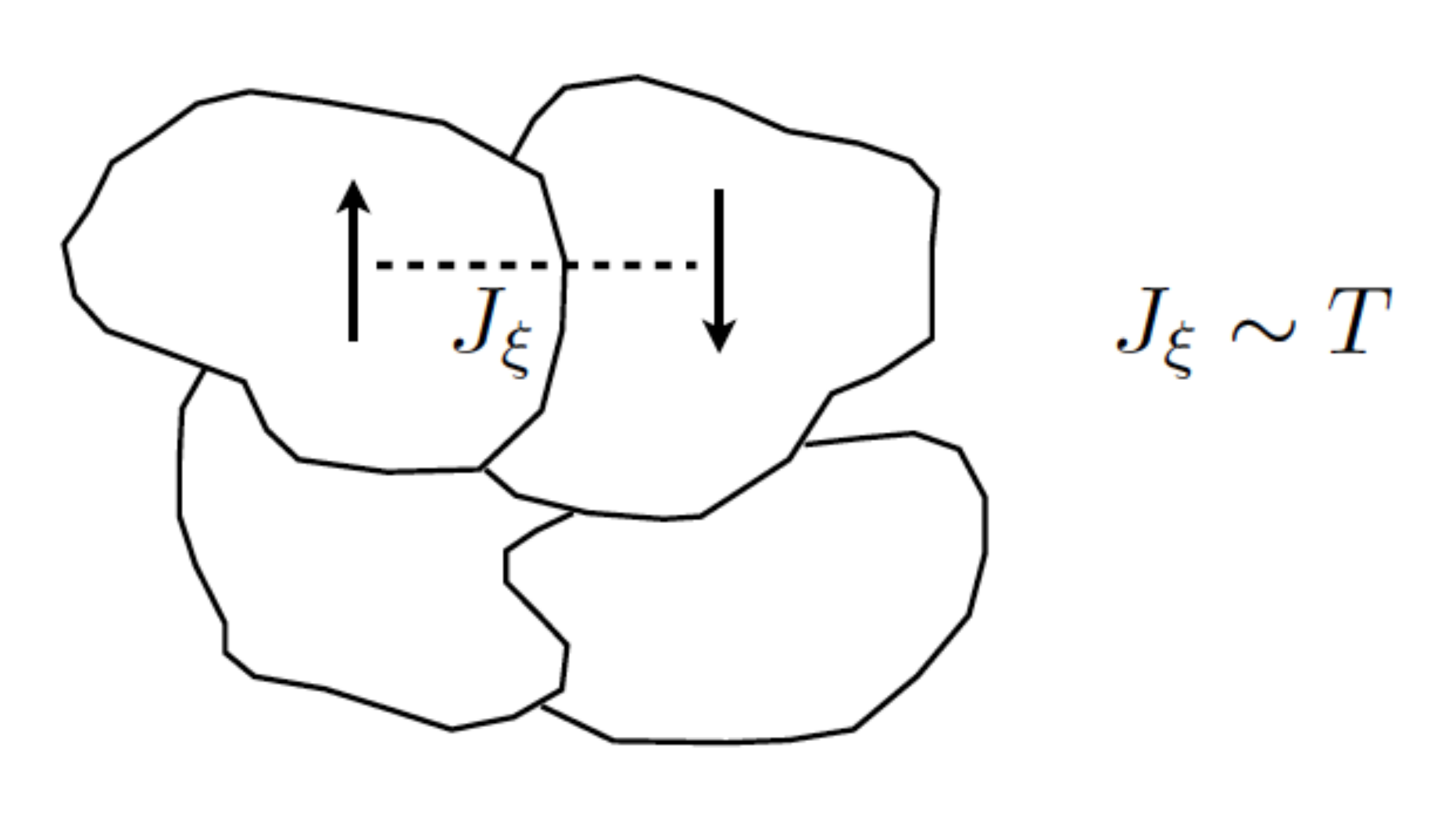}
}
\caption{Sketch of the renormalised effective model on the scale $\xi$ for the Ising model.}
\label{fig:sketchZTFP}
\end{figure}

\begin{figure}[H]
\centerline{
\includegraphics[width=0.7\textwidth]{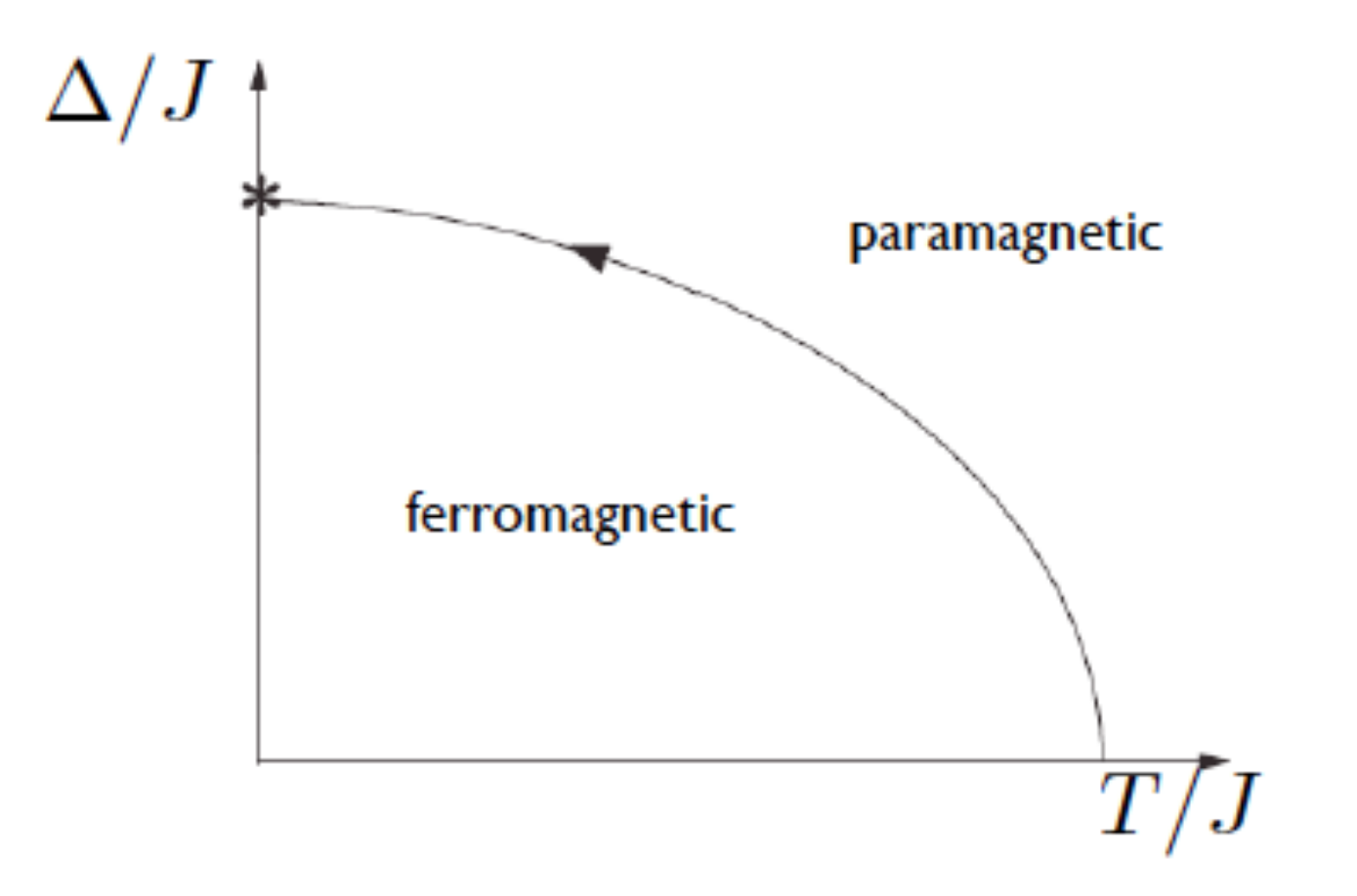}
}
\caption{Sketch of the phase diagram for the Random Field Ising model. The arrow on the transition line indicates that 
the RG flow is controlled by the zero temperature fixed point.}
\label{fig:RFIMPD}
\end{figure}

\begin{figure}[H]
\centerline{
\includegraphics[width=0.7\textwidth]{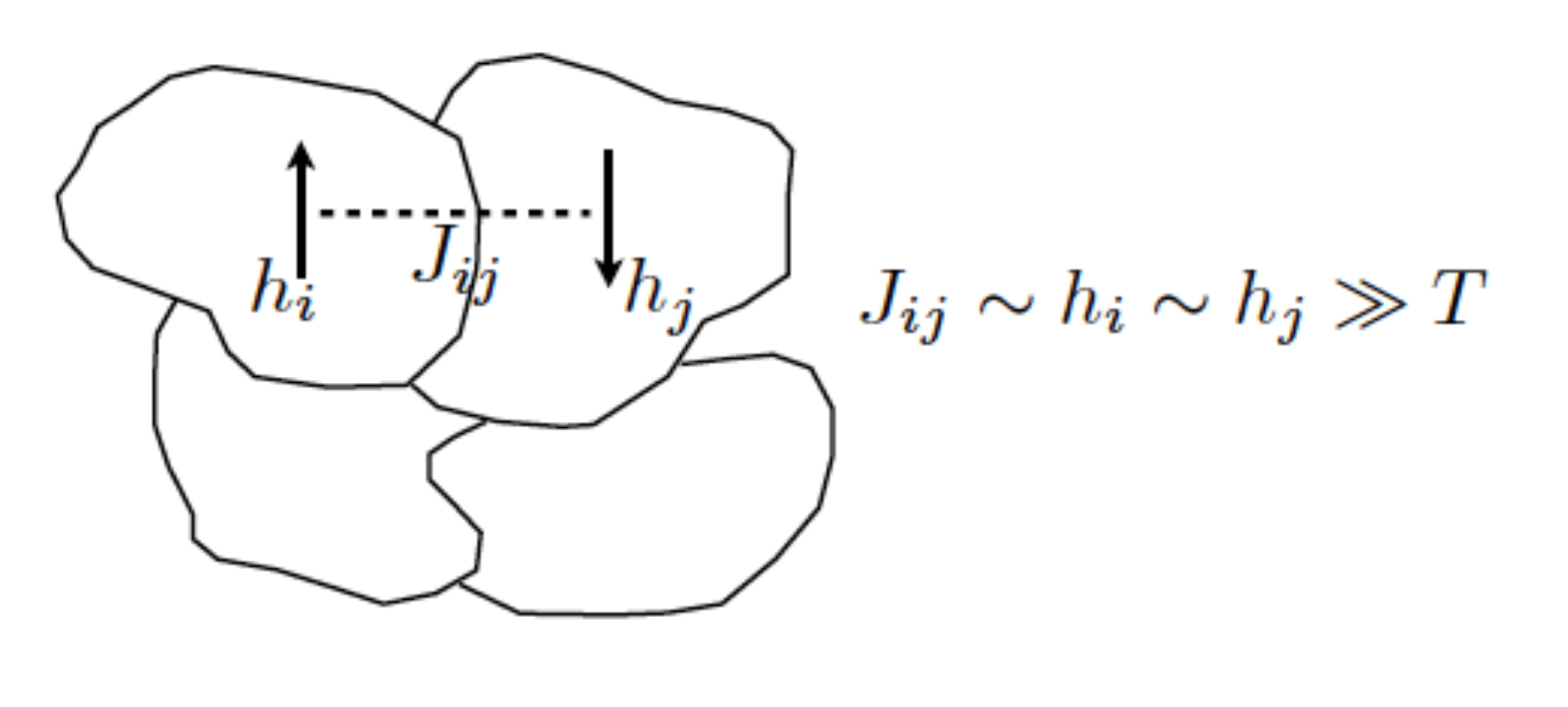}
}
\caption{Sketch of the renormalised effective model on the scale $\xi$ for the Random Field Ising model.}
\label{fig:RFIMEM}
\end{figure}

In the previous sections we studied the effects of quenched disorder on the equilibrium dynamics and
showed that even if the system is not ordered and far from the critical point the dynamics can become very slow.\\
Here we want instead to study the effect of the disorder on a critical point. From the point of view of the dynamical behaviour, 
one of the most remarkable effect of quenched disorder is to produce in some cases a critical equilibrium dynamics that is ultra-slow. In particular a divergence of relaxation times that is super-Arrhenius instead of being power law as in standard second order phase transition. This is due to a relationship between time and length scales that is exponential, 
$\tau\simeq \tau_0\exp(\xi(T)^\psi/T)$, instead of being power law ($\tau\propto \xi^z$). This phenomenon was called activated dynamic scaling \cite{fisher,fisher2} and is due to the existence of a new kind of renormalisation group fixed point that we shall now describe.  \\
Let us first recall that for standard second order phase transition, close to the critical point, the system is formed by correlated regions of linear size $\xi$. Each region can be in one of the low temperature phases, e.g. positively or negatively magnetised in the case of ferromagnets. In a (crude) real space renormalisation group sense, each correlated region 
can be represented by a coarse grained degree of freedom, associated to the order parameter. The value of this degree of 
freedom indicates in which low temperature phase the correlated region is. The interaction between correlated regions 
is represented by one (or more) coupling between these degrees of freedom. For instance, in the ferromagnetic case one associates to each region a spin and introduce a ferromagnetic coupling as interaction between regions, see Fig. \ref{fig:sketchZTFP}. This 
RG picture emerges naturally in real space renormalisation group treatments such as the Migdal-Kadanoff one \cite{bookRG}.
The fundamental property of second order phase transition is that the coarse grained energy scale $\Delta F_{\xi}$ obtained by renormalisation on the length-scale $\xi$ is of the order of the temperature\footnote{All these statements and the following ones are only true 
below the upper critical dimension where scaling holds.}. 
In the previous approximate description this means that $J_\xi\propto O(T)$. In more formal treatment one indeed finds that the renormalised theory on scales larger than $\xi$ is perturbative and high temperature like. This has several important consequences. The most important one is that correlated regions are "active", i.e. they flip rather easily, since 
$\tau\propto \exp(\Delta F_{\xi}/T)$. Actually, it is the pre-factor to the Arrhenius law that matters and gives the usual 
power law scaling between time and length characteristic of critical dynamics: $\tau \propto \xi^z$. Another remarkable consequence is that the singular part of the free energy, which scales as $|T-T_c|^{2-\alpha}$, 
can be obtained as $\Delta F_{\xi}/\xi^d\propto |T-T_c|^{\nu d}$. This directly implies the hyper-scaling relation 
$2-\alpha =\nu d$. \\
In presence of disorder this picture can break down. A new kind of fixed point can emerge, for which the typical energy scale on the length $\xi$, becomes much larger than the temperature. In this case 
$\Delta F_{\xi}$ is found to diverge as $\xi^\theta$ approaching the phase transition. 
In consequence the effective model on the scale 
$\xi$ is now characterized by couplings much larger than the temperature\footnote{In some analytical treatments where the energy scale is kept fixed under renormalisation, it is the temperature scale that is renormalised: $T_\xi\propto T/\xi^\theta$.}. Since what matters is the ratio between couplings and temperature one can describe this situation as infinite disorder couplings or zero temperature fixed point. One of the main consequence 
is that now the dynamics (for instance the Monte-Carlo dynamics) of the correlated regions becomes extremely slow. The relaxation time is given by the 
Arrhenius law applied to a correlated region: $\tau\propto \exp(\Delta F_{\xi}/T)$. This in turn implies a super-Arrenius divergence for the characteristic time-scale of critical dynamics: 
\[
\xi\propto \frac{1}{|T-T_c|^{\nu}} \qquad \Rightarrow \qquad \tau\propto \tau_0 \exp \left(\frac{K}{|T-T_c|^{\theta\nu}} \right)
\] 
An example of this behaviour is provided by the Random Field Ising model, whose Hamiltonian reads:
\[
H=-J\sum_{\langle i,j \rangle }\sigma_i^z\sigma_j^z-\sum_i h_i \sigma_i^z
\]
where $J$ is a ferromagnetic coupling and $h_i$ random fields corresponding to independent and identically distributed random variables of mean zero and variance $\Delta$. A sketch of the phase diagram of this model is presented in Fig. \ref{fig:RFIMPD}. 
The transition line separates the paramagnetic and the ferromagnetic phases. Except for $\Delta=0$ the critical behaviour 
is governed by a zero-temperature fixed point. The sketchy renormalisation group picture in terms of effective model on the scale 
$\xi$ is shown in Fig.  \ref{fig:RFIMEM}: correlated regions are described by Ising spins interacting via random (ferromagnetic) couplings and coupled to random external fields. 
As found in the Migdal-Kadanoff treatment \cite{MKRFIM}, which makes this RG picture concrete, both the coupling and the fields scale as $\xi^\theta\gg T$. More refined treatments, such as 
non-perturbative functional RG \cite{TarjusRFIM} and numerical simulations \cite{middleton} 
confirm and precise these results \cite{TarjusRFIM} and lead to the numerical values for the exponent: $\theta_{3D}\simeq 1.6$
and $\nu_{3D}\simeq 1.5$. As explained above, this implies an extremely fast growth of the relaxation time approaching the transition \cite{fisher2,tarjusbalog}:
$\tau \propto \tau_0 \exp \left(\frac{K}{|T-T_c|^{\theta_{3D} \nu_{3D}}} \right)$ with $\theta_{3D}\nu_{3D}\simeq 2.4$.

\section{Domain growth and coarsening in disordered systems}
\begin{figure}[H]
\centerline{
\includegraphics[width=0.6\textwidth]{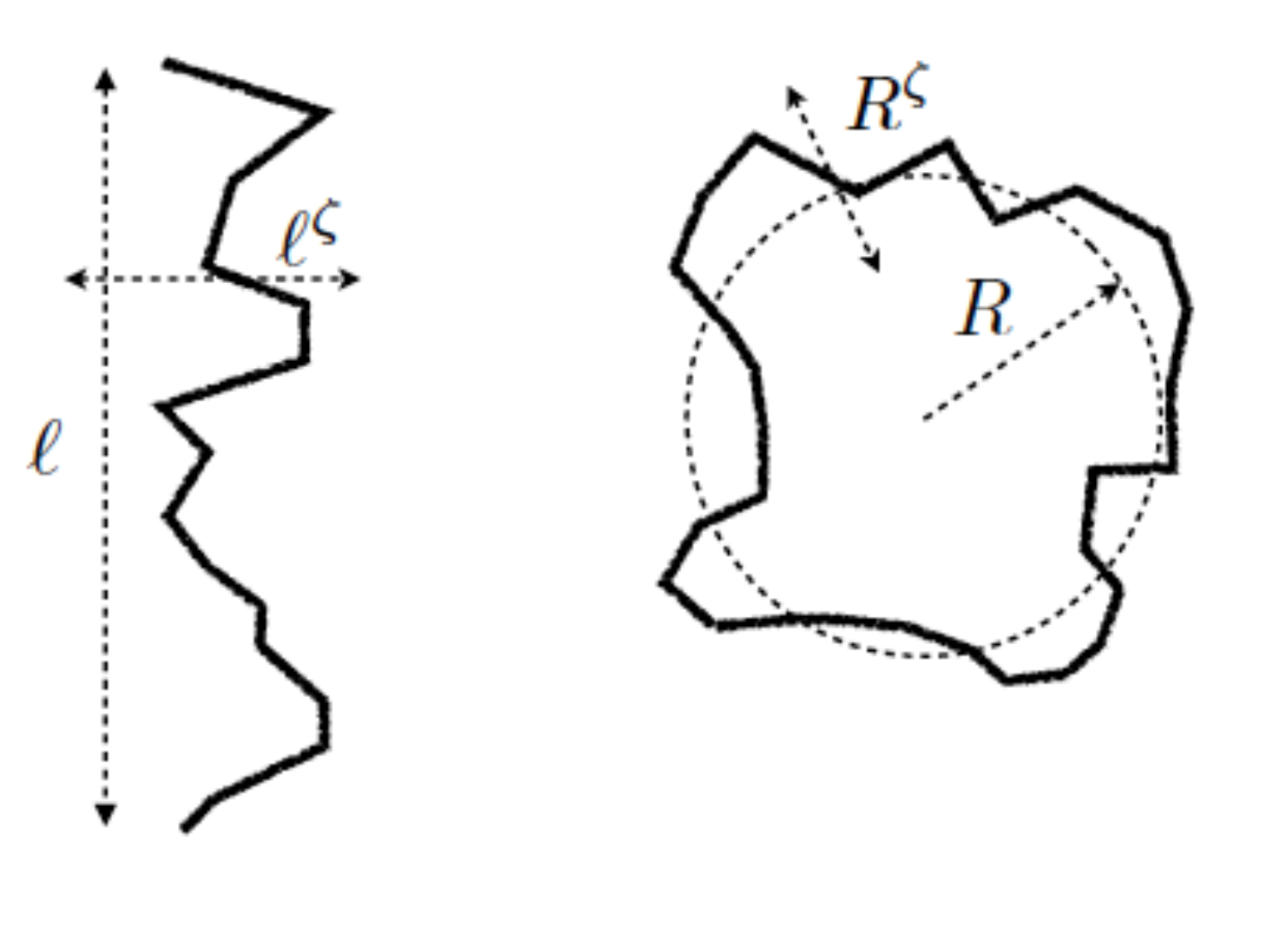}
}
\caption{Sketch of the wandering of domain walls in random ferromagnets.}
\label{fig:RM}
\end{figure}

\begin{figure}[H]
\centerline{
\includegraphics[width=0.55\textwidth]{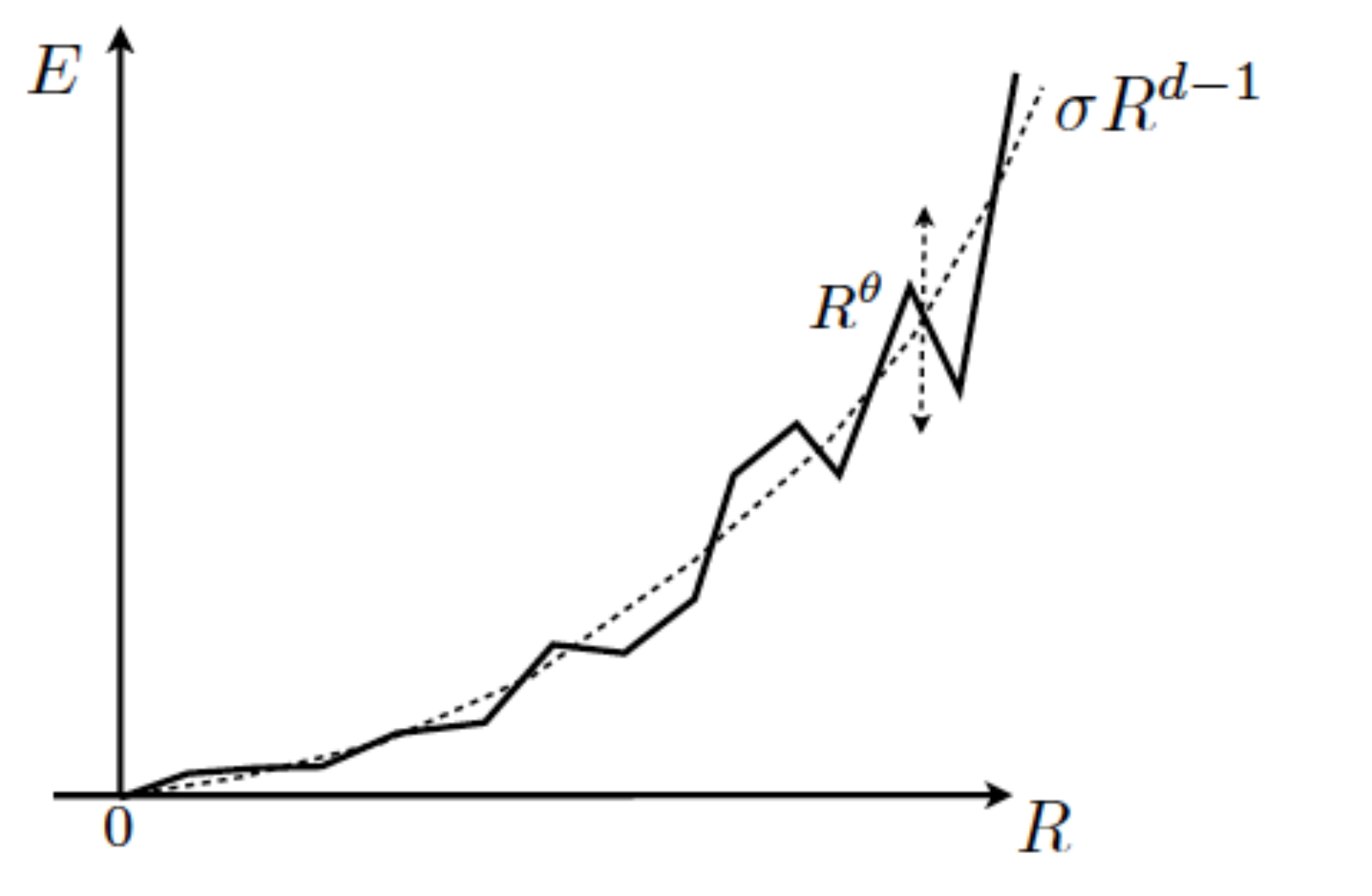}
}
\caption{Sketch of the energy as a function of the domain size: on top of the average contribution $\sigma R^{d-1}$ there is 
a fluctuating part of the order $R^\theta$ which gives rise to barriers whose heights increase as $R^\theta$.}
\label{fig:EDWR}
\end{figure}

Let us now consider the effect of disorder on the off-equilibrium dynamics due to thermal quenches. We consider, as in the previous chapter, quenches across phase transitions. The physics is very similar to the non-disordered case: there is a tendency to local equilibration, that takes place over a length-scale $\xi(t)$, but the system remains always out of equilibrium because of ageing dynamics (due to the slow motion of topological defects). 
What changes is the dynamics of the topological defects which now evolve in a disordered environment. Let us focus on the
case of domain growth. Without quenched disorder the motion of domain is curvature-driven as we explained. In presence
of disorder, instead, the motion of the domain boundary is governed by the competition between the tendency to reduce the interface energy 
and the gain obtained by wandering in profitable regions of the random environment, see \cite{reviewgiamarchi,reviewgiamarchi2}. \\
The study of domains boundary in random environment is another case in which the large scale physics is governed by a zero-temperature fixed point. Theory and numerical simulations \cite{reviewgiamarchi,reviewgiamarchi2} have established that in several cases, in particular in the case 
of the random bond and random field ferromagnets, the domain boundary is rough: on the scale $\ell$ it has transverse fluctuations
of the order of $\ell^\zeta$, see Fig. \ref{fig:RM}. 
The free-energy on the scale $\ell$ is given by two contributions:
\[
\Delta F_\ell=\sigma \ell^{d-1}+\Upsilon \ell^\theta
\] 
the first one is the surface tension and is a non fluctuating quantity, in the second one instead $\Upsilon$ 
is a random variable with a universal distribution (in the RG sense). Plotting the energy as a function of the domain radius 
one finds a curve similar to the one in Fig. \ref{fig:EDWR} which in average is the same than for non-disordered system but 
displays large scale barriers. A spherical domain of size $\ell$ has a tendency to shrink as for non-disordered
systems since in this way the interface energy is decreased. However, in order to do so, it has to go over barriers 
that scale as $\ell^\theta$. 
It is assumed that domain growth for disordered system can still be described by 
a scaling theory in terms of a growing length $\xi(t)$. 
However, now the relationship between time and length is no more power law, it is replaced by an activated dynamic relationship:
\[
\tau_\ell\propto \exp\left(\Upsilon \ell^{d-1}/T\right) \Rightarrow \xi(t)\propto (T\log t)^{1/\theta}
\]   
The scaling relationship (\ref{scaling}) reads in this case: 
\[
\langle O_x(t') O_y(t) \rangle=\frac{1}{|x-y|^\eta}f\left(\frac{|x-y|}{(T\log t)^{1/\theta}},\frac{\log t'}{\log t} \right)
\] 
In practice, because of the logarithms in the scaling laws domain growth in disordered system is a particularly slow process. 
It takes a very long-time to make increase the length $\xi(t)$ substantially.

\section{Infinite randomness fixed points}
The effect of quenched disorder on quantum dynamics also gives rise to new remarkable phenomena. This is actually a topic 
that is attracting a lot of attention recently and on which a lot of work is ongoing. Instead of giving a comprehensive presentation we shall briefly discuss what is one of the most influential concept and framework that actually mirrors what explained in the previous sections for classical dynamics.
The starting point of it was the study of random transverse Ising chains, for example the one with Hamiltonian:
\[
H=-\sum_i J_i \sigma_i^z \sigma_{i+1}^{z}-\frac{1}{2}\sum_i h_i \sigma_i^x
\] 
where $J_i>0,h_i>0$ are random quenched variables.  This model displays a phase transition between a disordered and a ferromagnetic ally ordered phase when $\Delta=\overline{\log h}$ approaches a critical value $\Delta_c$ from above. 
D. S. Fisher unveiled that the corresponding critical point is governed by a new kind of RG fixed point that is now called 
infinite randomness fixed point. By using the Dasgupta-Ma decimation procedure, a kind of real space RG where the most 
strongest couplings in the system are decimated sequentially, he showed that at criticality $\log J_i$ and $\log h_i$ grow
as $\sqrt \ell$ where $\ell$ is the length-scale over which the system is coarse-grained. This give rise to several remarkable phenomena such as strong difference between typical and average quantities and to an intricate critical behaviour. \\
The most 
important point for our discussion is that the RG behaviour actually resembles very much the one described above for zero-temperature fixed point. Indeed, in his seminal papers D. S. Fisher proposed to call it "tunnelling dynamics scaling" in analogy 
with the "activated dynamics scaling" discussed previously. Although for a long time this RG techniques 
and this kind of fixed point were used to investigate equilibrium properties of the ground state, already in the first papers 
it was pointed out that this fixed point should have a very important influence on the dynamical properties and possibly give rise to very slow dynamics. Very recently, this was explicitly shown in the context of the Many Body Localization transition (discussed by I. Aleiner at this school), see \cite{reviewMBL1,reviewMBL2}. This is certainly a research theme that will witness a lot of progress in the next few years. 

\chapter{Effective Temperatures}
The notion of effective temperatures is a recurrent theme in out of equilibrium physics. The notion of temperature, a single quantity that determines the state of the system, is so useful that researchers keep trying to find out whether it somehow holds also 
out of equilibrium. In this chapter we shall present several different notions of effective temperatures that have been found and 
discussed in the literature. 

\section{Metastable equilibrium}
The first notion of effective temperature we wish to discuss is also the closest one to the equilibrium temperature we are used to.
\begin{figure}[H]
\centerline{
\includegraphics[width=0.8\textwidth]{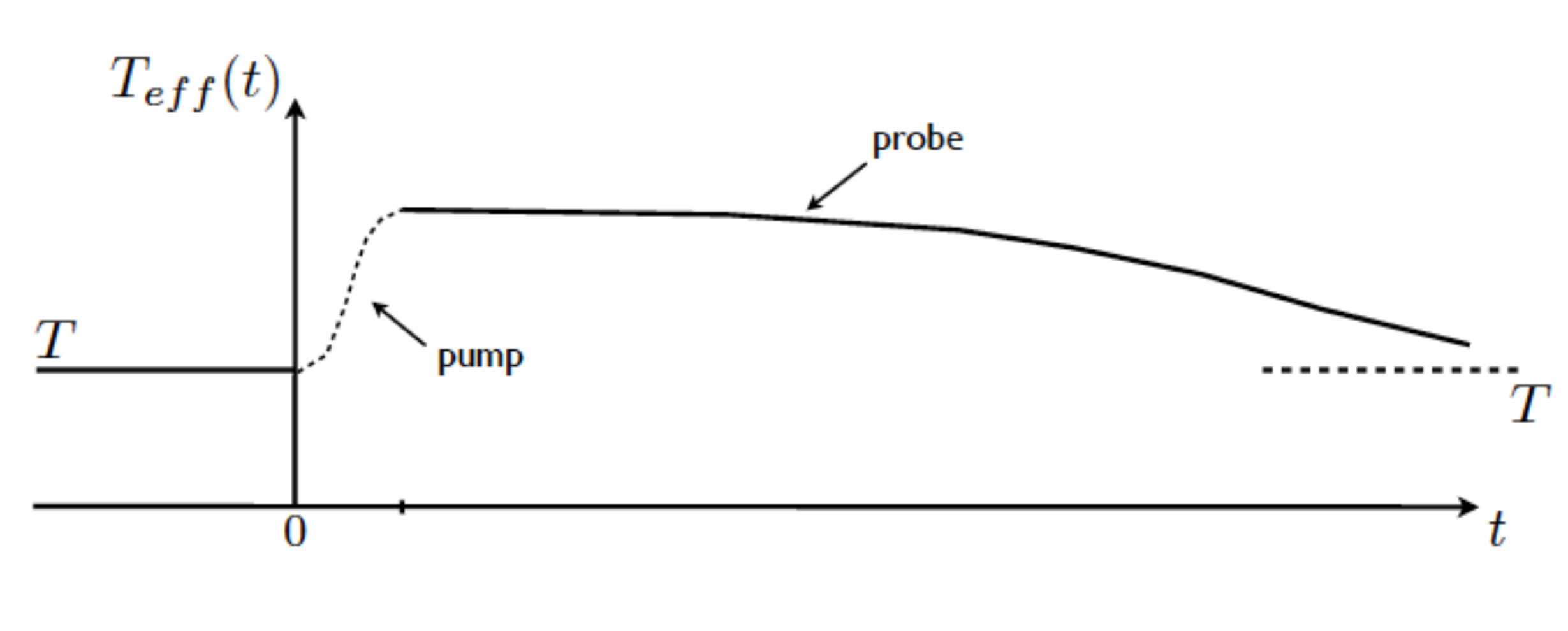}
}
\caption{Sketch of the evolution of the effective temperature of electrons in a pump and probe experiment. During the pump the system is truly out of equilibrium. After that, when is probed, electrons quickly reaches a metastable equilibrium with a well defined temperature that slowly evolve
toward the bath temperature on much larger time-scales.}
\label{fig:PP}
\end{figure}
Consider a system coupled to a thermal bath which is at temperature $T$. There are two important time-scales in the problem:
one is the relaxation time, $\tau_S$, of the degrees of freedom of the system in absence of the coupling to the bath, 
the other $\tau_B$ is the characteristic time over which the degrees of freedom of the bath evolve and exchange energy with the system.  By perturbing a system, initially at equilibrium at temperature $T$, one typically injects (or withdraws) energy from the system and puts it out of equilibrium. Subsequently, the system starts to evolve. If $\tau_S\ll \tau_B$ 
then on time-scale $\tau_S \ll t \ll \tau_B$ the system reaches a metastable equilibrium as if it were decoupled from the bath. 
Since energy is conserved---it is not exchanged on these time-scale---the system equilibrates to a temperature $T_S$ different from the one of the bath. This is the micro-canonical temperature corresponding to the new value of intensive energy reached after the perturbation.  
The effective temperature $T_S$ then slowly evolve and reaches $T$ only on larger time-scales, of the
order of $\tau_B$, see Fig. \ref{fig:PP}.\\
An example of this kind of situations was discussed by L. Perfetti at this school, see his lecture notes: the so-called pump and probe experiments that have been recently
the focus on an intense experimental research in the field of out of equilibrium hard condensed matter. In these cases
one pumps energy inside an electron system coupled to a phonon bath and then probe the following dynamics, for instance 
using a femto-second laser pulse. Since the characteristic time-scales of electrons and phonons are widely separated, $fs$ and $ ps$ respectively, electrons are indeed experimentally observed to equilibrate at temperature $T_S\gg T$. In Perfetti's notes, it is discussed an example in which electrons reach a meta-stable equilibrium state characterized by a Fermi-Dirac distribution at a temperature $T_S=2080K$ whereas the ambient temperature of the phonon bath is $T=130K$..

\section{Driven and Active Systems}
We focus now on systems permanently out of equilibrium. 
Some of them are generically coupled to an environment that provides dissipation and an external drive, for example 
magnetic systems through which an electronic current flow. Others constantly dissipate energy in order to evolve dynamically, 
like bacteria or other biological "degrees of freedom".    
These systems generically reach a steady state which cannot be described by equilibrium statistical mechanics means. There are a few cases however where despite the fact that
the system is driven or actively withdraws and then dissipates energy through the bath, a pseudo-equilibrium state with an effective temperature
is reached. We shall now explain how this comes about.\\
We take the general setting of quantum systems and consider the simplified case where all the effect of the environment and the driving can 
be condensed in additional self-energy terms, as we have already explained in chapter 2, see in particular eq. (\ref{SKFT}). In the absence of driving and if the environment is at equilibrium at temperature $T$ the additional self-energy terms must obey the quantum fluctuation-dissipation relation (\ref{QFDT}). For systems driven out of equilibrium there is no reason for the ratio $\nu(\omega)/\Im \eta(\omega)$ to acquire the QFDT form and, hence, no way to define an effective temperature. However, if the dynamics of the system takes place on time-scales much larger 
than the one typical of the drive and the environment, $\tau_E$, the only thing that matters is the behaviour of the additional self-energy terms at low frequency. Recalling that $\eta(\omega )$ and $\nu(\omega)$ entering in the Schwinger-Keldysh action (\ref{SKFT}) have the interpretation of 
a generalized friction and amplitude fluctuations respectively, it is not unreasonable that $\Im \eta(\omega)\simeq \eta_0\omega\tau_E$ and $\nu(\omega) \simeq \nu(0)$ for $\omega\ll 1/\tau_E$, as it indeed happens in many cases. If these small frequency behaviours hold then for $\omega \ll 1/\tau_E$:
\[
\frac{\nu(\omega)}{\Im \eta(\omega)}\rightarrow \frac{\nu(0)}{\eta_0 \omega\tau_E}=\frac{2T_{eff}}{\omega}
\]
This implies that the effect of the environment and the driving is {\it as if} the slow degrees of freedom of the system were only coupled to a classical thermal bath at temperature $T_{eff}$. As a theoretical example where this phenomenon happens we consider the model of two dimensional
itinerant magnet placed between two noninteracting leads studied by A. Mitra and A. Millis \cite{mitra}. 
\begin{figure}[H]
\centerline{
\includegraphics[width=0.8\textwidth]{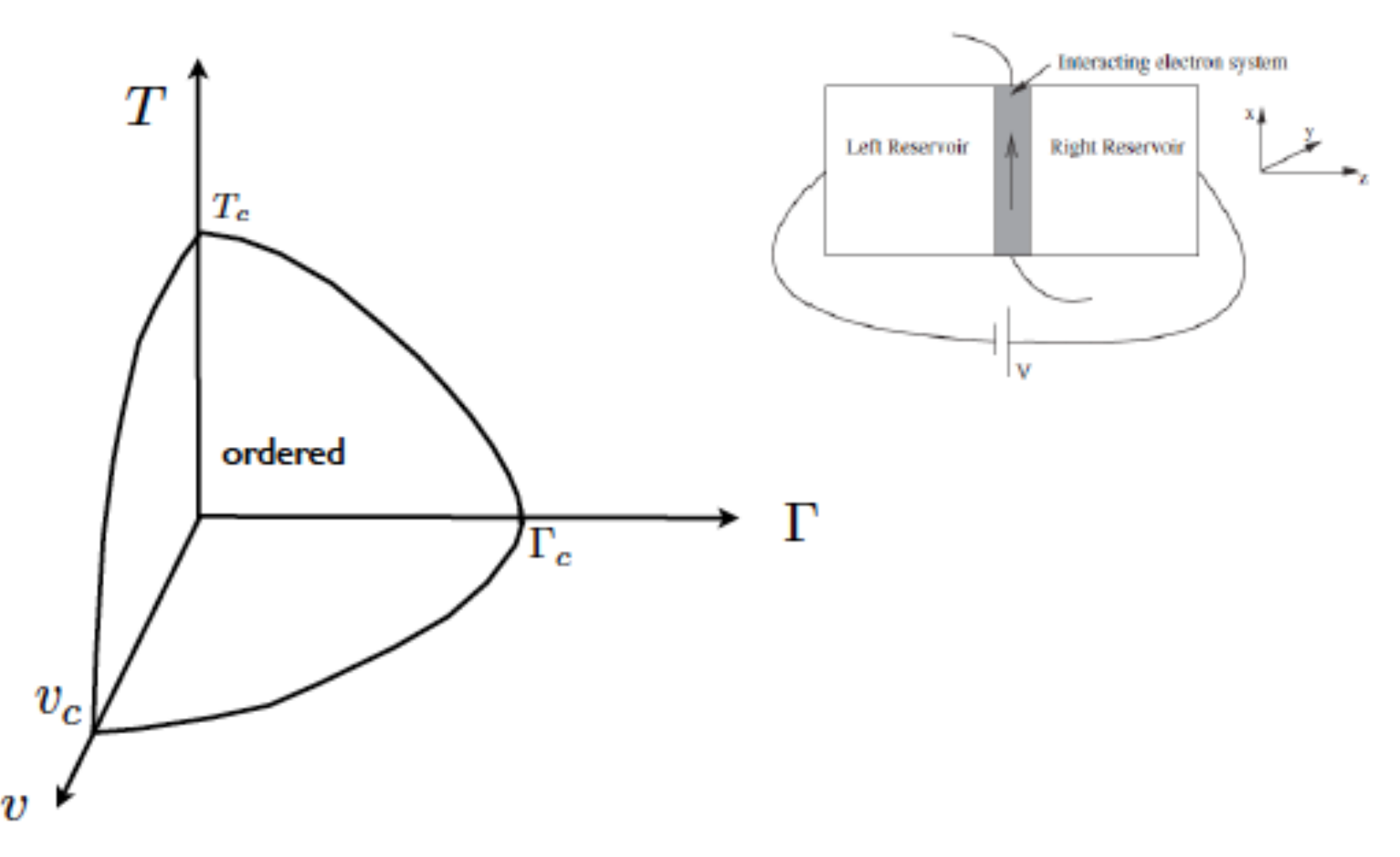}
}
\caption{Right: schematic view of itinerant magnet placed between two noninteracting leads (from [Mitra and Millis 2006]). Right: Schematic phase diagram.}
\label{fig:AM}
\end{figure}
The schematic phase diagram for this non-equilibrium system is shown in Fig. \ref{fig:AM}: it depicts a phase transition out of equilibrium between a magnetically ordered and a disordered state. 
A priori, this transition should be unrelated to the equilibrium one. Instead, the mechanism explained above induces a classical pseudo-equilibrium state for the slow critical degrees of freedom even when the quantum system is driven out of equilibrium by the voltage and is applied to a zero temperature bath: as far as universal quantities are concerned, voltage
acts as a temperature as explained in \cite{mitra}. In consequence the critical behaviour everywhere along the transition dome (except for $v=0, T=0$) corresponds to classical critical 
slowing down despite the fact that the system is strongly out of equilibrium. 
Similarly, studying the slow out of equilibrium dynamics obtained changing 
suddenly the voltage from $v_i>v_c$ to $v_f<v_c$, one finds that the system shows slow classical coarsening dynamics \cite{aron} even at zero temperature and in the regime of strong quantum fluctuations. 
The reason is again the one 
explained above: the slow degrees of freedom the domain walls in this case evolve as if the whole effect of the environment, the driving and the 
slow degrees of freedom was to provide a thermal bath at temperature $T_{eff}$.\\
Another interesting example is provided by active systems, a research topic that received recently a lot of attention in connection with the properties of living matter. In these cases, elementary degrees of freedom move because they withdraw and then dissipate energy from the environment. When the typical relaxation time-scales of the system become much longer then the bath ones a phenomenon identical to the one discussed above takes place and a well-defined effective temperature emerges \cite{BKTEFF}.  

\section{Glassy systems}

\begin{figure}[H]
\centerline{
\includegraphics[width=0.7\textwidth]{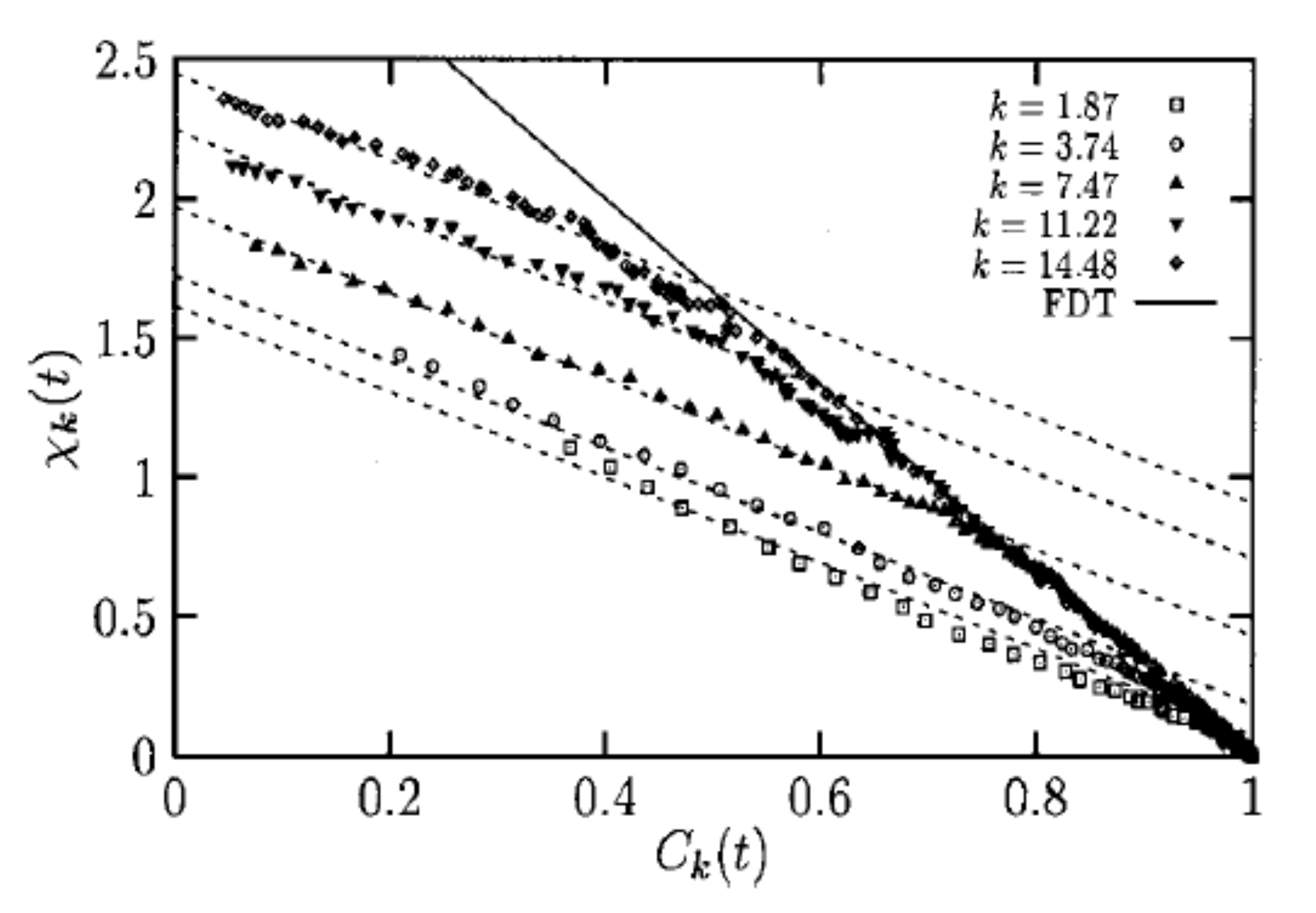}
}
\caption{Single particle density fluctuations at wave-vector $k$ in a sheared super-cooled liquids (from numerical simulations 
[Berthier and Barrat 2002]
). 
Remarkably, all different values of $k$ are characterized by the same value of $T_{eff}$.}
\label{fig:FDR}
\end{figure}

We end this chapter by making an exception to what stated in the introduction of the previous chapter and 
discussing briefly effective temperatures for glassy systems. Spin-glasses and structural glasses show an incredibly slow dynamics. 
After thermal quenches they display an ageing dynamics richer than the coarsening one discussed in the previous chapters. 
As for coarsening, the ageing dynamics is characterized by a separation of time-scales: short ones on which rapid degrees of freedom 
equilibrate and large ones on which the slow degrees of freedom display out of equilibrium behaviour. As we explained in chapter 2, one 
of the main signature of equilibrium dynamics is the fluctuation-dissipation relation (FDR) between correlation and response functions. The exact 
solution of the ageing dynamics of mean-field spin glasses \cite{cuku} suggested to look for generalisations of FDR out of equilibrium. 
Pragmatically one can define an effective temperature from the generalised FD relationship:
\begin{equation}\label{FR}
-\frac{1}{T_{eff}}\partial_t C(t,t')=R(t,t')
\end{equation}
where $C(t,t')$ and $R(t,t')$ are respectively the correlation and response function for a given observable. For an equilibrium system the previous relation is verified with $T_{eff}$ equal to the equilibrium temperature and $C,R$ that are just a function of $t-t'$. Instead, during ageing both functions depend on $t$ and $t'$ and using the previous equation to define $T_{eff}$ one generically finds an effective temperature that depends on $t,t'$.\\ It is remarkable that what can seem just a formal definition of temperature was shown in the theoretical study of spin-glasses and structural glasses to have a precise meaning at long times: $T_{eff}$ is the effective temperature of the slow degrees of freedom, the ones that display ageing dynamics \cite{cukupe}. Physically, this means that by probing the temperature of the system, one finds $T$ or $T_{eff}$, depending whether the
time-scale of the thermometer are short or long. 
This was obtained first in the study of mean-field glass models and then found in numerical simulations and a few experiments. 
This notion of effective temperature was generalised also to (slowly) driven glassy system, e.g. sheared super-cooled liquids,
where again it was found first analytically and then in simulations that slow degrees of freedom have an effective temperature different (generically higher) than the equilibrium one.  
In Fig. \ref{fig:FDR} we show the "classic" FDR plot in which the integrated response $\chi(t,t')=\int_t^{t'}ds R(s,t')$ is plotted parametrically as 
a function of $C(t,t')$. The standard equilibrium FDR would lead to $\chi=-\frac{1}{T}(C(t,t')-C(t,t))$. By normalising the equal time correlation function to one, in equilibrium we would then find a straight line with slope $-1/T$ crossing the $C$ axis at one. Instead, the result found in simulations, theory and a few experiments is the one shown in Fig. \ref{fig:FDR}: for large values of $C$ one indeed finds a slope $-1/T$, but below the value of $C$ corresponding to the plateau value in the plots of $C(t,t+\tau)$ one finds another slope equal to $-1/T_{eff}$. This indeed shows that 
the formal definition described above makes sense, since $T_{eff}$ depends on $t,t'$ only through $C$. The large values of $C$ corresponds to 
rapid degrees of freedom. These are the ones that relax from one to the plateau value in Fig. \ref{fig:corraging} and they are equilibrated at the bath temperature $T$ as expected. Instead the slow degrees of freedom, responsible for ageing, which correspond to values of $C$ lower than the plateau value are characterized by an effective temperature\footnote{Spin-glasses are more complicated from this point of view. They are characterized by a hierarchy of $T_{eff}$, one for each values of $C$, see \cite{leticialeshouches}.} which is higher than $T$. 

The theoretical explanation of this phenomenon cannot be presented here briefly. Hence, we refer to specific reviews on the subject \cite{leticialeshouches}. 
We conclude by mentioning that although there is a theoretical framework that suggests and explain effective temperature in glassy systems, 
and despite several simulations providing supporting evidences, numerical and experimental investigations are still going on in order to determine 
in a conclusive way whether this notion of effective temperature hold for realistic systems \cite{kurchannature}. The main issue is whether $T_{eff}$ really is observable independent, as a true temperature should be \cite{gnan}.


\begin{thebibliography}{99}

 \bibitem[Amit and Martin-Mayor 2005]{bookRG}
D. J. Amit and V. Martin-mayor, {\it Field Theory, the Renormalization Group, And Critical Phenomena: Graphs To Computers},
World Scientific (2005).


\bibitem[Aron et al. 2009]{aron}
C. Aron, G. Biroli, L. F. Cugliandolo, 
Phys. Rev. Lett. {\bf 102}, 050404 (2009). 

\bibitem[Aron et al. 2010]{aronS}  
C. Aron, G. Biroli and L.F. Cugliandolo, J.Stat.Mech.1011:P11018 (2010)

\bibitem[Balog and Tarjus 2015]{tarjusbalog}
I. Balog, G. Tarjus, ArXiv:1501.05770, {\it 
Activated dynamic scaling in the random-field Ising model: a nonperturbative functional renormalization group approach}.



\bibitem[Barthel and U. Schollw\"ock 2008]{BS} T. Barthel and U. Schollw\"ock, Phys. Rev. Lett. {\bf 100}, 100601
(2008).

\bibitem[Basko et al. 2006]{reviewMBL1}
D.M. Basko, I. Aleiner, B.A. Altshuler, Annals of Physics {\bf 321} (2006) 1126.


\bibitem[Berthier and Barrat 2002]{BB}
L. Berthier, J.-L. Barrat, 
J. Chem. Phys. {\bf 116}, (2002).

\bibitem[Berthier and Kurchan 2013]{BKTEFF}
L. Berthier, J. Kurchan, Nature Physics {\bf 9} 310 (2013). 


\bibitem[Biroli and Kurchan 2001]{biku} Biroli G. and Kurchan J. (2001), Phys. Rev.
E 64, 016101.


\bibitem[Biroli and Parcollet 2002]{biroliparcollet}
G. Biroli and O. Parcollet, Physical Review B, {\bf 65}, 094414 (2002).


\bibitem[Biroli et al. 2008]{birolikollath}
G. Biroli, C. Kollath, A.M. L\"auchli. Phys. Rev. Lett. {\bf 105} (25), 250401.


\bibitem[Biroli et al. 2010]{sicilia}
G. Biroli, L.F. Cugliandolo, A. Sicilia,
Phys. Rev. E {\bf 81}, 050101(R) (2010).


\bibitem[Sciolla and Biroli 2013]{DT}
B. Sciolla and G. Biroli, Phys. Rev. B 201110R (2013).


\bibitem[Bovier et al. 2000]{bovier}Bovier A, Eckhoff M, Gayrard V and Klein, M (2000);
J. Phys. A. Math. Gen 33   L447;
Comm. Math. Phys.  228  219 (2002);
Journal. Europ. Math Soc.  6  399 (2004).

\bibitem[Bray and Rodgers 1988]{braygriffiths}
A. J. Bray and G. J. Rodgers, Phys. Rev. B {\bf 38}, 9252 (1988).


\bibitem[Bray 2002]{bray}
A. J. Bray, Adv. Physics {\bf 51} 481 (2002).

\bibitem[Calabrese and Gambassi 2005]{calabresegambassi}
P. Calabrese, A. Gambassi
Journal of Physics A: Mathematical and General {\bf 38}, R133 (2005).


\bibitem[Cao and Machta 1993]{MKRFIM}
M. S. Cao and J. Machta, Phys. Rev. B {\bf 48}, 3177 (1993).

\bibitem[Cardy]{cardybook} J. Cardy, {\it Scaling and Renormalization in Statistical Physics},
Cambridge Lecture Notes in Physics 1996.

\bibitem[Casetti 1997]{casetti}
L. Casetti, M. Cerruti-Sola, M. Pettini, and E. G. D. Cohen
Phys. Rev. E {\bf 55}, 6566 (1997)

\bibitem[Castiglione et al. 2008]{vulpiani} P. Castiglione, M. Falcioni, A. Lesne, A. Vulpiani, {\it Chaos and Coarse Graining in Statistical Mechanics},
Cambridge Univ. Press 2008.

\bibitem[Crooks 1998]{Crooks} Crooks G.E. (1998), J. Stat. Phys. 90, 1481.



\bibitem[Cugliandolo 2002]{leticialeshouches}
L.F. Cugliandolo, {\it Dynamics of Glassy Systems}, 	Lecture notes, Les Houches, July 2002. 



\bibitem[Cugliandolo and Kurchan 1994]{cuku}
L. F. Cugliandolo, J. Kurchan, J. Phys. A {\bf 27}, 5749 (1994).

\bibitem[Cugliandolo et al. 1997]{cukupe}
L. F. Cugliandolo, J. Kurchan, L. Peliti
Phys. Rev. E {\bf 55}, 3898 (1997).


\bibitem[Deutsch 1991]{DR1}J. M. Deutsch, Phys. Rev. A {\bf 43}, 2046 (1991).



\bibitem[Doyon and Andrei 2006]{doyon}
B. Doyon and N. Andrei
Phys. Rev. B {\bf 73} 245326 (2006).

\bibitem[Esposito et al. 2009]{esposito} M. Esposito, U. Harbola, and S. Mukamel,
Rev. Mod. Phys. {\bf 81} 1665 (2009).

\bibitem[Fisher 1986]{fisher2}D. S. Fisher,
Phys. Rev. Lett. {\bf 56}, 416 (1986).

\bibitem [Gardiner and Zoller 2000]{gardiner} C.W. Gardiner, P. Zoller, {\it Quantum noise}, Springer, Berlin 2000.    

\bibitem[Gaveau and Schulman 1998]{gaveauschulman}  Gaveau B. and  Schulman L.S. (1998), { Jour. Math. Phys}
{\bf  39}  1517;  Gaveau B. and  Schulman L.S. (1996), {\ Jour. Math. Phys}
{\bf  37}, 3897;  Phys. Lett.  A229 347.
 

\bibitem[Giamarchi 2009]{reviewgiamarchi}
T. Giamarchi, Encyclopedia of Complexity and Systems Science, 2019 (2009).

\bibitem[Gozzi 1984]{gozzi}
E. Gozzi Phys.Lett. B {\bf 143} 183 (1984).

\bibitem[Griffiths 1969]{griffiths}
R.B.Griffiths, Phys. Rev. Lett. {\bf 23}, 17 (1969).

\bibitem[Henley 2004]{RK2} C. Henley, J. Physics. Cond Matt. {\bf 16} S891 (2004).

\bibitem[Jarzynski 1997]{Jarzynski}  Jarzynski C. (1997), 
Phys. Rev. Lett. 78, 2690; \\
Phys. Rev. E 56, 5018.

\bibitem[Kamenev 2011]{kamenev} A. Kamenev (2011), 
{\it Field theory on non-equilibrium systems}, Cambridge U. Press.
  
\bibitem[Kockelkoren and Chat\'e 2002]{chate}
J. Kockelkoren, H. Chat\'e, Physica D {\bf 168} 80 (2002).  
  
\bibitem[Kurchan 2005]{kurchannature}
J. Kurchan, Nature {\bf 433}, 222 (2005)

\bibitem[Kurchan 2008]{kurchan} J. Kurchan (2008)
{\it Six out of equilibrium lectures}, Summer School in Les Houches. 
  
  
  \bibitem[Leggett et al. 1987]{caldeiraleggett}
A. J. Leggett, S. Chakravarty, A. T. Dorsey, M. P. A. Fisher, A. Garg and W. Zwerger, Rev.
Mod. Phys. {\bf 59}, 1 (1987).

\bibitem[Mahan 2000]{mahan} G.D. Mahan (2000),
{\it Many Particle Systems}, Springer. 

\bibitem[Martinelli 1998]{martinelli}
F. Martinelli, Journal of Statistical Physics {\bf 92} 337 (1998).

\bibitem[Mazur and Montroll 1960]{mazur} P. Mazur and E. Montroll, J. Math. Phys. {\bf 1}, 70 (1960).


\bibitem[Middleton and Fisher 2002]{middleton}
A. A. Middleton and D. S. Fisher,
Phys. Rev. B {\bf 65}, 134411 (2002).

\bibitem[Mitra and Millis 2006]{mitra}
A. Mitra, S. Takei, Y. B. Kim, A.J. Millis, Phys. Rev. Lett. {\bf 97}, 236808 (2006); 
A. Mitra, A. J. Millis
Phys. Rev. B {\bf 77} 220404 (2008); A. Mitra, A. J. Millis
Phys. Rev. B {\bf 84} 054458 (2011).

\bibitem[Nandkishore and Huse 2014]{reviewMBL2}
R.Nandkishore, D. A. Huse, arXiv:1404.0686.

\bibitem[Ninarello et al. 2014]{gnan}
A. S. Ninarello, N. Gnan, and F. Sciortino, 
J. Chem. Phys. {\bf 141}, 194507 (2014).

\bibitem[Rigol et al. 2008]{rigol} M. Rigol, V. Dunjko, and M. Olshanii, Nature {\bf 452}, 854 (2008). 


\bibitem[Rokhsar and Kivelson 1988]{RK}
D. Rokhsar and S. Kivelson Phys. Rev. Lett. {\bf 61} 2376 (1988). 



\bibitem[Sachdev 2011]{sachdev}
S. Sachdev, {\it Quantum Phase Transitions}, Cambridge Univ. Press 2011. 


\bibitem[Seifert 2008]{seifert} U. Seifert,
Eur. Phys. J. B,  423 {\bf 64} 2008. 


\bibitem[Sicilia et al. 2007]{cugliandolobray}
A. Sicilia, J. J. Arenzon, A. J. Bray, L. F. Cugliandolo,
Phys. Rev. Lett. {\bf 98}, 145701 (2007); Phys. Rev. E {\bf 76}, 061116 (2007)



\bibitem[Srednicki 1994]{DR2} M. Srednicki, Phys. Rev. E {\bf 50}, 888 (1994).



\bibitem[Tissier and Tarjus 2012]{TarjusRFIM}
M. Tissier, G. Tarjus
Phys. Rev. B {\bf 85}, 104203 (2012)

\bibitem[Villain 1984]{fisher}
J. Villain, Phys. Rev. Lett. {\bf 52}, 1543 (1984).




\bibitem[Weiss 2008] {weiss}U. Weiss, {\it Quantum dissipative systems}, World Scientific, Singapore 2008.

\bibitem[Wiese and Le Doussal 2007]{reviewgiamarchi2}
K. J. Wiese, P. Le Doussal, Markov Processes Relat. Fields {\bf 13} 777 (2007). 

\bibitem[Zinn-Justin 1996]{Zinn} Zinn-Justin J. (1996),
 {\it Quantum field theory and critical phenomena}, Oxford U. Press.
 

\bibitem[Zurek 1996]{KZ1}
W.H. Zurek, Phys. Rep. {\bf 276} 177 (1996)


\bibitem[Zwanzig 2001]{zwanzig} R. Zwanzig, {\it Nonequilibrium statistical mechanics}, Oxford University Press, Oxford 2001.

























\end{thebibliography}
\end{document}